\documentclass[amsmath,superscriptaddress,showpacs,prb,twocolumn,longbibliography]{revtex4-1}
\usepackage{bbm}
\usepackage{bm}
\usepackage{enumerate}
\usepackage{graphicx}
\usepackage[dvips]{epsfig}
\usepackage{epsf}
\usepackage[latin1]{inputenc}
\usepackage{amsmath}
\usepackage{amsfonts}
\usepackage{amssymb}

\usepackage{xcolor}

\usepackage{units}

\bibpunct{[}{]}{,}{n}{}{}

\makeatletter
\newcommand{\Rmnum}[1]{\expandafter\@slowromancap\romannumeral #1@}
\makeatletter

\begin{document}
\title{Magnonic Analogue of Edelstein Effect in Antiferromagnetic Insulators}
\author{Bo Li}
\affiliation{Department of Physics and Astronomy and Nebraska Center for Materials and Nanoscience, University of Nebraska, Lincoln, Nebraska 68588, USA}

\author{Alexander Mook}
\affiliation{Institut f\"ur Physik, Martin-Luther-Universit\"at Halle-Wittenberg, D-06099 Halle (Saale), Germany}
\affiliation{Department of Physics, University of Basel, Klingelbergstrasse 82, CH-4056 Basel, Switzerland}

\author{Aldo Raeliarijaona}
\affiliation{Department of Physics and Astronomy and Nebraska Center for Materials and Nanoscience, University of Nebraska, Lincoln, Nebraska 68588, USA}

\author{Alexey A. Kovalev}
\affiliation{Department of Physics and Astronomy and Nebraska Center for Materials and Nanoscience, University of Nebraska, Lincoln, Nebraska 68588, USA}
\begin{abstract}
We investigate the nonequilibrium spin polarization due to a temperature gradient in antiferromagnetic insulators, which is the magnonic analogue of the inverse spin-galvanic effect of electrons. We derive a linear response theory of a temperature-gradient-induced spin polarization for collinear and noncollinear antiferromagnets, which comprises both extrinsic and intrinsic contributions. 
We apply our theory to several noncentrosymmetric antiferromagnetic insulators, i.e., to a one-dimensional antiferromagnetic spin chain, a single layer of kagome noncollinear antiferromagnet, e.g., $\text{KFe}_3(\text{OH})_6(\text{SO}_4)_2$, and a
noncollinear breathing pyrochlore antiferromagnet, e.g., LiGaCr$_4$O$_8$. The shapes of our numerically evaluated response tensors agree with those implied by the magnetic symmetry. Assuming a realistic temperature gradient of $\unit[10]{K/mm}$, we find two-dimensional spin densities of up to $\sim \unit[10^6]{\hbar/cm^2}$ and three-dimensional bulk spin densities of up to $\sim \unit[10^{14}]{\hbar/cm^3}$, encouraging an experimental detection.
\end{abstract}
\maketitle

\noindent

\section{Introduction}
Generation of nonequilibrium spin imbalance is increasingly important for the current spintronics research~\cite{RevModPhys.76.323}, especially in the context of nonequilibrium torques~\cite{RevModPhys.91.035004}. In metallic and semiconductor materials, spin-orbit coupling (SOC) facilitates the interplay between the orbital and spin degrees of freedom, which allows feasible electric manipulation of spins, e.g., for technological applications. One consequence of such coupling is the
inverse spin-galvanic effect~\cite{aronov1989nuclear,EDELSTEIN1990233,Ganichev} which attracted considerable interest in recent years~\cite{PhysRevB.75.155323,doi:10.1063/1.1833565,PhysRevLett.96.186605,PhysRevLett.97.126603,Gambardella3175,PhysRevB.67.033104,PhysRevLett.112.096601,PhysRevLett.112.096601,PhysRevB.89.245443,PhysRevB.93.195440,PhysRevB.95.205424,Ganichev2002,PhysRevLett.104.146802}. The nonequilibrium spin polarization contains an extrinsic part dependent on the transport relaxation time and an intrinsic part independent of the relaxation time~\cite{RevModPhys.91.035004}, and it can lead to spin-orbit torques. Both field-like and damping-like spin-orbit torques can arise due to the nonequilibrium spin polarization at interfaces between magnetic and nonmagnetic materials~\cite{Chernyshov.OverbyNP2009,MihaiMiron.GaudinNM2010,Miron.Garello2011,Pesin2012,Qaiumzadeh.DuinePRB2015,Ado.TretiakovPRB2017,PhysRevMaterials.3.011401}.

In ferromagnetic and antiferromagnetic insulators, magnons -- the quantum quasiparticles carrying energy and spin -- can mediate various transport phenomena. The Dzyaloshinskii-Moriya interaction (DMI) \cite{Dzyaloshinsky58, Moriya60} in such systems can lead to magnon spin-momentum locking \cite{PhysRevLett.119.107205}, magnon-mediated magnetization torques~\cite{PhysRevB.90.224403,PhysRevB.93.161106,PhysRevB.95.165106}, and magnonic thermal Hall~\cite{PhysRevLett.104.066403,Onose297,PhysRevLett.106.197202,PhysRevB.84.184406,PhysRevB.89.054420,PhysRevB.98.094419,PhysRevB.99.054409,PhysRevB.95.014422,PhysRevLett.121.097203,PhysRevB.99.014427,PhysRevB.94.174444,PhysRevLett.122.057204} and spin Nernst effects~\cite{PhysRevB.93.161106,PhysRevLett.117.217203,PhysRevLett.117.217202,PhysRevB.96.134425,2019arXiv190710567L,PhysRevB.100.100401,PhysRevB.98.035424,PhysRevB.97.174407,PhysRevB.97.140401,PhysRevB.96.224414,PhysRevB.99.224433}. In~Ref.~\cite{2019arXiv190710567L}, two of us speculated about a possibility of magnon-mediated spin polarization in insulating antiferromagnets lacking inversion symmetry. 

In this work, we study the magnonic analogue of the Edelstein effect by considering  antiferromagnetic insulators. The spin nonconservation in such systems can be caused by noncollinear spin order or anisotropic exchange. We consider a linear response to the temperature gradient replicated by a pseudo gravitational potential~\cite{PhysRev.135.A1505} in the magnon Hamiltonian. The final result for the magnonic spin polarization is separated into the extrinsic and intrinsic contributions. We apply our theory to several models and discuss relevant material candidates. In 1D, an antiferromagnetic spin chain with anisotropic nearest exchange and Rashba-like DMI serves as a toy model exhibiting both intrinsic and extrinsic contributions to the magnonic analogue of the Edelstein effect. In 2D and 3D, we concentrate on realistic noncollinear antiferromagnets on the kagome and breathing pyrochlore lattices. From the magnetic point group, we establish the response tensor shapes which agree with our numerical results.

The paper is organized as follows. In Sec.~\ref{sec:HamiltonianAndEigenstates}, we discuss the Holstein-Primakoff transformation of magnons in noncollinear antiferromagnets, introduce spin density operator for magnons, and discuss the diagonalization procedure. In Sec.~\ref{sec:LinearResponseTheory}, we derive the expression for the magnonic spin polarization as a linear response to the temperature gradient. In Sec.~\ref{sec:SymmetryConstraints}, we discuss the symmetry constraints on the response tensor. In Sec.~\ref{sec:Models}, we apply our theory to an antiferromagnetic spin chain and to noncollinear antiferromagnets on the kagome and breathing pyrochlore lattices. We also estimate the nonequilibrium spin density using real material parameters. In Sec.~\ref{sec:ComputerExperiments}, we perform atomistic spin dynamics simulations and compare with our results from the previous section. Finally, we conclude our discussion in Sec.~\ref{sec:Conclusion} with a summary and an outlook. Appendices contain more detailed information about our derivations.

\section{Hamiltonian and Eigenstates}
\label{sec:HamiltonianAndEigenstates}

We consider a general Hamiltonian of the form:
\begin{equation}\label{eq:GeneralHamiltonian}
    H=\sum_{i,j} \left[ J_{ij}^{\alpha\beta} S^\alpha_i S^\beta_j+\mathbf{D}_{ij}\cdot(\mathbf{S}_i\times\mathbf{S}_j) \right] + \sum_i H_i,
\end{equation}	
where $i,j$ label different lattice sites and $\alpha$, $\beta$ stand for different spin vector components, i.e., $x,y,z$. Moreover, $J_{ij}^{\alpha\beta}$ is the symmetric exchange energy between $\alpha$, $\beta$ spin components on two sites $i$ and $j$, while antisymmetric exchange is described by the DMI vector $\mathbf{D}_{ij}$ between spins on sites $i$ and $j$. Effects of the on-site anisotropy and magnetic field may also be included in our analysis via the last term, $H_i=K_i(\mathbf{S}_i\cdot\hat{n}_i)^2+(\mathbf{S}_i\cdot \mathbf{B})$.

By performing the Holstein-Primakoff transformation~\cite{Holstein1940} in the limit of large $S$, we obtain up to the leading order, $S_i^{x} \approx \sqrt{\frac{S_i}{2}}(a_i^\dagger +  a_i)$, $S_i^{y} \approx \mathrm{i} \sqrt{\frac{S_i}{2}} (a_i^\dagger - a_i)$, and $S^z_i=S_i-a_i^\dagger a_i$. Keeping the leading order terms, we obtain the bilinear Hamiltonian written in magnon particle-hole space as
\begin{equation}\label{eq:Hrealspace}
    H=\frac{1}{2}\int d\mathbf{r}\Psi^\dagger(\mathbf{r})\mathcal{H}\Psi(\mathbf{r}),
\end{equation}
where $\Psi(\mathbf{r})=(a_1(\mathbf{r}), \ldots,  a_N(\mathbf{r}), a_1^\dagger(\mathbf{r}), \ldots, a_N^\dagger(\mathbf{r}))^T$, with $N$ being  the number of atoms in each unit cell. The corresponding  Hamiltonian in the momentum space is
\begin{equation}\label{eq:Hmomentumspace}
H=\frac{1}{2}\sum_\mathbf{k}\Psi^\dagger_\mathbf{k}\mathcal{H}_\mathbf{k}\Psi_\mathbf{k},
\end{equation}
where $\Psi_\mathbf{k}=(a_{1,\mathbf{k}}, \ldots, a_{N,\mathbf{k}}, a^\dagger_{1,-\mathbf{k}}, \ldots, a^\dagger_{N,-\mathbf{k}})^T$. The Hamiltonian can be diagonalized by the paraunitary transformation matrix $T_\mathbf{k}$,
\begin{equation}\label{eq:Diagonalization}
T_{\mathbf k}^\dagger {\mathcal H}_\mathbf{k}T_\mathbf{k}=\mathcal{E}_\mathbf{k},
\end{equation}
where the eigenenergy matrix contains all eigenvalues $\mathcal{E}_\mathbf{k}=\text{Diag}(\varepsilon_{1,\mathbf{k}}, \ldots, \varepsilon_{N,\mathbf{k}}, \varepsilon_{1,-\mathbf{k}} ,\ldots,  \varepsilon_{N,-\mathbf{k}})$. The transformation matrix satisfies the paraunitary normalization relations, $
T_\mathbf{k}^\dagger \sigma_3 T_\mathbf{k}=\sigma_3$
and $T_\mathbf{k} \sigma_3 T^\dagger_\mathbf{k}=\sigma_3$, where here and henceforth $\sigma_i$ ($i=1,2,3$) stands for the Pauli matrices acting in the particle-hole space. The particle-hole space Hamiltonian can be regarded as a pseudo-Hermitian Hamiltonian, with the eigenequation \cite{PhysRevX.7.041045}
\begin{eqnarray}\label{eq:Eigenequation}
  &&\sigma_3 H_\mathbf{k}|u_{n,\mathbf{k}}\rangle=\bar{\varepsilon}_{n,\mathbf{k}}|u_{n,\mathbf{k}}\rangle,
  \end{eqnarray}  
where $|u_{n,\mathbf{k}}\rangle_l=(T_\mathbf{k})_{ln}$, which satisfies the normalization relation $\langle u_{n,\mathbf{k}}|\sigma_3|u_{m,\mathbf{k}}\rangle=(\sigma_3)_{nm}$.
Moreover, the magnon basis possesses the particle-hole symmetry (PHS) $\Psi^\dagger_{\mathbf{k}}=(\sigma_1\Psi_{-\mathbf{k}})^T$ so that the Hamiltonian obeys $\sigma_1 H_\mathbf{k}\sigma_1=H_{-\mathbf{k}}^\ast$, which leads to $\bar{\varepsilon}_{n+N,\mathbf{k}}=-\bar{\varepsilon}_{n,-\mathbf{k}}$ and $|u_{n,\mathbf{k}}\rangle=e^{i\phi_n}\sigma_1|u_{n+N,-\mathbf{k}}\rangle^\ast$, where $\phi_n$ is a redundant phase factor. 

 The spin density of magnons calculated up to quadratic terms leads to spin density matrix  $\hat{S}_\mu=-\frac{1}{2}\sigma_0\otimes\text{Diag}(\left< S^\mu_1\right>/S_1, \cdots ,\left< S^\mu_N\right>/S_N)$, where $\mu=x,y,z$, $\sigma_0$ describes the particle-hole space, and averages of spins (in general different form each other) within a unit cell have been taken in equilibrium.  
The spin density operator of magnons up to quadratic terms becomes 
\begin{eqnarray}\label{eq:Spindensity}
    S_\mu=\frac{1}{V}\sum_\mathbf{k}\Psi_\mathbf{k}^\dagger \hat{S}_\mu\Psi_\mathbf{k}.
\end{eqnarray}
 We note that PHS implies an equality $\langle u_{n,\mathbf{k}}|\hat{S}_\mu|u_{n,\mathbf{k}}\rangle=\langle u_{n+N,-\mathbf{k}}|\hat{S}_\mu|u_{n+N,-\mathbf{k}}\rangle$.

 \section{Linear Response Theory}
 \label{sec:LinearResponseTheory}

In this section, we perform linear response calculations of the nonequilibrium spin density with respect to a temperature gradient $\nabla_\nu T$, i.e.,
\begin{eqnarray}
    \label{eq:Totalresponse}
    \langle S_\mu\rangle =\chi_{\mu\nu} \nabla_\nu T =\left(\chi^\mathrm{ex}_{\mu\nu}+\chi^\mathrm{in}_{\mu\nu} \right) \nabla_\nu T,
\end{eqnarray}
where  we separated the response tensor $\chi_{\mu \nu}$ into extrinsic, $\chi^\mathrm{ex}_{\mu\nu}$, and intrinsic, $\chi^\mathrm{in}_{\mu\nu}$, parts.

We introduce a perturbation corresponding to a pseudo-gravitational potential $\phi(\mathbf{r})$: 
\begin{equation}\label{eq:pseudograviation}
    H^\prime=\frac{1}{4}\int d\mathbf{r}\Psi^\dagger(\mathbf{r})(\mathcal{H}\phi(\mathbf{r})+\phi(\mathbf{r})\mathcal{H})\Psi(\mathbf{r}),
\end{equation}
where $\phi(\mathbf{r})=-T(\mathbf{r})/T$. Up to the linear order, the spatial gradients of this potential replicate the presence of the temperature gradient in the system.
In addition,  the pseudo-gravitational potential also amends
the spin density operator \cite{PhysRev.135.A1505, PhysRevB.99.024404}. This can be seen by considering a response to magnetic field in the presence of perturbation \eqref{eq:pseudograviation}. The total macroscopic spin density operator becomes
\begin{eqnarray}\label{eq:Spincorrection}
    S_\mu=\frac{1}{V}\int d\mathbf{r}\Psi^\dagger (\mathbf{r})
    \left(\hat{S}_\mu+\frac{\phi}{2} \hat{S}_\mu+\hat{S}_\mu\frac{\phi}{2} \right) \Psi(\mathbf{r}).
\end{eqnarray}
Thus, the nonequilibrium spin density contains two parts:
\begin{eqnarray}\label{eq:Totalspin}
    \langle S_{\mu}\rangle_\mathrm{tot}=\langle S_\mu\rangle_\mathrm{neq}+\langle \delta S_\mu\rangle_\mathrm{eq} = (K_{\mu\nu}+D_{\mu\nu}) \nabla_{\nu}\phi,
\end{eqnarray}
where the term proportional to $K_{\mu\nu}$ corresponds to the unperturbed spin density operator and it can be calculated within the Kubo linear response formalism. The dipole contribution, $D_{\mu\nu}$, is evaluated with respect to the equilibrium state as it originates from the correction to the spin density in Eq.~ \eqref{eq:Spincorrection} containing the temperature gradient.

We first calculate $K_{\mu\nu}$ within the Kubo linear response formalism~\cite{PhysRevB.93.161106, PhysRevLett.117.217203} in which the spin accumulation is given by  
\begin{equation}\label{eq:Linearresponse}
	\langle S_\mu\rangle_{\text{neq}}=\lim_{\omega\to 0}\frac{1}{i\omega}[\Pi_{\mu\nu}(\omega)-\Pi_{\mu\nu}(0)]\nabla_\nu\phi,
\end{equation}
where
\begin{align}
    \Pi_{\mu\nu}(i\omega_m)=-\int_{0}^{1/{k_BT}}d\tau e^{i\omega_m\tau}\langle T_\tau S_\mu(\tau) J_\nu^q(0)\rangle,    
\end{align}
and $\omega_m$ is the bosonic Matsubara frequency. The $\nu$ component of the macroscopic heat current, $J_\nu^q = \frac{1}{V} \int \mathrm{d} \mathbf{r} j^q_\nu(\mathbf{r})$, is derived from the heat current density $\boldsymbol{j}^q=\frac{1}{4}\Psi^\dagger(\mathbf{r})(\mathcal{H}\sigma_3\mathbf{v}+\mathbf{v}\sigma_3\mathcal{H})\Psi(\mathbf{r})$, with velocity $\mathbf{v}=i[H,\mathbf{r}]$. The heat current density can be inferred from the continuity equation, i.e., $\dot{\rho}_E+\bm{\nabla}\cdot\boldsymbol{j}^q=0$, with $\rho_E$ being the energy density of the system. In Appendix~\ref{sec:AppLinRep}, we provide the detailed calculation of the response tensor $K_{\mu\nu}$ divided into intraband and interband parts: $K_{\mu\nu}=K^{\text{intra}}_{\mu\nu}+K^{\text{inter}}_{\mu\nu}$, whose explicit forms read
\begin{eqnarray}
    &&K_{\mu\nu}^{\text{intra}}=\frac{1}{V}\sum\limits_\mathbf{k}\sum_{n=1}^{2N}\frac{1}{\Gamma_n}(\mathcal{J}_{\nu,\mathbf{k}})_{nn}(\mathcal{S}_{\mu,\mathbf{k}})_{nn}\partial_\varepsilon n_\mathrm{B}[\bar{\varepsilon}_{\mathbf{k},n}],
 \label{eq:Intraband}\\
    && K_{\mu\nu}^{\text{inter}}=\frac{4}{V}\sum\limits_\mathbf{k}\sum\limits_{m\neq n}\frac{\text{Im}[(\sigma_3\mathcal{S}_{\mu,\mathbf{k}})_{nm}(\sigma_3\mathcal{J}_{\nu,\mathbf{k}})_{mn}]n_\mathrm{B}[\bar{\varepsilon}_{\mathbf{k},n}]}{(\bar{\varepsilon}_{\mathbf{k},n}-\bar{\varepsilon}_{\mathbf{k},m})^2},\label{eq:Interband}\nonumber\\
\end{eqnarray}
where $n_\mathrm{B}(x)=1/(e^{x/k_BT}-1)$ is the Bose-Einstein distribution function, and
we used notations $\mathcal{S}_{\mu,\mathbf{k}}=T^\dagger_\mathbf{k}\hat{S}_{\mu}T_\mathbf{k}$, $\bm{\mathcal{J}}_{\mathbf{k}}=T^\dagger_\mathbf{k}\mathbf{J}_{\mathbf{k}}^qT_\mathbf{k}$, and $\mathbf{J}_{\mathbf{k}}^q=\frac{1}{4}(\mathcal{H}_\mathbf{k}\sigma_3\mathbf{v}_{\mathbf{k}}+\mathbf{v}_{\mathbf{k}}\sigma_3\mathcal{H}_\mathbf{k})$ with $\mathbf{v}_{\mathbf{k}}=\frac{\partial \mathcal{H}_\mathbf{k}}{\partial\mathbf{k}}$. As can be seen from Eq.~\eqref{eq:Intraband}, the phenomenological spectrum broadening, given by $\Gamma_n$, is crucial for the intraband component, whereas it does not enter the intrinsic contribution.
Plugging $\mathcal{J}_{\nu,\mathbf{k}}=\frac{1}{4}(\mathcal{E}_\mathbf{k}\sigma_3\tilde{v}_{\nu,\mathbf{k}}+\tilde{v}_{\nu,\mathbf{k}}\sigma_3\mathcal{E}_\mathbf{k})$ with $\tilde{v}_{\nu,\mathbf{k}}=T_\mathbf{k}^\dagger v_{\nu} T_\mathbf{k}$ into Eq.~\eqref{eq:Intraband} (see details in Appendix~\ref{sec:AppLinRep}), we obtain the intraband (extrinsic) response tensor:
\begin{eqnarray}\label{eq:Intraband1}
    \chi^\mathrm{ex}_{\mu\nu}
    =\frac{1}{VT}
    \sum_{\mathbf{k},n=1}^N
    \frac{1}{\Gamma_n}(\mathcal{S}_{\mu,\mathbf{k}})_{nn}
    v_{n\mathbf{k},\nu}
    \varepsilon_{n,\mathbf{k}} 
    \left[-\frac{\partial n_\mathrm{B}(\varepsilon_{n,\mathbf{k}})}{\partial\varepsilon} \right].
\end{eqnarray}
This result can be also obtained from the Boltzmann transport theory with the relaxation time $\tau_n=1/(2\Gamma_n)$.
The interband contribution in Eq.~\eqref{eq:Interband} can be reorganized as
\begin{eqnarray}\label{eq:Interband1}
    K_{\mu\nu}^{\text{inter}}=\frac{1}{V}
    \sum_{\mathbf{k},n=1}^{2N}
    \left[-(\Omega^S_{n,\mathbf{k}})_{\mu\nu}\bar{\varepsilon}_{n,\mathbf{k}}-(m^S_{n,\mathbf{k}})_{\mu\nu} \right] n_\mathrm{B}(\bar{\varepsilon}_{n,\mathbf{k}}),\nonumber\\
\end{eqnarray}
where
\begin{eqnarray}\label{eq:Berrycurvature}
    &&(\Omega^S_{n,\mathbf{k}})_{\mu\nu}=\sum_{m(\neq n)}\frac{2\text{Im}[(\sigma_3 \mathcal{S}_{\mu,\mathbf{k}})_{nm}(\sigma_3 \tilde{v}_{\nu,\mathbf{k}})_{mn}]}{(\bar{\varepsilon}_{n,\mathbf{k}}-\bar{\varepsilon}_{m,\mathbf{k}})^2},\nonumber\\
    &&(m^S_{n,\mathbf{k}})_{\mu\nu}=\sum_{m(\neq n)}\frac{-\text{Im}[(\sigma_3 \mathcal{S}_{\mu,\mathbf{k}})_{nm}(\sigma_3 \tilde{v}_{\nu,\mathbf{k}})_{mn}]}{(\bar{\varepsilon}_{n,\mathbf{k}}-\bar{\varepsilon}_{m,\mathbf{k}})}.\nonumber\\
\end{eqnarray}
Here $(\Omega^S_{n,\mathbf{k}})_{\mu\nu}$ satisfies a relation $(\Omega^S_{n,\mathbf{k}})_{\mu\nu}=(\Omega^S_{n+N,-\mathbf{k}})_{\mu\nu}$ and a sum rule $\sum_{n=1}^{2N}(\Omega^S_{n,\mathbf{k}})_{\mu\nu}=0$.

The expression for $K_{\mu\nu}^{\text{inter}}$ is not yet the final result for the intrinsic response. We now show that it can be conveniently combined with the dipole contribution
\begin{eqnarray}\label{eq:Dipolecontribution}
    D_{\mu\nu}=\left\langle\frac{1}{V}\int d\mathbf{r}\Psi^\dagger(\mathbf{r})\hat{S}_\mu r_{\nu}\Psi(\mathbf{r}) \right\rangle_\mathrm{eq},
\end{eqnarray}
where we used that $[\hat{S}_\nu,r_\nu]=0$. 
To calculate this term, we explicitly introduce a perturbation corresponding to an external magnetic field $\mathbf{B}(\mathbf{r})$ into Hamiltonian $H$~\cite{PhysRevLett.99.197202,PhysRevB.99.024404}:
\begin{eqnarray}
    \hat{{\mathcal H}}_B=-[\mathbf{B}(\mathbf{r})\cdot\hat{\mathbf{S}}+\hat{\mathbf{S}}\cdot\mathbf{B}(\mathbf{r})],
\end{eqnarray} 
where $\mathbf{B}(\mathbf{r})$ varies slowly in space, i.e., on a length scale much larger than the lattice constant. 
The dipole moment can then be found from a thermodynamic relation \cite{PhysRevB.99.024404}
\begin{eqnarray}\label{eq:Dipolecontribution1}
    D_{\mu\nu}=-\lim_{\mathbf{B}\rightarrow 0}\frac{\partial \Omega}{\partial(\partial_{r_\nu}B_\mu)},
\end{eqnarray}
where $\Omega$ is the thermodynamic grand potential of the system and the limit of vanishing magnetic field has to be taken.
Using the Maxwell relation 
\begin{eqnarray}
    \left(\frac{\partial D_{\mu\nu}}{\partial T}\right)_{\mathbf{B},\partial_\mathbf{r}\mathbf{B}}=\left[\frac{\partial S}{\partial(\partial_{r_\nu}B_\mu)}\right]_{T,\mathbf{B}},
\end{eqnarray}
we introduce an auxiliary quantity $\tilde{D}_{\mu\nu} = - \frac{\partial K}{\partial(\partial_{r_\nu}B_\mu)}$,  where $K=\Omega+TS$ and
\begin{eqnarray}\label{Maxwell1}
\tilde{D}_{\mu\nu}=\frac{\partial (\beta D_{\mu\nu})}{\partial\beta}.
\end{eqnarray}
From the auxiliary quantity $\tilde{D}_{\mu\nu}$ we can calculate $D_{\mu\nu}$. The former is calculated using the perturbation theory applied to
\begin{align}
    K(\mathbf{r})=\frac{1}{2}\sum_{\mathbf{k},n=1}^{2N}(\sigma_3)_{nn}g(\bar{\varepsilon}_{n,\mathbf{k}})\langle\psi_{n,\mathbf{k}}(\mathbf{r})|\hat{K}|\psi_{n,\mathbf{k}}(\mathbf{r})\rangle,    
\end{align}
where $|\psi_{n,\mathbf{k}}(\mathbf{r})\rangle=e^{i\mathbf{k}\cdot\mathbf{r}}|u_{n,\mathbf{k}}\rangle$.
For a perturbation $\mathbf{B}(\mathbf{r})=B/q\sin(\mathbf{q}\cdot\mathbf{r})\hat{\mathbf{e}}_\mu$, with $\mathbf{q}=q\hat{\mathbf{e}}_\nu$, we obtain
\begin{eqnarray}\label{Auxiliary}
\tilde{D}_{\mu\nu}=\lim_{\mathbf{q}\rightarrow 0}\frac{-2}{VB}\int d\mathbf{x}\delta K(\mathbf{r})\cos(\mathbf{q}\cdot\mathbf{r}),
\end{eqnarray}
where only the leading order correction $\delta K(\mathbf{r})$ due to the magnetic field is considered. It is obtained from the expansion:
\begin{widetext}
\begin{eqnarray}\label{Variation}
    \delta K(\mathbf{r})&&=\frac{1}{2}\sum_{n\mathbf{k}}\delta g(\bar{\varepsilon}_{n\mathbf{k}})(\sigma_3)_{nn}\langle\psi_{n\mathbf{k}}|\hat{K}_0|\psi_{n\mathbf{k}}\rangle-g(\bar{\varepsilon}_{n\mathbf{k}})(\sigma_3)_{nn}\langle\psi_{n\mathbf{k}}|\hat{{\mathcal H}}_B|\psi_{n\mathbf{k}}\rangle\nonumber\\
    &&+g(\bar{\varepsilon}_{n\mathbf{k}})(\sigma_3)_{nn}
    \left(\langle\delta\psi_{n\mathbf{k}}|\hat{K}_0|\psi_{n\mathbf{k}}\rangle+\langle\psi_{n\mathbf{k}}|\hat{K}_0|\delta\psi_{n\mathbf{k}}\rangle
    \right),
\end{eqnarray}
with
\begin{eqnarray}
    |\delta\psi_{n\mathbf{k}}\rangle=\sum_{m\neq n}\frac{iB}{2q}(\sigma_3)_{mm}
    \left[e^{i(\mathbf{k}+\mathbf{q})\cdot\mathbf{r}}|u_{m,\mathbf{k}+\mathbf{q}}\rangle \frac{\langle u_{m,\mathbf{k}+\mathbf{q}}|(S_{\mu,\mathbf{k}}+S_{\mu,\mathbf{k}+\mathbf{q}})|u_{n,\mathbf{k}}\rangle}{\bar{\varepsilon}_{n\mathbf{k}}-\bar{\varepsilon}_{m,\mathbf{k}+\mathbf{q}}}-(\mathbf{q}\rightarrow-\mathbf{q}) \right],
\end{eqnarray}
\end{widetext}
where $S_{\mu,\mathbf{k}}=e^{-i\mathbf{k}\cdot\mathbf{r}}\hat{S}_\mu e^{i\mathbf{k}\cdot\mathbf{r}}=\hat{S}_\mu$.
After substituting Eq.~\eqref{Variation} into Eq.~\eqref{Auxiliary} we find
\begin{eqnarray}\label{Auxiliary1}
    \tilde{D}_{\mu\nu}&&=\frac{1}{V}\sum_{n\mathbf{k}}g(\bar{\varepsilon}_{n,\mathbf{k}})\bar{\varepsilon}_{n\mathbf{k}}(\Omega^S_{n,\mathbf{k}})_{\mu\nu}+[g(\bar{\varepsilon}_{n,\mathbf{k}})+g^\prime(\bar{\varepsilon}_{n,\mathbf{k}})\nonumber\\
    &&\times\bar{\varepsilon}_{n,\mathbf{k}}](m^S_{n,\mathbf{k}})_{\mu\nu},
\end{eqnarray}
where for quasi-equilibrium magnons with non-zero chemical potential we should have $\bar{\varepsilon}_{n,\mathbf{k}} \rightarrow \bar{\varepsilon}_{n,\mathbf{k}}-\mu$.  
Utilizing this expression as well as Eq.~\eqref{Maxwell1}, we obtain the dipole contribution:
\begin{eqnarray}
    D_{\mu\nu}&&=\frac{1}{V}\sum_{n\mathbf{k}} \left[(\Omega^S_{n,\mathbf{k}})_{\mu\nu}\int_0^{\bar{\varepsilon}_{n\mathbf{k}}} d\eta g(\eta) +(m^S_{n,\mathbf{k}})_{\mu\nu} g(\bar{\varepsilon}_{n,\mathbf{k}}) \right].\nonumber\\
\end{eqnarray}
This result has to be combined with the Kubo part in Eq.~\eqref{eq:Interband1} to give us the total intrinsic contribution: 
\begin{eqnarray}\label{eq:Interband3}
\chi^\mathrm{in}_{\mu\nu}
=\frac{2k_B}{V} \sum_{n=1}^N\sum_\mathbf{k}(\Omega^S_{n,\mathbf{k}})_{\mu\nu} c_1[n_\mathrm{B}(\varepsilon_{n,\mathbf{k}})],
\end{eqnarray}
where we used notation $c_1(x)=(1+x)\ln(1+x)-x\ln(x)$. Note that we have expressed Eq.~\eqref{eq:Interband3} in particle space by utilizing the properties of $(\Omega^S_{n\mathbf{k}})_{\mu\nu}$~\cite{2019arXiv190710567L}. 

Equations \eqref{eq:Intraband1} and \eqref{eq:Interband3} are the main results of this section. These formulas apply as long as the noninteracting approximation is meaningful, e.g., at low temperatures.
In Sec.~\ref{sec:Models}, we use these formulas to make numerical predictions of the nonequilibrium spin density for several relevant models, including material candidates.

\section{Symmetry Constraints}
\label{sec:SymmetryConstraints}
In this section, we discuss constraints on the magnon response tensor, $\chi_{\mu\nu}$, posed by the symmetries.
To generate the nonequilibrium spin density with magnons one needs a system in which spin is not conserved locally or globally, at least for one direction of the spin polarization. This is often the case in non-collinear antiferromagnets or in systems with Dzyaloshinskii-Moriya interactions. For example, for inversion symmetric systems spin density is \emph{globally} conserved \cite{2019arXiv190710567L}.
To see this, note that inversion symmetry implies $\mathcal{H}_\mathbf{k}=\mathcal{H}_{-\mathbf{k}}$, which leads to $T_{\mathbf{k}}=T_{-\mathbf{k}}$, $\mathcal{E}_{\mathbf{k}}=\mathcal{E}_{-\mathbf{k}}$ and $\mathbf{v}_{n,\mathbf{k}}=-\mathbf{v}_{n,-\mathbf{k}}$. Substituting these relations into Eq.~\eqref{eq:Intraband1} results in $\chi^\mathrm{ex}_{\mu\nu}=-\chi^\mathrm{ex}_{\mu\nu}=0$. Furthermore, inversion symmetry also enforces the relation $(\Omega^S_{n,\mathbf{k}})_{\mu\nu}=-(\Omega^S_{n,-\mathbf{k}})_{\mu\nu}$, which results in $\chi^\mathrm{in}_{\mu\nu}=-\chi^\mathrm{in}_{\mu\nu}=0$, that is, in a vanishing intrinsic response. Below, in Sec.~\ref{sec:Models}, we show several examples of collinear and non-collinear systems in which spin can be generated.

In general, the response tensor will be constrained by symmetry operations of a specific material under consideration.
The constraining relations can be readily found within the framework of linear response theory \cite{PhysRevB.92.155138,PhysRevB.95.014403}. Assuming that a system respects a symmetry operation represented by $g$, we find for an arbitrary operator $\hat{A}$ that $\langle g(\psi_{n\mathbf{k}})| \hat{A}|g(\psi_{m\mathbf{k}})\rangle= \langle\psi_{ng(\mathbf{k})}| g^{-1}\hat{A}g|\psi_{mg(\mathbf{k})}\rangle $ when the operation is unitary, and $\langle g(\psi_{n\mathbf{k}})| \hat{A}|g(\psi_{m\mathbf{k}})\rangle=\langle\psi_{ng(\mathbf{k})}| g^{-1}\hat{A}g|\psi_{mg(\mathbf{k})}\rangle^\ast $, when the operation is antiunitary. Operators transform as $g^{-1}\hat{v}_ig=\sum_{j}R^v_{ij}\hat{v}_j$ and $g^{-1}\hat{S}_ig=\sum_{j}R^s_{ij}\hat{S}_j$, where $R^{v/s}$ is the corresponding matrix representation of $g$ with respect to the Cartesian components $v_j$ or $\hat{S}_j$. We find $R^v=\pm R$ and $R^{s}=\pm\det(R)R$ where $\pm$ refers to unitary ($+$) or antiunitary ($-$) symmetries, respectively. 
Under the above premises, the following symmetry requirements on elements of the response tensor arise:
\begin{eqnarray}\label{eq:Symmetry3}
    \chi^\mathrm{ex}_{\mu\nu}&&=\det(R)R_{\mu i}R_{\nu j}\chi^\mathrm{ex}_{ij} \nonumber\\
    \chi^\mathrm{in}_{\mu\nu}&&=\pm\det(R)R_{\mu i}R_{\nu j}\chi^\mathrm{in}_{ij},
\end{eqnarray}
where $\pm$ corresponds to unitary and antiunitary symmetry operations, respectively. Later on, we show that these two relations result in different shapes of the response tensors, which is useful for distinguishing extrinsic and intrinsic contributions.
Notice that tensors $\chi^\mathrm{ex}_{\mu\nu}$ and $\chi^\mathrm{in}_{\mu\nu}$ transform differently under antiunitary operations which is a consequence of a complex factor in the expression for $(\Omega^S_{n,\mathbf{k}})_{\mu\nu}$ corresponding to taking the imaginary part in Eq.~\eqref{eq:Berrycurvature}.
Given the transformation properties of velocity and spin, one finds that $\chi^\mathrm{ex}_{\mu\nu}$ is even and $\chi^\mathrm{in}_{\mu\nu}$ is odd under the time-reversal transformation. Consequently, a reversal of the magnetic ordering causes $\chi^\mathrm{in}_{\mu\nu}$ to flip sign while $\chi^\mathrm{ex}_{\mu\nu}$ is invariant under such transformation:
\begin{subequations}
\begin{align}
    \chi^\mathrm{in}_{\mu\nu} [\{ \mathbf{S}_i \}] &= -  \chi^\mathrm{in}_{\mu\nu}[\{ -\mathbf{S}_i \}], \\
    \chi^\mathrm{ex}_{\mu\nu} [\{ \mathbf{S}_i \}] &=  \chi^\mathrm{ex}_{\mu\nu}[\{ -\mathbf{S}_i \}]. \label{eq:symmetry-ex-edel}
\end{align}
\end{subequations}
Thus, it is possible to disentangle extrinsic from intrinsic contributions by measuring the nonequilibrium spin density for two antiferromagnetically ordered states related by the time reversal transformation. Such approach has been used in studies of the spin Hall effect \cite{Kimata2019}.

\section{Models}
\label{sec:Models}

In this section, we apply our theory to specific models. To obtain some intuition, we first focus on a toy model of collinear antiferromagnetic spin chain with anisotropic exchange and inversion asymmetry resulting in Rashba-type DMI. We then focus on more realistic noncollinear kagome and breathing pyrochlore antiferromagnets, for which we use material parameters established in the literature. To satisfy the requirement of inversion asymmetry, we assume that the kagome antiferromagnet can have interfacial inversion asymmetry, e.g., due to thin film geometry in contact with another material. The breathing pyrochlore antiferromagnet has bulk inversion asymmetry.
For details of the Holstein-Primakoff transformations and explicit expressions of the magnon Hamiltonians, we refer the reader to Appendix~\ref{sec:AppModels}.  


\subsection{Antiferromagnetic Spin Chain}
\label{sec:ExampleAFChain}

\begin{figure}
	\begin{tabular}{ccc}
        \centerline{\includegraphics[width=0.8\columnwidth]{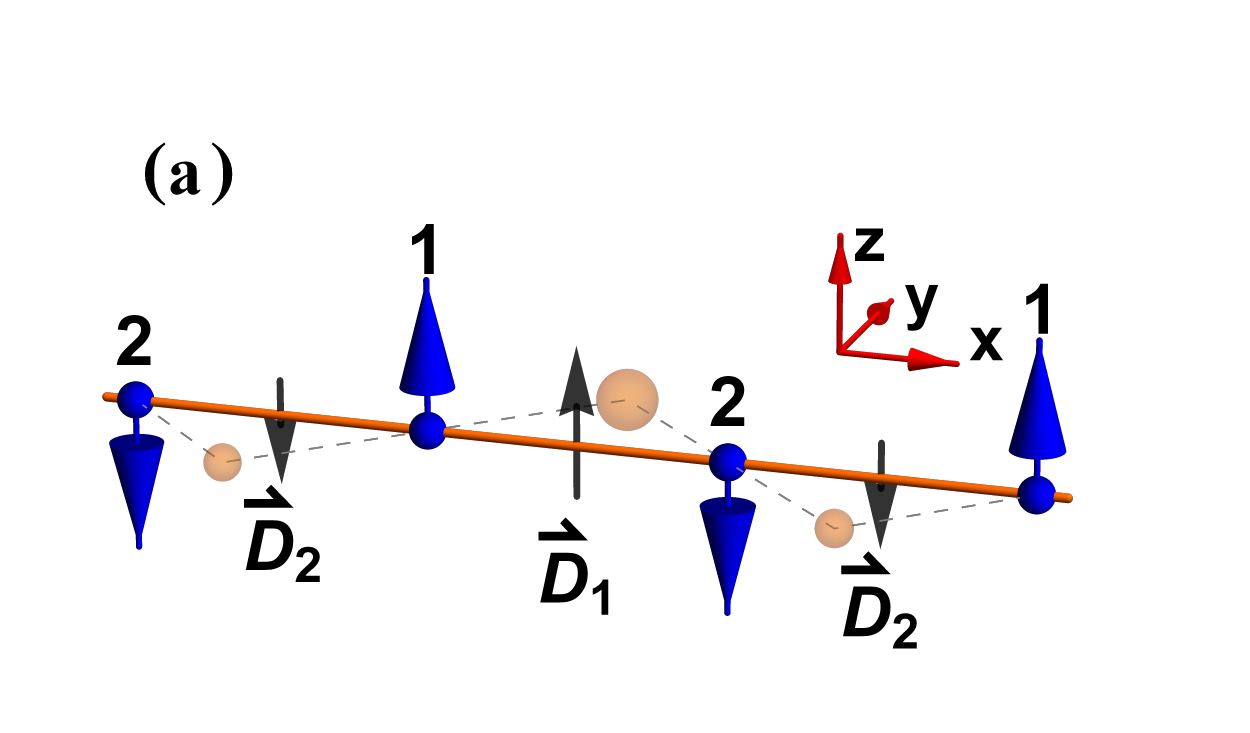}}\\
		\centerline{\includegraphics[width=0.9\columnwidth]{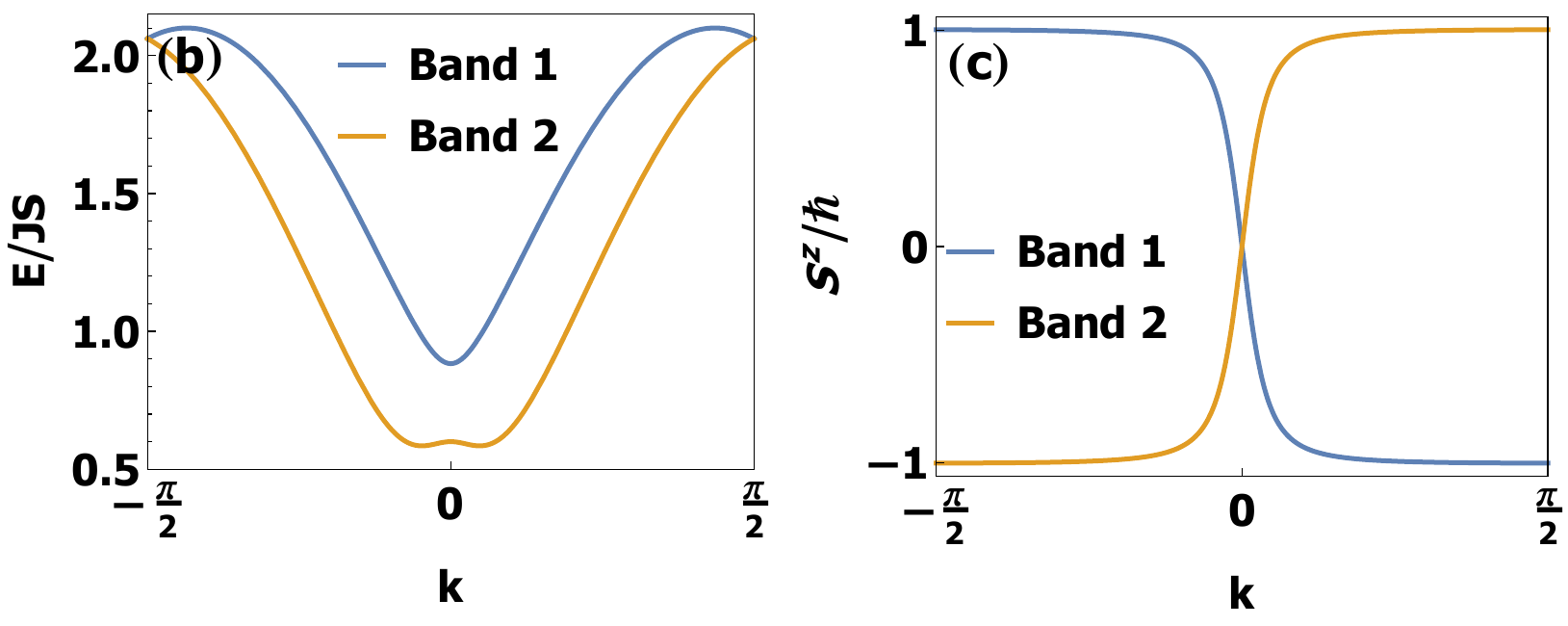}}\\
		\centerline{\includegraphics[width=0.9\columnwidth]{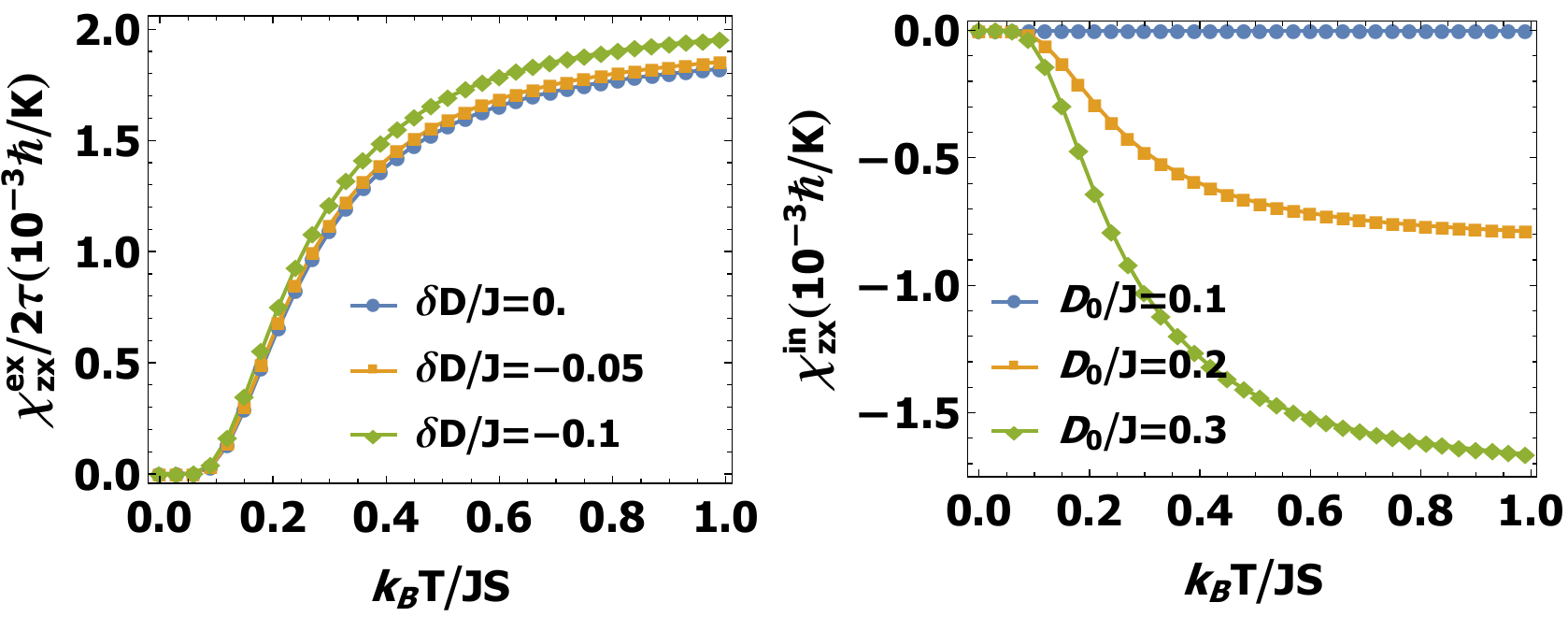}}
	\end{tabular}
	\protect\caption{(Color online) (a) Spin order and DMI vectors in the antiferromagnet spin chain model. (b) and (c) Magnon dispersion and magnon spin expectation value in the 1D Brillouin zone. We used $D_0/J=0.2$, $\delta D/J=-0.1$. (d) and (e) Extrinsic and intrinsic response coefficients. In (d), $\tau=JS/(2\Gamma_n)$ is the
dimensionless magnon lifetime ($\hbar$ is set to one). Parameters read $\lambda=1.05$, $\gamma=0.95$, $J=2$meV, $S=3/2$, and $D_0/J=0.2$.}
	\label{fig:Spinchain}
\end{figure}

As a simple model, we first consider the antiferromagnetic spin chain shown in Fig.~\ref{fig:Spinchain}(a).
Similar to Eq.~\eqref{eq:GeneralHamiltonian}, the Hamiltonian
\begin{eqnarray}
    H=&&\sum_{i}\sum_{\nu=\pm 1}[J(\gamma S^x_{1,i}S^x_{2,i+\nu}+S^y_{1,i}S^y_{2,i+\nu}+\lambda S^z_{1,i}S^z_{2,i+\nu})\nonumber\\
    &&+D_{12}^\nu\mathbf{e}_z\cdot(\mathbf{S}_{1,i}\times\mathbf{S}_{2,i+\nu})],
    \label{eq:spinHam-chain}
\end{eqnarray}
contains the anisotropic symmetric exchange interaction, which is given in terms of an energy $J$ and dimensionless factors $\gamma$ and $\lambda$, and the antisymmetric exchange interaction described by DMI vectors along $z$ direction.
We choose $\gamma \le 1$ and $\lambda \ge 1$, such that the collinear state with N\'eel vector along $z$ direction is the classical magnetic ground state. For $\gamma \neq 1$, the anisotropy causes the magnons to experience the effect of ``squeezing'' \cite{PhysRevB.96.020411}.
Note that $\lambda$ has to be larger than a critical value to avoid the spins from canting due to DMI. The DMI strength is set to $D^{+}_{12}=D_1$ and $D^{-}_{12}=D_2$, where $\nu = \pm$ refers to the direction of the bond [$+$ for going from the left to the right in Fig.~\ref{fig:Spinchain}(a)]. 

It is convenient to reparameterize the DMI as $D_0=(D_1+D_2)/2J$ and $\delta_D=(D_1-D_2)/2J$. The staggered contribution to DMI is necessary for the model to exhibit both intrinsic as well as extrinsic effects. To see this, observe that only in the absence of the inversion symmetry we can have $D_0 \ne 0$. However, when $\delta_D=0$, the system still holds a $\mathcal{T}\ast\mathcal{M}_x$ symmetry, where $\mathcal{T}$ is time reversal and $\mathcal{M}_x$ is the mirror symmetry with respect to the $y-z$ plane passing through the atoms. Applying the corresponding Cartesian representation matrix $R=\text{Diag}\{-1,1,1\}$ of $\mathcal{T}\ast\mathcal{M}_x$ to Eq.~\eqref{eq:Symmetry3}, the intrinsic part $\chi^\mathrm{in}_{zx}$ is rendered zero. Therefore, we set $\delta_D \ne 0$ to ensure the appearance of intrinsic contributions.  

In Fig.~\ref{fig:Spinchain}(b), we show the magnon band structure. The degeneracy of spin-up and -down modes is lifted by the DMI and $\gamma \ne 1$. On top of that, since $\gamma \ne 1$ spin is not conserved and we observe the magnon spin-momentum locking~\cite{PhysRevLett.119.107205} as shown in Fig.~\ref{fig:Spinchain}(c), which is in agreement with Ref.~\cite{PhysRevB.96.020411}. This is in contrast to the usual case of uniaxial collinear AFMs that features two eigenmodes with opposite spin quanta $\pm\hbar$. Figs.~\ref{fig:Spinchain}(d) and (e) show the extrinsic and intrinsic response coefficient, respectively. For the calculation of the extrinsic response, we regarded the broadening as a constant, $\Gamma_n=\hbar/2\tau$, where $\tau $ is the magnon lifetime~\cite{Ritzmann2015Model}. In Figs.~\ref{fig:Spinchain}(d) and (e), the extrinsic spin accumulation dominates.

\begin{figure*}
		\begin{tabular}{ccc}
			\includegraphics[width=0.95\textwidth]{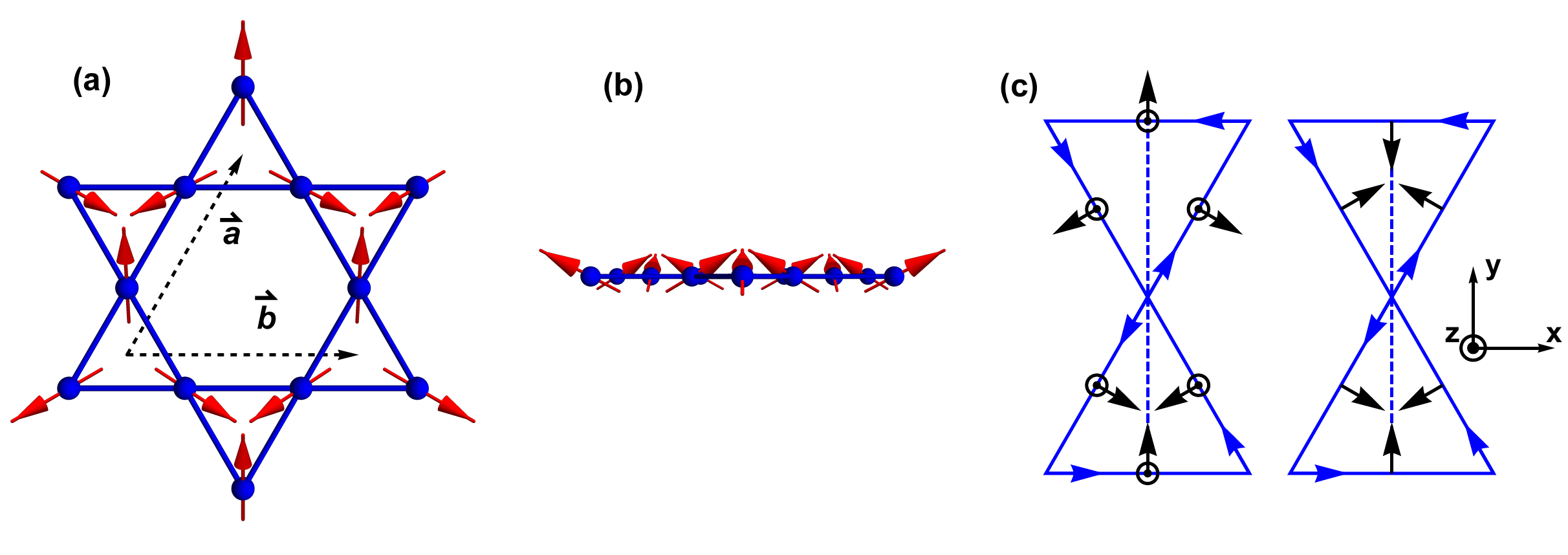}\\	\includegraphics[width=1.\textwidth]{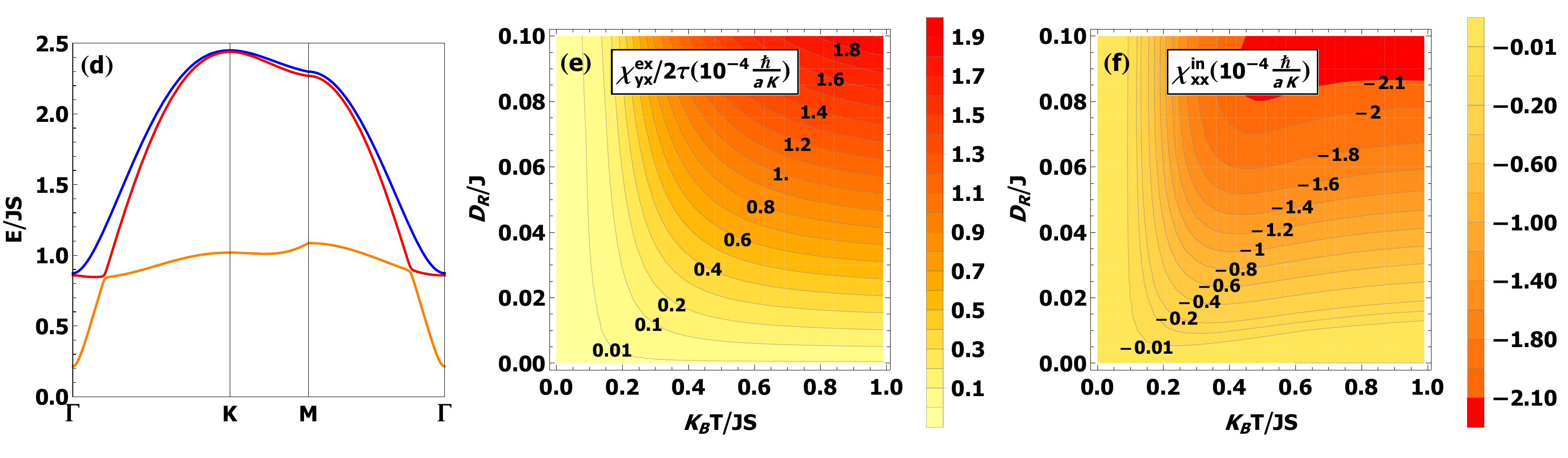}
		\end{tabular}
		\caption{(Color online) Noncollinear antiferromagnetic PVC order on the kagome lattice. (a) and (b) Ground state spin configuration from above and front view. Lattice vectors are denoted by $\vec{a}$ and $\vec{b}$. (c) Left: intrinsic DMI vectors; right: Rashba DMI vectors. Arrows along the bonds indicate ordering of sites in DMI terms. (d) Magnon dispersion with $D_R/J=0.06$. (e), (f) Extrinsic and intrinsic response tensor elements $\chi^\mathrm{ex}_{yx}$ and $\chi^\mathrm{in}_{xx}$, respectively. $\tau$ is the dimensionless magnon lifetime and $a$ denotes the lattice constant. We used the material parameters of $\text{KFe}_3(\text{OH})_6(\text{SO}_4)_2$: $J_1=3.18 \text{meV}$, $J_2=0.11\text{meV}$, $|D_p|/J_1=0.062$, $D_z/J_1=-0.062$ and $S=5/2$.}
		\label{fig:KagomeAF}
	\end{figure*}

To obtain an intuitive understanding of the extrinsic contributions, we recall the usual electronic Edelstein effect scenario in a Rashba system. Upon shifting the spin-momentum locked Fermi circles in reciprocal space due to application of an electric field, electronic states with a particular spin polarization are more occupied than those with opposite spin polarization (e.g., see Fig.~13 of Ref.~\cite{RevModPhys.89.025006}). Consequently, this redistribution leads to a nonzero macroscopic spin density in nonequilibrium. A similar explanation can by given for the magnonic case. First, we consider the band 2 [cf.~Fig.~\ref{fig:Spinchain}(b)]. According to Fig.~\ref{fig:Spinchain}(c), magnons in band 2 have a positive (negative) spin for negative (positive) momentum $k$, which corresponds to magnon spin-momentum locking discussed in Ref.~\cite{PhysRevLett.119.107205}. Upon application of the temperature gradient (or the pseudo-gravitational potential) we redistribute magnons from $k$ to $-k$ (or \textit{vice versa}, depending on the direction of the gradient), causing an excess of magnons with positive spin. Although there is some cancellation between the lower and upper band, the different thermal occupation ensures that there is a nonzero resulting net spin density in nonequilibrium.
There is no such simple picture for the intrinsic contributions, which arise due to interband mixing~\cite{RevModPhys.91.035004}.

\subsection{Kagome Antiferromagnet}

In several real materials, spin nonconservation naturally emerges due to noncollinear antiferromagnetism. For example, noncollinear antiferromagnets (NAFMs) exist in layered quasi-two-dimensional kagome and triangular magnetic structures, and  in three-dimensional pyrochlore magnetic structures. We first take the kagome antiferromagnet in the so-called $\mathbf{q}=0$ phase with positive vector chirality (PVC) \cite{PhysRevLett.113.237202,PhysRevB.92.144415,PhysRevB.95.094427}, which is depicted in Fig.~\ref{fig:KagomeAF}(a), as an example.

The spin Hamiltonian under consideration is
\begin{eqnarray}
H=\sum_{\langle ij\rangle}J_1\mathbf{S}_i\cdot\mathbf{S}_j+\mathbf{D}_{ij}\cdot(\mathbf{S}_i\times\mathbf{S}_j)+\sum_{\langle\langle ij\rangle\rangle}J_2 \mathbf{S}_i\cdot\mathbf{S}_j,\nonumber\\
\end{eqnarray}
where the three terms describe the nearest-neighbor exchange with $J_1>0$, DMI, and the second-nearest neighbor exchange with $J_2 > 0$. The DMI vector $\mathbf{D}_{ij}$ is composed of intrinsic DMI and extrinsic Rashba-DMI, i.e., $\mathbf{D}_{ij}=\mathbf{D}_{\mathrm{in}}+\mathbf{D}_{R}$. The intrinsic DMI $\mathbf{D}_{\mathrm{in}}=\mathbf{D}_p+D_{z,ij}\hat{z}$ has out-of-plane contributions $D_{z,ij}$ as well as in-plane contributions $\mathbf{D}_{p}=D_p\hat{n}_{ij}$ along $\hat{n}_{ij}$. The DMI vectors are arranged as shown in the left part of Fig.~\ref{fig:KagomeAF}(c). 
Accounting for the antiferromagnetic exchange interactions and only for the intrinsic DMI, the classical ground state is the $120^\circ$-ordered antiferromagnetic state [cf.~Fig.~\ref{fig:KagomeAF}(a)] with a small out-of-plane canting, with an angle given by $\eta=\frac{1}{2}\tan^{-1}(\frac{-2D_p}{\sqrt{3(J_1+J_2)}-D_z})$ [cf.~Fig.~\ref{fig:KagomeAF}(b)]. Thus, there is a weak ferromagnetic moment in $z$ direction and the texture is not fully compensated. Here, we are concentrating on nonequilibrium spin densities in $x$ and $y$ direction, along which the texture is compensated.

Although nonzero $\mathbf{D}_p$ breaks the mirror symmetry of the kagome lattice, the system is still inversion symmetric. Thus, we need the Rashba-like DMI described by $\mathbf{D}_{R}$ that we envision to arise in an inversion-symmetry breaking environment, as caused, e.g., by putting a single kagome layer on a substrate. The vector $\mathbf{D}_{R}$ lies in the kagome plane and has directions similar to $\mathbf{D}_p$, but with the crucial difference that its directions are always pointing in the same direction relative to the bond [compare the left and right part of Fig.~\ref{fig:KagomeAF}(c)]. We also note that a large Rashba-DMI can twist the system into a spiral state. We confirmed numerically that this does not happen for $|\mathbf{D}_R|/J<0.06$ using computational package SpinW~\cite{Toth_2015}.

\begin{table}
	\centering
	\caption{The shape of spin polarization response tensors enforced by magnetic point goup symmetry for selected noncollinear antiferromagnets.} 
	\label{table:Symmetry}
	\begin{tabular}{c c c c}
		\hline\hline 
		Structure & Extrinsic & Intrinsic \\ [0.5ex] 
		\hline 
		Kagome(PVC,SVC)&$\left(\begin{array}{cc}
		0&-\chi^\mathrm{ex}_{yx}\\\chi^\mathrm{ex}_{yx}&0
		\end{array}\right)$ & $\left(\begin{array}{cc}
		\chi^\mathrm{in}_{xx}&0\\0&\chi^\mathrm{in}_{xx}
		\end{array}\right)$
		\\
		Kagome(NVC) &$\left(\begin{array}{cc}
		0&\chi^\mathrm{ex}_{xy}\\\chi^\mathrm{ex}_{yx}&0
		\end{array}\right)$ & $\left(\begin{array}{cc}
		\chi^\mathrm{in}_{xx}&0\\0&\chi^\mathrm{in}_{yy}
		\end{array}\right)$
		\\
		\\
		\hline
		Triangle &$\left(\begin{array}{cc}
		0&-\chi^\mathrm{ex}_{yx}\\\chi^\mathrm{ex}_{yx}&0
		\end{array}\right)$ & $\left(\begin{array}{cc}
		\chi^\mathrm{in}_{xx}&0\\0&\chi^\mathrm{in}_{xx}
		\end{array}\right)$
		\\
		\\
		\hline
		Pyrochlore (AIAO) &$\left(\begin{array}{ccc}
		0&0&0\\0&0&0\\0&0&0
		\end{array}\right)$ & $\left(\begin{array}{ccc}
		\chi^\mathrm{in}_{xx}&0&0\\0&\chi^\mathrm{in}_{xx}&0\\0&0&\chi^{in}_{xx}
		\end{array}\right)$
		\\[1ex]
		\hline 
	\end{tabular}
\end{table}

The kagome NAFM described above exhibits two symmetries: (i) the mirror reflection with respect to the $y-z$ plane plus time-reversal, $g_1=\mathcal{M}_{x}\mathcal{T}$, and (ii) the threefold rotation about the $z$ axis, $g_2=\mathcal{C}_{3z}$.
According to Eq.~\eqref{eq:Symmetry3}, these two symmetries fix the extrinsic and intrinsic response tensors to the forms in Table~\ref{table:Symmetry} (Kagome PVC), where only the in-plane spin polarizations are allowed. 

Based on what we have discussed so far, we propose potassium iron jarosite $\text{KFe}_3(\text{OH})_6(\text{SO}_4)_2$ as a candidate material. Concentrating on a single kagome layer of this material and assuming that the mirror symmetry is broken due to a proper environment, the magnon dispersion is given in Fig.~\ref{fig:KagomeAF}(d). We used parameters $J_1=3.18 \text{meV}$, $J_2=0.11\text{meV}$, $|D_p|/J_1=0.062$, $D_z/J_1=-0.062$ and $S=5/2$ \cite{PhysRevB.98.094419,PhysRevLett.96.247201}. The spin density response is captured by virtue of Eqs.~\eqref{eq:Intraband1} and \eqref{eq:Interband3}. The results for the extrinsic, $\chi^\mathrm{ex}_{yx}$, and intrinsic contributions, $\chi^\mathrm{in}_{xx}$, are shown in Figs.~\ref{fig:KagomeAF} (e) and (f), respectively. The effect becomes stronger as we increase Rashba-DMI. The contributions $\chi^\mathrm{ex}_{xx}$ and $\chi^\mathrm{in}_{yx}$ are zero in agreement with tensor shapes in Table~\ref{table:Symmetry}. 

Approximating the magnon band broadening $\Gamma_n \sim \hbar/2\tau$ as a constant, with a magnon lifetime $\tau\sim \unit[10^{-10}]{s}$, and using a lattice constant  $a=\unit[10^{-9}]{m}$, a Rashba-DMI $D_R=0.06J$, a temperature gradient $\partial_x T= \unit[10]{ K/mm}$ \cite{PhysRevB.92.020410}, and a temperature $T=0.5JS$ [which corresponds to a temperature $\approx \unit[46]{K}$ for $\text{KFe}_3(\text{OH})_6(\text{SO}_4)_2$] we obtain the extrinsic part of the temperature-gradient-induced spin density $\langle S^y\rangle_\mathrm{ex}\simeq\unit[5\times10^{6}]{\hbar/cm^2} $; and the intrinsic part $\langle S^x\rangle_\mathrm{in}\simeq  \unit[2\times10^{5}]{\hbar/cm^2}$.
With larger temperature gradients, the extrinsic contribution can be made comparable to spin densities generated by the electronic Edelstein effect \cite{PhysRevB.93.195440}, which are of the order of $10^{8}\sim \unit[10^{10}]{\hbar/cm^2}$.

\subsection{Breathing Pyrochlore Antiferromagnets}

\begin{figure}
	\begin{tabular}{cc}
		\includegraphics[width=0.85\columnwidth]{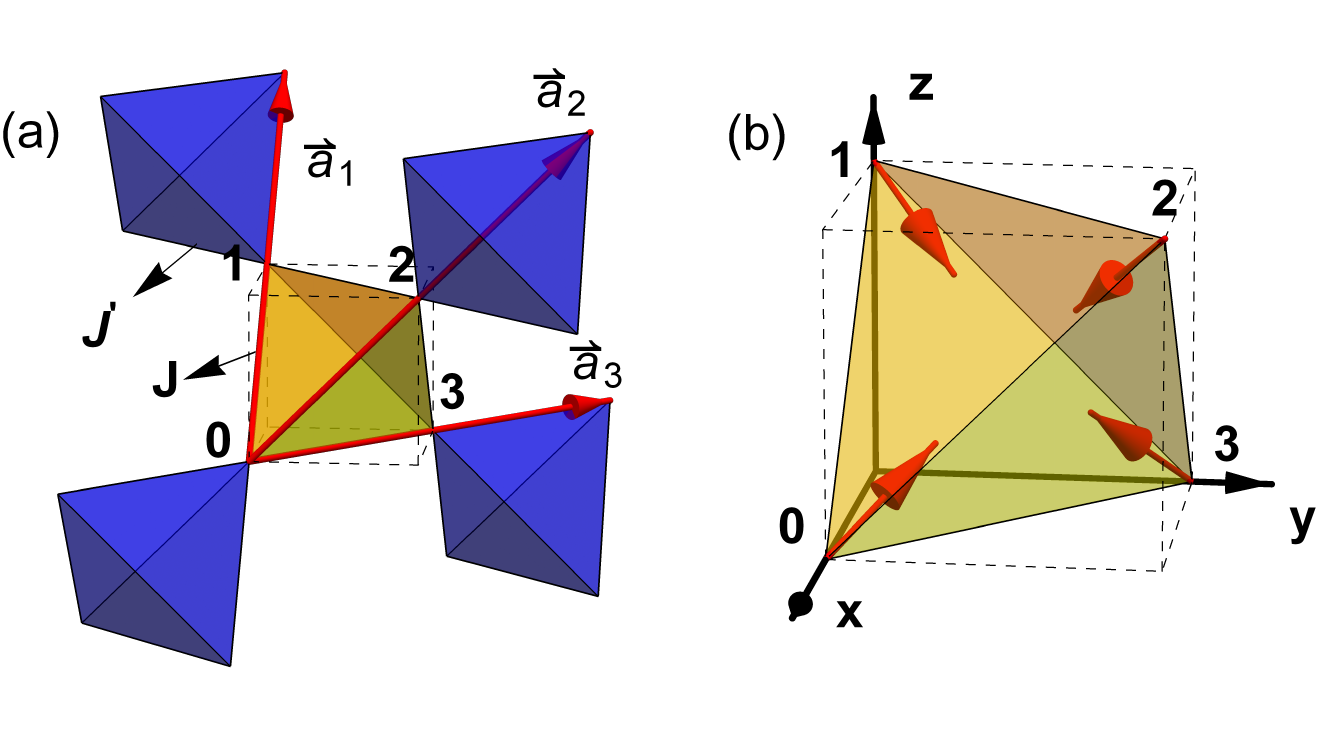}\\
		\includegraphics[width=1.\columnwidth]{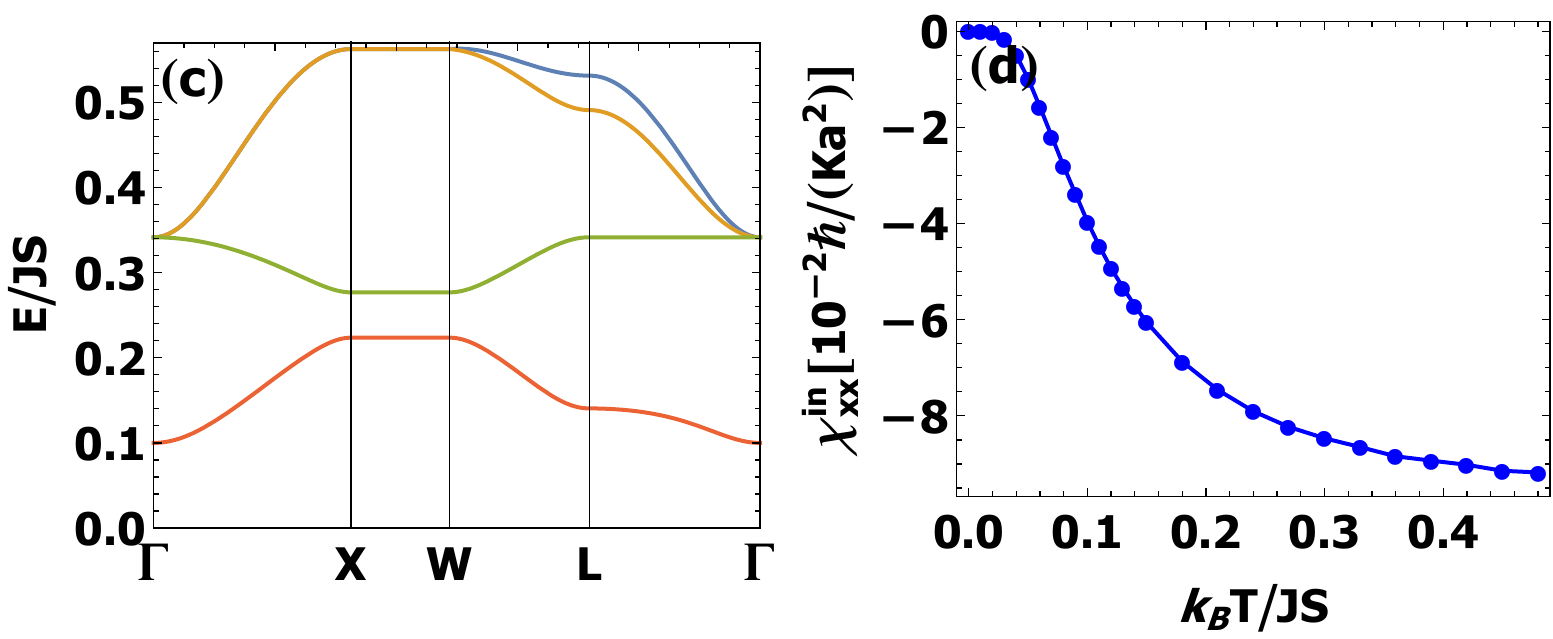}
	\end{tabular}
	\caption{(Color online).(a) Breathing pyrochlore lattice with indicated lattice vectors $\vec{a}_i$ ($i=1,2,3$) and nearest-neighbor exchange in up-pointing (blue, $J^\prime$) and down-pointing (yellow, $J$) tetrahedra. (b) Spin order in the all-in-all-out configuration. (c) Magnon band structure. (d) The intrinsic response $\chi_{xx}^\mathrm{in}$, with $a$ denoting the lattice constant. Parameters read $J\approx 50 K$ (4.3 meV), $J^\prime/J=0.6$, $D/J=-0.2$, $S=3/2$ to mimic the material LiGaCr$_4$O$_8$.}
	\label{fig:Pyrochlore}
\end{figure}

The 3D pyrochlore lattices, which consist of corner-sharing tetrahedra, are well-known for exhibiting noncollinear spin structures. Here, to break bulk inversion symmetry, we concentrate on the so-called ``breathing'' pyrochlore antiferromagnets that possess different exchange interaction in up-pointing (u) and down-pointing (d) tetrahedra [see Fig.~\ref{fig:Pyrochlore}(a)]. The minimal Heisenberg model is \cite{Li2016}
\begin{eqnarray}
    H= J\sum_{\langle i,j \rangle\in u}\mathbf{S}_i\cdot\mathbf{S}_j
     + J^\prime \sum_{\langle i,j \rangle\in d}\mathbf{S}_i\cdot\mathbf{S}_j
     + D \sum_i(\mathbf{S}_i\cdot\hat{\mathbf{z}}_i)^2\nonumber\\
\end{eqnarray} 
The first two terms describe the antiferromagnetic exchange interactions in up-pointing and down-pointing tetrahedra, respectively. The last term comprises easy-axis anisotropy ($D<0$), with $\hat{\mathbf{z}}_i$ being a unit vector pointing either towards or away from the tetrahedon's center of gravity. This model can be energetically optimized to different spin configurations depending on the values of $J^\prime/J$, and $D/J$ \cite{Li2016,PhysRevB.98.045109}, but here we only concentrate on the all-in--all-out (AIAO) order depicted in Fig.~\ref{fig:Pyrochlore}(b), in which all spins of a single tetrahedron are either pointing inward [yellow tetrahedron in Fig.~\ref{fig:Pyrochlore}(b)] or outward (blue tetrahedra).

The AIAO order respects the magnetic point group $\bar{T}_d=\text{Span}\{C_3,C_2,\mathcal{T}*\sigma_d,\mathcal{T}*S_4\}$  \cite{PhysRevB.97.115162,2017arXiv171208170H}. Here, we give the representative generators of these symmetries: $C_3$ is the three-fold rotation with respect to $[1,1,1]$ axis; $C_2$ is two-fold rotation about $[1,0,0]$ axis; $\mathcal{T}*\sigma_d$ is time-reversal followed by the reflection about $(\bar{1},1,0)$ plane; and $\mathcal{T}*S_4$ is time-reversal followed by the combination of the four-fold rotation about $[1,0,0]$ and the reflection about $(1,0,0)$.   We find that this symmetry constraint eliminates any extrinsic response and enforces the intrinsic response tensor to be proportional to a unit matrix, see Table~\ref{table:Symmetry}. In Fig.~\ref{fig:Pyrochlore} (c), we plot the dispersion of the four magnon bands for the AIAO phase with $J\approx50 K$ (4.3 meV) and $J^\prime/J=0.6$, which is the breathing ratio of LiGaCr$_4$O$_8$ \cite{PhysRevLett.110.097203}. We used $D/J=-0.2$ to stabilize the AIAO order. 
In Fig.~\ref{fig:Pyrochlore} (d), we show the intrinsic response $\chi^\mathrm{in}_{xx} = \chi^\mathrm{in}_{yy} = \chi^\mathrm{in}_{zz}$, which are the only nonzero tensor elements, in agreement with the symmetry analysis. If we assume $\partial_x T = \unit[10]{ K/mm}$, $T=0.12JS$, and $a\sim \unit[10^{-9}]{m}$, the intrinsic spin accumulation is estimated to be $\langle S^x\rangle_\mathrm{in} \simeq 5\times\unit[10^{14}]{\hbar/cm^3}$. We can compare this result with the electronic Edelstein effect by converting its 2D spin density to a bulk density: $\langle S \rangle^\mathrm{2D}_{\text{electron}} / a \sim \unit[10^{15}]-\unit[10^{17}]{ \hbar/cm^3 }$. Thus, the intrinsic contribution in breathing pyrochlores is comparable with the electronic Edelstein effect. We believe that this result is detectable in experiment either by transport measurements similar to those used for detection of the inverse spin Hall effect, by magnetooptical Kerr microscopy, or by magnetic sensing based on the nitrogen-vacancy (NV) centres~\cite{pub.1100161418}.

\section{Computer Experiments}
\label{sec:ComputerExperiments}

To better understand the nonequilibrium spin density brought about by the magnonic counterpart to the Edelstein effect, we use atomistic spin dynamics simulations. We describe spin dynamics using the stochastic Landau-Lifshitz-Gilbert (sLLG) equation
\begin{align}
	\dot{\mathbf{S}}_i = -\frac{\gamma}{\mu \left(1+\alpha^2\right)} \left[ \mathbf{S}_i \times \mathbf{B}_i + \alpha \mathbf{S}_i \times \left(\mathbf{S}_i \times \mathbf{B}_i \right) \right], \label{eq:sLLG}
\end{align}
comprising the damped precession of $\mathbf{S}_i$ about its local field $\mathbf{B}_i = \mathbf{b}_i - \partial H/\partial \mathbf{S}_i$. The stochastic field $\mathbf{b}_i(t) = \sqrt{2 \alpha k_\mathrm{B}T \mu / (\gamma \Delta t)} \, \mathbf{G}(t)$ simulates thermal noise \cite{Brown1963,evans2014atomistic}. $\mathbf{G}(t)$ is a three-dimensional Gaussian random number distribution with zero mean.
$\alpha$, $\gamma$, and $\mu = 2 \mu_\mathrm{B} \sqrt{S(S+1)}$ are the dimensionless Gilbert damping, the gyromagnetic ratio, and the modulus of the magnetic moment at each lattice site, respectively. The numerical integration of Eq.~\eqref{eq:sLLG} is done by the Heun method \cite{evans2014atomistic} with time steps $\Delta t \le \unit[1]{fs}$. 

We consider the antiferromagnetic spin chain introduced in Sec.~\ref{sec:ExampleAFChain} and study this model in a nonequilibrium situation. As was shown in Sec.~\ref{sec:ExampleAFChain}, the extrinsic contribution to the nonequilibrium spin density dominates over the intrinsic contribution for the spin chain model. Thus, we focus on the extrinsic contributions and set $\delta D = 0$, rendering intrinsic contributions zero by symmetry.

We simulate a spin chain of $N=480$ spins with spin Hamiltonian as in Eq.~\eqref{eq:spinHam-chain}. First, to characterize the chain in terms of magnon variables, i.e., in terms of (i) the magnon dispersion and (ii) the magnon spin, we calculate the dynamical structure factor
\begin{align}
    \mathcal{F}(k,\omega) = \frac{1}{\sqrt{2 \pi} N} \sum_{i,j} \mathrm{e}^{\mathrm{i} k (x_i-x_j)} \int_{-\infty}^\infty \mathrm{e}^{\mathrm{i} \omega t} \left\langle S_i^+(t) S_j^-(0) \right\rangle \, \mathrm{d} t,
\end{align}
i.e., the time and space Fourier transform of the spin-spin time-correlation function. $x_i$ denotes the $x$ coordinate of the $i$th spin and $S_i^\pm = S_i^x \pm \mathrm{i} S_i^y$. 

\begin{figure}
  \centering
  \includegraphics[width=1.0\columnwidth]{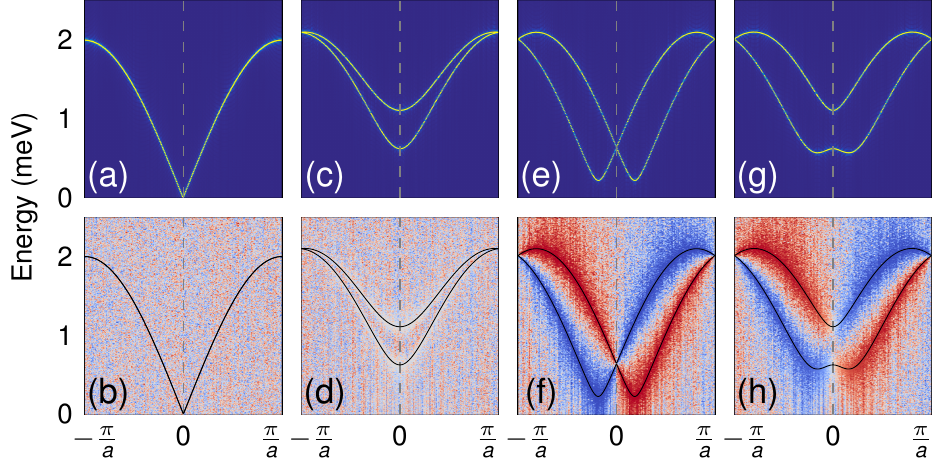}
  \caption{Magnon spectra of the antiferromagnetic spin chain as obtained from numerical simulations for selected parameters; top row: dynamical structure factor; bottom row: the spin of magnons or Stokes parameter ratio $\sigma(k,\omega)$ (red: negative; gray: zero; blue: positive). Black solid lines show the analytically obtained magnon dispersion (within linear spin-wave theory). Parameters read $J=\unit[1]{meV}$, and (a,b) $\lambda = 1$, $D=0$, $\gamma=1$, (c,d) $\lambda = 1.05$, $D=0$, $\gamma=0.9$, (e,f) $\lambda = 1.05$, $D=\unit[0.3]{meV}$, $\gamma=1$, and (g,h) $\lambda = 1.05$, $D=\unit[0.3]{meV}$, $\gamma=0.9$. A small simulation temperature $T=\unit[0.01]{K}$ and Gilbert damping $\alpha = 0.001$ were chosen to reduce lifetime broadening.} 
  \label{fig:magnon-spectra}             
\end{figure}

The numerically determined magnon spectra for the spin chain are shown in Fig.~\ref{fig:magnon-spectra}(a), (c), (e), and (g); they agree with those obtained analytically in the previous section [shown as black lines in Fig.~\ref{fig:magnon-spectra}(b), (d), (f), and (h)]. In Fig.~\ref{fig:magnon-spectra}(a), we depict the dispersion of the isotropic antiferromagnetic spin chain ($\lambda = 1$, $D=0$, $\gamma =1$) with the two degenerate linear Goldstone modes. This degeneracy is lifted in the presence of spin-nonconserving anisotropies $\lambda >1$ and $\gamma <1$ [cf.~Fig.~\ref{fig:magnon-spectra}(c)]. In Fig.~\ref{fig:magnon-spectra}(e), we show the Rashba-like spin-split dispersion in the presence of nonzero DMI and $\lambda>1$, and in Fig.~\ref{fig:magnon-spectra}(g) the dispersion in the presence of both anisotropies as well as DMI, for which the band degeneracy at $k=0$ is lifted [as compared to (e)].

The magnon spin is extracted by computing the Stokes parameters $I(k,\omega) = |\mathcal{S}^x|^2 + |\mathcal{S}^y|^2$ and $V(k,\omega) = -2 \mathrm{Im} ( \mathcal{S}^x \mathcal{S}^{y,\ast} )$ \cite{Barker2016}, where $\boldsymbol{\mathcal{S}}=\boldsymbol{\mathcal{S}}(k,\omega)$ is the space and time Fourier transform of the spin configuration $\{\mathbf{S}_i(t)\}$. The quantity $\sigma(k,\omega) = V(k,\omega)/I(k,\omega)$ measures the ratio of circular to total polarization intensity; its sign reveals the sign of the magnon spin. There is no feature of $\sigma(k,\omega)$ in Fig.~\ref{fig:magnon-spectra}(b), in agreement with the previous section. In contrast, $\sigma(k,\omega)$ becomes zero (gray color) in Fig.~\ref{fig:magnon-spectra}(d), indicating that the magnon spin is suppressed due to ellipticity or ``squeezing'', which is in agreement with Ref.~\cite{PhysRevB.96.020411}. Without squeezing but nonzero DMI we identify spin-up and spin-down magnons by the antisymmetric blue-red features in Fig.~\ref{fig:magnon-spectra}(f). In the presence of squeezing and DMI this asymmetric feature survives [panel (h)] and shows that the spin expectation value continuously goes through zero upon crossing $k=0$, an observation which is in agreement with Fig.~\ref{fig:Spinchain}(c).

In the previous section, we obtained a nonzero magnonic spin polarization for the case in Fig.~\ref{fig:magnon-spectra}(g) and (h) [which are respectively reminiscent of Fig.~\ref{fig:Spinchain}(b) and (c)], but zero effect for the other cases. We will now put this prediction to the test.

To do so, direct nonequilibrium simulations with an imprinted temperature gradient were performed. The spin chain was separated into three parts of equal length ($160$ spins each). The terminating parts have temperature $T \pm \Delta T/2$, while the temperature in the central part linearly interpolates between the two ends. Following this temperature profile, a heat bath with temperature $T_i$ is assigned to each spin $i$. After establishing a steady state in this nonequilibrium situation, the spin density $\langle \mathbf{S} \rangle = \frac{1}{160} \sum_{i=161}^{320} \langle \mathbf{S}_i \rangle$ of the central chain segment is measured and normalized to the number of spins in this segment. 

There is an additional technicality of the simulation:
Since two neighboring spins in the central chain segment experience slightly different temperatures ($T_i \ne T_{i+1}$), their net moment does not compensate exactly. Repeating this argument for all spins of the central segment, we conclude that there is a tiny net magnetization simply due to the temperature dependence of the sublattice magnetizations. The sign of this artificial magnetization is determined by the direction of the first spin at the cold end of the central segment. This artificial effect would superimpose with the magnon analogue of the Edelstein effect. Thus, to avoid the non-Edelstein contribution, we simulate two uncoupled spin chains with opposite spin textures in parallel. The non-Edelstein contributions are exactly opposite, because the sublattice magnetization is reversed, and sum to zero. In contrast, the extrinsic Edelstein contributions are time-reversal even as shown in Eq.~\eqref{eq:symmetry-ex-edel} and do \emph{not} cancel out.

\begin{figure}
  \centering
  \includegraphics[width=1.0\columnwidth]{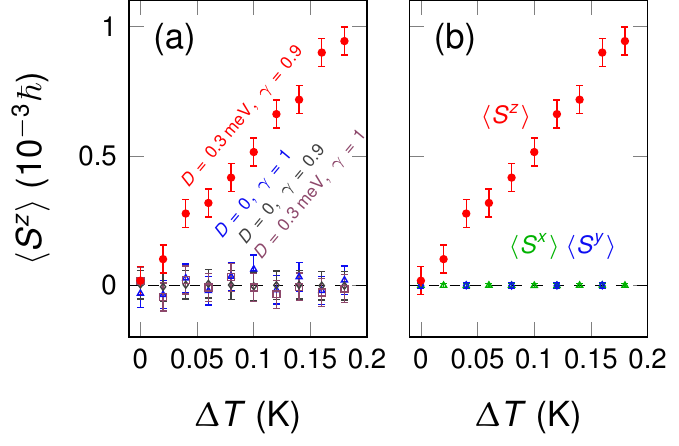}
  \caption{Results from direct nonequilibrium spin dynamics simulations of the thermally induced magnonic analogue of the Edelstein effect in an anisotropic antiferromagnetic spin chain; parameters read $J=\unit[1]{meV}$, $\lambda = 1.02$, and $\alpha = 10^{-4}$. (a) Nonequilibrium spin density $\langle S^z \rangle$ in dependence on temperature difference $\Delta T$ for selected parameter combinations. (b) $\langle S^i \rangle$ ($i=x,y,z$) in dependence on $\Delta T$. An average temperature of $T=\unit[0.2]{K}$ was used for all simulations.} 
  \label{fig:SimulationChainResults}             
\end{figure} 

Our simulation results are presented in Fig.~\ref{fig:SimulationChainResults}.
The $z$ spin accumulation $\langle S^z \rangle$ is zero in equilibrium [$\Delta T = 0$ in Fig.~\ref{fig:SimulationChainResults}(a)], as expected for an antiferromagnet in zero magnetic field. It stays zero in nonequilibrium ($\Delta T \ne 0$), if either DMI or squeezing (or both) are absent [compare brown, blue, and purple marks in Fig.~\ref{fig:SimulationChainResults}(a)]. However, it becomes nonzero if DMI and squeezing are present (red marks), in full agreement with theory.

The other Cartesian components of the spin density, i.e, $\langle S^x \rangle$ and $\langle S^y \rangle$ are zero even in nonequilibrium [blue and green marks in Fig.~\ref{fig:SimulationChainResults}(b)]. This is not surprising, because no magnon state has a nonzero $x$ or $y$ spin. Thus, a nonequilibrium state cannot give rise to spin density of those components. In contrast, $\langle S^z \rangle$ increases approximately linearly with the external force $\Delta T$. 

We note in passing other results that are not explicitly shown. We found that (i) reversing $D$ reverses $\langle S^z \rangle$ due to the reversion of the magnon spin, (ii) increasing $\lambda$ increases the spin wave gap, leading to a decreasing $\langle S^z \rangle$, and (iii) increasing the Gilbert damping $\alpha$ diminishes the $\langle S^z \rangle$, because the magnon transport lifetime decreases.

Overall, we find excellent qualitative agreement with theory (Sec.~\ref{sec:ExampleAFChain}). However, we mention that we cannot compare numbers, because the classical white noise used to model temperature bath results in a Rayleigh-Jeans distribution rather than in the true Bose-Einstein distribution. Thus, the simulation suffers from the classical equipartition and does not account for the quantum freezing of degrees of freedom as temperature goes to zero.

\section{Conclusion}
\label{sec:Conclusion}

\begin{figure}
	\includegraphics[width=1.\columnwidth]{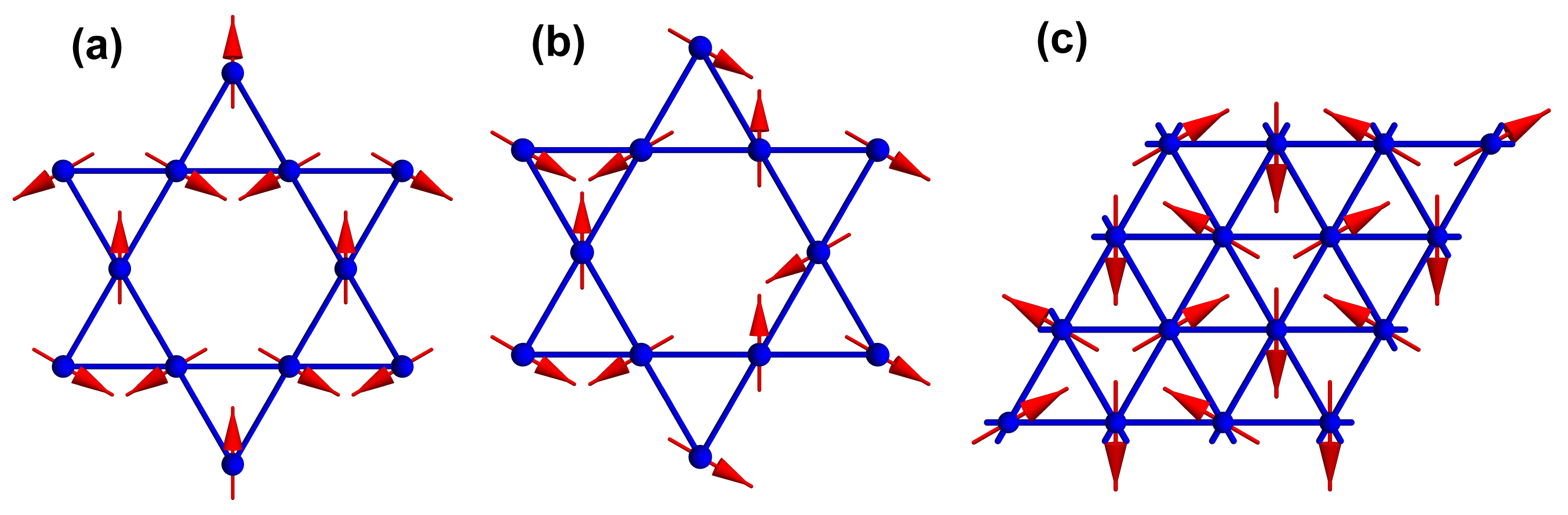}
	\caption{(Color online). (a), (b) Noncollinear spin textures on the kagome lattice, with (a) negative vector chirality (NVC) and (b) staggered vector chirality (SVC). (c) Noncollinear antiferromagnetic ground state on the 2D triangular lattice.}
	\label{fig:2D_NAFMs}
\end{figure}

We have shown that a temperature gradient can induce a nonequilibrium spin density due to magnonic transport in antiferromagnetic insulators with inversion asymmetry and spin non-conservation. Our linear response theory revealed both intrinsic and extrinsic contributions that behave differently under time reversal. Consequently, these two contributions correspond to different elements of the response tensor, which can facilitate their experimental disentanglement, e.g., in the presence of magnetic domains. Our proposal can be realized in (quasi-)2D and 3D noncollinear antiferromagnets, for which we find sizable effects in realistic material candidates.
Our predictions can be tested by transport measurements similar to
those used for detection of the inverse spin Hall effect, by magnetooptical Kerr microscopy, or by magnetic sensing
based on the nitrogen-vacancy (NV) centres.
Given the omnipresence of inversion-symmetry-breaking interfaces (or surfaces) in experimental setups, observation of the magnonic analogue of the Edelstein effect can stimulate further developments in the field of spintronics.
In particular, with the important role played by the electronic Edelstein effect in modern spintronics in mind, we hope to have stimulated experimental research on the magnonic analogue of the Edelstein effect.  

The abundance of antiferromagnetic materials holds great promise for the identification of well-suited experimental candidates.
In kagome NAFMs, the coplanar magnetic order can exhibit three types of vector chiralities: positive, negative, and staggered, which are respectively abbreviated by PVC, NVC, and SVC~\cite{doi:10.1080/00018732.2018.1571986,PhysRevB.95.094427} and depicted in Figs.~\ref{fig:KagomeAF}(a), \ref{fig:2D_NAFMs}(a), and \ref{fig:2D_NAFMs}(b). Their distinct magnetic symmetries cause distinct magnonic spin polarization responses, which are tabulated in Table~\ref{table:Symmetry}. 
Besides kagome magnets, quasi-2D triangular antiferromagnets [cf.~Fig.~\ref{fig:2D_NAFMs}(c)] with the $120^\circ$ spin order~\cite{PhysRevB.79.144416,Gvozdikova_2011} could be suitable candidates. Such systems as RbFe(MoO$_4$)$_2$ \cite{PhysRevB.67.094434} and Ba$_3$NiNb$_2$O$_9$ \cite{PhysRevLett.109.257205} share symmetries with the PVC kagome NAFMs, resulting in identical response tensor shapes [cf.~Table~\ref{table:Symmetry}]. Similar to kagome NAFMs, the 3D breathing pyrochlores can exhibit magnetic orders different from the AIAO order~\cite{Li2016,PhysRevB.98.045109}, which changes their magnetic symmetries and, thus, the expected response tensor shapes. Experimentally, the breathing pyrochlore  materials Ba$_3$Yb$_2$Zn$_5$O$_{11}$\cite{PhysRevLett.116.257204,PhysRevB.93.220407},  LiInCr$_4$O$_8$ \cite{PhysRevLett.113.227204} have been studied, all of which may be considered for a proof-of-principle study of our predictions.
\begin{acknowledgments}
We thank S.~Sandhoefner for helpful discussions. This work was supported by DOE Early Career Award No. DE-SC0014189. 
\end{acknowledgments}


\appendix
\begin{widetext}	
	
\section{Linear Response for Antiferromagnets}
\label{sec:AppLinRep}

\subsection{General Theory}
For the $\mu$ component of a spatially averaged observable $A_\mu=\frac{1}{V}\int d\mathbf{r}\Psi^\dagger(\mathbf{r})\hat{A}_\mu\Psi(\mathbf{r})$, the non-equilibrium response to a temperature gradient is
\begin{equation}
    \langle A_\mu\rangle_{\text{ne}}=\lim_{\omega\to 0}\frac{1}{i\omega}[\Pi_{\mu\nu}(\omega)-\Pi_{\mu\nu}(0)]\nabla_\nu\phi,
\end{equation}
where the correlator in frequency space is defined as  
\begin{eqnarray}
	\Pi_{\mu\nu}(i\omega_m)=-\int_{0}^{\beta}d\tau e^{i\omega_m\tau}\langle T_\tau A_\mu(\tau)J_\nu^q(0)\rangle.
\end{eqnarray}
In momentum space, $A_\mu=\frac{1}{V}\sum_{\mathbf{k}}\Psi^\dagger_\mathbf{k} A_{\mu,\mathbf{k}}\Psi_\mathbf{k}$ and $J^q_\nu=\sum_{\mathbf{k}}\Psi^\dagger_\mathbf{k}J^q_{\nu,\mathbf{k}}\Psi_\mathbf{k}$, with $J_{\nu,\mathbf{k}}^q=\frac{1}{4}(\mathcal{H}_\mathbf{k}\sigma_3\mathbf{v}_{\nu,\mathbf{k}}+\mathbf{v}_{\nu,\mathbf{k}}\sigma_3\mathcal{H}_\mathbf{k})$. 
Here, $J_\nu^q$ comes from $\frac{\partial H^\prime}{\partial t}=\frac{i}{\hbar}[H,H^\prime]=J^q_\nu\nabla_\nu\phi$, see the supplementary of Refs. \cite{PhysRevB.93.161106,PhysRevLett.117.217203}. Plugging in above expressions, the correlation tensor can be presented as
\begin{eqnarray}\label{A3}
		\Pi_{\mu\nu}(i\omega_m)&=&-\frac{1}{V}\sum\limits_{\mathbf{k},\mathbf{k}^\prime}\int_{0}^{\beta}d\tau e^{i\omega_m\tau}\langle\Psi^\dagger_\mathbf{k}(\tau)A_{\mu,\mathbf{k}}\Psi_\mathbf{k}(\tau)\Psi_{\mathbf{k}^\prime}^\dagger J^q_{\nu,\mathbf{k}^\prime}\Psi_{\mathbf{k}^\prime}\rangle\nonumber\\
		&=&-\frac{1}{V}\sum\limits_{\mathbf{k},\mathbf{k}^\prime}\int_{0}^{\beta}d\tau e^{i\omega_m\tau}(A_{\mu,\mathbf{k}})_{\alpha\gamma}(J^q_{\nu,\mathbf{k}\prime})_{\rho\sigma}\langle T_\tau\Psi^\dagger_{\mathbf{k},\alpha}(\tau)\Psi_{\mathbf{k},\gamma}(\tau)\Psi^\dagger_{\mathbf{k}^\prime,\rho}(0)\Psi_{\mathbf{k}^\prime,\sigma}(0)\rangle.
\end{eqnarray}
According to Wick's theorem,
\begin{eqnarray}\label{A4}
			&&\langle T_\tau\Psi^\dagger_{\mathbf{k},\alpha}(\tau)\Psi_{\mathbf{k},\gamma}(\tau)\Psi^\dagger_{\mathbf{k}^\prime,\rho}(0)\Psi_{\mathbf{k}^\prime,\sigma}(0)\rangle_\mathrm{connected}\nonumber\\
			&=&\langle T_\tau\Psi_{\mathbf{k}^\prime,\sigma}(0)\Psi_{\mathbf{k},\alpha}^\dagger(\tau)\rangle\langle T_\tau\Psi_{\mathbf{k},\gamma}(\tau)\Psi_{\mathbf{k}\prime,\rho}^\dagger(0)\rangle+\langle T_\tau\Psi_{\mathbf{k},\alpha}^\dagger(\tau)\Psi_{\mathbf{k}\prime,\rho}^\dagger(0)\rangle  \langle T_\tau\Psi_{\mathbf{k}^\prime,\gamma}(\tau)\Psi_{\mathbf{k},\sigma}(0)\rangle.
\end{eqnarray}
Here, the second anomalous term can be shown to be equivalent to the first term. First, we note that the basis $\Psi_\mathbf{k}$ obeys the particle-hole symmetry, $\Psi_\mathbf{k}=(\Psi_{-\mathbf{k}}^\dagger\sigma_1)^T$, which leads to the relation
\begin{eqnarray}\label{A5}
    A_\mu&=&\frac{1}{V}\sum\limits_{\mathbf{k},\alpha\beta}\Psi_{\mathbf{k},\alpha}^\dagger(A_{\mu,\mathbf{k}})_{\alpha\beta}\Psi_{\mathbf{k},\beta}=\frac{1}{V}\sum\limits_{\mathbf{k},\lambda\gamma}\Psi_{-\mathbf{k},\lambda}^\dagger(\sigma_1A_{\mu,\mathbf{k}}^T\sigma_1)_{\lambda\gamma}\Psi_{-\mathbf{k},\gamma}.
\end{eqnarray}
Hence, we gain the relation: $\sigma_1A_{\mathbf{k},\mu}^T\sigma_1=A_{\mu,-\mathbf{k}}$, which will be used repeatedly in the later proof. Second, the systematic linear response analysis needs a plain expression of the particle-hole space Green function, whose definition is $\mathcal{G}(\mathbf{k},\tau;\mathbf{k}^\prime,0)_{i,j}\equiv\mathcal{G}(\mathbf{k},\mathbf{k}^\prime;\tau)_{i,j}\equiv-\langle T_\tau\Psi_{\mathbf{k},i}(\tau)\Psi_{\mathbf{k}^\prime,j}^\dagger(0)\rangle$. We derive the Green function expression by virtue of its equation of motion,
\begin{eqnarray}\label{A6}
    \partial_\tau\mathcal{G}(\mathbf{k},\mathbf{k}^\prime;\tau)_{\alpha\beta}=-\delta(\tau)\sigma_{3,\alpha\beta}\delta_{\mathbf{k},\mathbf{k}^\prime}-(\sigma_3 \mathcal{H}_\mathbf{k})_{\alpha\gamma}\mathcal{G}(\mathbf{k},\mathbf{k}^\prime;\tau)_{\gamma\beta},
\end{eqnarray}
where we used the relation
\begin{eqnarray}
    \partial_\tau\Psi_{\mathbf{k},\alpha}(\tau)=[H,\Psi_{\mathbf{k},\alpha}(\tau)]
    =-\frac{1}{2}(\sigma_3\mathcal{H}_\mathbf{k})_{\alpha\gamma}\Psi_{\mathbf{k},\gamma}+\frac{i}{2}\Psi^\dagger_{-\mathbf{k},\gamma}(\mathcal{H}_{-\mathbf{k}}\sigma_2)_{\gamma\alpha}=-(\sigma_3\mathcal{H}_\mathbf{k})_{\alpha\gamma}\Psi_{\mathbf{k},\gamma}.
\end{eqnarray}
The equation of motion [Eq.~\eqref{A6}] in matrix form reads
\begin{equation}
    (\partial_\tau+\sigma_3 \mathcal{H}_\mathbf{k})\mathcal{G}(\mathbf{k},\mathbf{k}^\prime;\tau)=-\sigma_3\delta(\tau)\delta_{\mathbf{k},\mathbf{k}^\prime},
\end{equation}
so that $\mathcal{G}(\mathbf{k},\mathbf{k}^\prime;\tau)=\frac{-\sigma_3\delta(\tau)\delta_{\mathbf{k},\mathbf{k}^\prime}}{\partial_\tau+\sigma_3 \mathcal{H}_\mathbf{k}}$ and $\mathcal{G}(\mathbf{k},
 \mathbf{k}^\prime;ik_n)=\frac{\sigma_3}{ik_n-\sigma_3 \mathcal{H}_\mathbf{k}}\delta_{\mathbf{k},\mathbf{k}^\prime}$ in frequency-momentum space.\\

Now we show that the anomalous term in Eq.~\eqref{A4} can be alternatively expressed, with the help of particle-hole symmetry, in form of Green function
\begin{eqnarray}
    \langle T_\tau\Psi_{\mathbf{k},\alpha}^\dagger(\tau)\Psi_{\mathbf{k}\prime,\rho}^\dagger(0)\rangle&=&\langle T_\tau\sigma_{1,\alpha\delta}\Psi_{-\mathbf{k},\delta}(\tau)\Psi_{\mathbf{k}\prime,\rho}^\dagger(0)\rangle=-\sigma_{1,\alpha\delta}\mathcal{G}(-\mathbf{k},\mathbf{k}^\prime;\tau)_{\delta\rho},\nonumber\\
    \langle T_\tau\Psi_{\mathbf{k}^\prime,\gamma}(\tau)\Psi_{\mathbf{k},\sigma}(0)\rangle&=&\langle T_\tau\Psi_{\mathbf{k},\gamma}(\tau)\Psi_{-\mathbf{k}^\prime,\mu}^\dagger(0)\sigma_{1,\mu\sigma}\rangle\nonumber=-\mathcal{G}(\mathbf{k},-\mathbf{k}^\prime;\tau)_{\gamma\mu}\sigma_{1,\mu\sigma}.
\end{eqnarray}
Therefore, Eq.~\eqref{A4} and the correlation tensor in Eq.~\eqref{A3} are rewritten in terms of Green function as
\begin{eqnarray}
    &&\langle T_\tau\Psi^\dagger_{\mathbf{k},\alpha}(\tau)\Psi_{\mathbf{k},\gamma}(\tau)\Psi^\dagger_{\mathbf{k}^\prime,\rho}(0)\Psi_{\mathbf{k}^\prime,\sigma}(0)\rangle=\mathcal{G}_{\sigma\alpha}(\mathbf{k}^\prime,\mathbf{k};-\tau)\mathcal{G}_{\gamma\rho}(\mathbf{k},\mathbf{k}^\prime;\tau)+[\sigma_1\mathcal{G}(-\mathbf{k},\mathbf{k}^\prime;\tau)]_{\alpha\rho}[\mathcal{G}(\mathbf{k},-\mathbf{k}^\prime;\tau)\sigma_1]_{\gamma\sigma},
\end{eqnarray}
and
\begin{eqnarray}\label{A10}
    \Pi_{\mu\nu}(i\omega_m)
    =-\frac{1}{V}\sum\limits_{\mathbf{k},\mathbf{k}^\prime}\int_{0}^{\beta}d\tau e^{i\omega_m\tau}(A_{\mu,\mathbf{k}})_{\alpha\gamma}(J^q_{\nu,\mathbf{k}\prime})_{\rho\sigma}\{\mathcal{G}_{\sigma\alpha}(\mathbf{k}^\prime,\mathbf{k};-\tau)\mathcal{G}_{\gamma\rho}(\mathbf{k},\mathbf{k}^\prime;\tau)+[\sigma_1\mathcal{G}(-\mathbf{k},\mathbf{k}^\prime;\tau)]_{\alpha\rho}[\mathcal{G}(\mathbf{k},-\mathbf{k}^\prime;\tau)\sigma_1]_{\gamma\sigma}\},\nonumber\\
\end{eqnarray}
respectively.
Furthermore, with the aid of the Green function relation $\mathcal{G}(-\mathbf{k},\tau)
 =-\sigma_1\mathcal{G}(\mathbf{k},-\tau)^T\sigma_1$, we can prove the equivalence of the first and second part on the right hand side of Eq.~\eqref{A10}. As a result, the correlation function becomes
\begin{eqnarray}
	\Pi_{\mu\nu}(i\omega_m)=-\frac{2}{V}\sum\limits_{\mathbf{k}}\int_{0}^{\beta}d\tau e^{i\omega_m\tau}\text{tr}[A_{\mu,\mathbf{k}}\mathcal{G}(\mathbf{k},\tau)J_{\nu,\mathbf{k}}^q \mathcal{G}(\mathbf{k};-\tau)],
\end{eqnarray}
where $\mathcal{G}(\mathbf{k},\tau)=\frac{\sigma_3}{ik_n-\sigma_3 \mathcal{H}_\mathbf{k}}$. Let's transform the Green function to frequency space with $\mathcal{G}(\mathbf{k};\tau)=\frac{1}{\beta}\sum\limits_{iq_n}e^{-iq_n\tau}\mathcal{G}(\mathbf{k};iq_n)$, then
\begin{eqnarray}
    \Pi_{\mu\nu}(i\omega_m)
    &=&\frac{2}{V}\sum\limits_{\mathbf{k}}\int_{-\infty}^{+\infty}\frac{d\omega_1}{2\pi}\frac{d\omega_2}{2\pi}\text{tr}[A_{\mu,\mathbf{k}} S(\mathbf{k},\omega_1)J_{\nu,\mathbf{k}}^q  S(\mathbf{k},\omega_2)]\frac{n_\mathrm{B}(\omega_1)-n_\mathrm{B}(\omega_2)}{\omega_1-\omega_2-i\omega_m}.
\end{eqnarray}
Here, we performed the Matsubara summation and utilized $\mathcal{G}(\mathbf{k};ik_n)= \int_{-\infty}^{+\infty}\frac{d\omega}{2\pi}\frac{S(\mathbf{k},\omega)}{ik_n-\omega}$, with $S(\mathbf{k},\omega)$ being the spectral function. Going back to the real time space and taking the zero frequency limit, we obtain the response tensor
\begin{eqnarray}\label{A13}
    K_{\mu\nu}&=&-i\frac{\partial \Pi_{\mu\nu}(\omega+i0^+)}{\partial\omega}|_{\omega\to 0}\nonumber\\
    &=&\frac{2}{V}\sum\limits_{\mathbf{k}}\int_{-\infty}^{+\infty}\frac{d\varepsilon}{2\pi}n_\mathrm{B}(\varepsilon)\text{tr}[(G^R-G^A)(A_{\mu,\mathbf{k}}\frac{\partial G^R}{\partial\varepsilon}J^q_{\nu,\mathbf{k}}-J^q_{\nu,\mathbf{k}}\frac{\partial G^A}{\partial\varepsilon}A_{\mu,\mathbf{k}})],
\end{eqnarray}
where we used the relation
\begin{align}
    \int_{-\infty}^{\infty}\frac{d\omega}{2\pi}\frac{S(k,\omega)}{(\varepsilon-\omega\pm i0^+)^2}=-\frac{\partial}{\partial\varepsilon}\int_{-\infty}^{\infty}\frac{d\omega}{2\pi}\frac{S(k,\omega)}{\varepsilon-\omega\pm i0^+}=-\frac{\partial G^{R/A}}{\partial\varepsilon}    
\end{align}
and the expression $S(\mathbf{k},\varepsilon)=i(G^R-G^A)$.

\subsection{In the Eigenstate Basis}
To distinguish the intraband and interband contributions, we rewrite the response tensor in Eq.~\eqref{A13} in the eigenstate basis via the transformation $\Psi_\mathbf{k}=T_\mathbf{k}\Gamma_\mathbf{k}$. By definition, we have the Green function transformation $ \mathcal{G}(\mathbf{k};\tau)=T_\mathbf{k}g(\mathbf{k},\tau)T_\mathbf{k}^\dagger$, where $g(\mathbf{k},\tau)= -\langle T_\tau\Gamma_\mathbf{k}(\tau)\Gamma_\mathbf{k}^\dagger(0)\rangle$ and $g^{R/A}(\mathbf{k},\varepsilon)=\frac{\sigma_3}{\varepsilon-\sigma_3\mathcal{E}_\mathbf{k}\pm i0^+}$. After this transformation, we obtain
\begin{equation}
    K_{\mu\nu}=\frac{2}{V}\sum_{\mathbf{k}}\int_{-\infty}^{+\infty}\frac{d\varepsilon}{2\pi}n_\mathrm{B}(\varepsilon)\text{tr}[(g^R-g^A)(\mathcal{A}_{\mu,\mathbf{k}}\frac{\partial g^R}{\partial\varepsilon}\mathcal{J}_{\nu,\mathbf{k}}-\mathcal{J}_{\nu,\mathbf{k}}\frac{\partial g^A}{\partial\varepsilon}\mathcal{A}_{\mu,\mathbf{k}})],
\end{equation}
where $\mathcal{J}_{\nu,\mathbf{k}}=T^\dagger_\mathbf{k}J_{\nu,\mathbf{k}}^qT_\mathbf{k}$ and $\mathcal{A}_{\mu,\mathbf{k}}=T_\mathbf{k}^\dagger A_{\mu,\mathbf{k}}T_\mathbf{k}$. We split the expression into two parts: intraband and interband contributions. Owing to the hermitian conjugate property of operators, we write the response tensor elements as
\begin{eqnarray}
     K_{\mu\nu}&=&\frac{2}{V}\sum\limits_{\mathbf{k}}\sum\limits_{mn}\int_{-\infty}^{+\infty}\frac{d\varepsilon}{2\pi}n_\mathrm{B}(\varepsilon)[(g_m^R-g_m^A)((\mathcal{A}_{\mu,\mathbf{k}})_{mn}\frac{\partial g^R_n}{\partial\varepsilon}(\mathcal{J}_{\nu,\mathbf{k}})_{nm}-(\mathcal{J}_{\nu,\mathbf{k}})_{mn}\frac{\partial g^A_n}{\partial\varepsilon}(\mathcal{A}_{\mu,\mathbf{k}})_{nm})]\nonumber\\
    &=&\frac{2i}{V}\sum\limits_\mathbf{k}\sum\limits_{mn}(\mathcal{A}_{\mu,\mathbf{k}})_{mn}(\mathcal{J}_{\nu,\mathbf{k}})_{nm}\frac{\sigma_{3,mm}\sigma_{3,nn}[n_\mathrm{B}((\sigma_3\mathcal{E}_\mathbf{k})_{mm})-n_\mathrm{B}((\sigma_3\mathcal{E}_\mathbf{k})_{nn})]}{[(\sigma_3\mathcal{E}_\mathbf{k})_{mm}-(\sigma_3\mathcal{E}_\mathbf{k})_{nn}+i 0^+]^2},
\end{eqnarray}
where we took the approximation $g_m^R-g_m^A=i2\text{Im}(g_m^R)=-i2\pi\sigma_{3,mm}\delta[\varepsilon-(\sigma_3\mathcal{E}_\mathbf{k})_{mm}]$. If we incorporate the magnon spectrum broadening $\Gamma_m$ into the Green function, i.e., $g^R_m(\varepsilon)=\frac{\sigma_{3,mm}}{\varepsilon-(\sigma_3\mathcal{E}_\mathbf{k})_{mm}+i\Gamma_m}$, the response tensor can be naturally divided into two parts, $K_{\mu\nu}=K^{\text{intra}}_{\mu\nu}+K^{\text{inter}}_{\mu\nu}$, where
\begin{equation}\label{Intraband}
    K_{\mu\nu}^{\text{intra}}=\frac{1}{V}\sum\limits_\mathbf{k}\sum\limits_{n}\frac{1}{\Gamma_n}(\mathcal{J}_{\mathbf{k},\nu})_{nn}(\mathcal{A}_{\mu,\mathbf{k}})_{nn}\partial_\varepsilon n_\mathrm{B}[(\sigma_3\mathcal{E}_\mathbf{k})_{nn}],
\end{equation}
and
\begin{eqnarray}\label{Interband}
    && K_{\mu\nu}^{\text{inter}}=\frac{2i}{V}\sum\limits_\mathbf{k}\sum\limits_{m\neq n}(\mathcal{A}_{\mu,\mathbf{k}})_{mn}(\mathcal{J}_{\nu,\mathbf{k}})_{nm}\frac{\sigma_{3,mm}\sigma_{3,nn}[n_\mathrm{B}((\sigma_3\mathcal{E}_\mathbf{k})_{mm})-n_\mathrm{B}((\sigma_3\mathcal{E}_\mathbf{k})_{nn})]}{[(\sigma_3\mathcal{E}_\mathbf{k})_{mm}-(\sigma_3\mathcal{E}_\mathbf{k})_{nn}]^2}.
\end{eqnarray}
The limit $\Gamma_n \to 0$ for $K_{\mu\nu}^{\text{inter}}$ is taken here.
In consideration of $A_\mu^\dagger=A_\mu$ and $(J^q_\nu)^\dagger=J_\nu^q$, Eq.~\eqref{Interband} can be transformed to
\begin{eqnarray}\label{Interband1}
    && K_{\mu\nu}^{\text{inter}}=\frac{4}{V}\sum\limits_\mathbf{k}\sum\limits_{m\neq n}\frac{\text{Im}[(\sigma_3\mathcal{A}_{\mu,\mathbf{k}})_{nm}(\sigma_3\mathcal{J}_{\mathbf{k},\nu})_{mn}]n_\mathrm{B}[(\sigma_3\mathcal{E}_\mathbf{k})_{nn}]}{[(\sigma_3\mathcal{E}_\mathbf{k})_{mm}-(\sigma_3\mathcal{E}_\mathbf{k})_{nn}]^2}.
\end{eqnarray}
The intraband response  Eq.~\eqref{eq:Intraband} in the main text can be recovered if we consider $\mathcal{J}_{\mathbf{k},\nu}=\frac{1}{4}(\mathcal{E}_\mathbf{k}\sigma_3\tilde{v}_{\mathbf{k},\nu}+\tilde{v}_{\mathbf{k},\nu}\sigma_3\mathcal{E}_\mathbf{k})$ whose diagonal components read
\begin{equation}
    (\mathcal{J}_{\mathbf{k},\nu})_{nn}=\frac{1}{2}(\sigma_3\mathcal{E}_\mathbf{k})_{nn}(\tilde{v}_{\nu,\mathbf{k}})_{nn},
\end{equation}
where
\begin{eqnarray}
    \tilde{v}_{\mathbf{k},\nu}
    =\partial_{k_\nu}\mathcal{E}_\mathbf{k}-(\partial_{k_\nu} T_\mathbf{k}^\dagger)\mathcal{H}_\mathbf{k}T_\mathbf{k}-T_\mathbf{k}^\dagger \mathcal{H}_\mathbf{k}(\partial_{k_\nu} T_\mathbf{k}).
\end{eqnarray}
From the paraunitary relation of $T_\mathbf{k}$ and $\partial_{k_\nu}(T_\mathbf{k}\sigma_3T_\mathbf{k}^\dagger)=0$, we get
$\partial_{k_\nu} T_\mathbf{k}^\dagger=-T_\mathbf{k}\sigma_3(\partial_{k_\nu} T_\mathbf{k})\sigma_3 T_\mathbf{k}^\dagger$. From $T_\mathbf{k}^\dagger \mathcal{H}_\mathbf{k}T_\mathbf{k}=\mathcal{E}_\mathbf{k}$ and $(T_\mathbf{k})^{-1}=\sigma_3T_\mathbf{k}^\dagger\sigma_3$, we have $T_\mathbf{k}^\dagger \mathcal{H}_\mathbf{k}=\mathcal{E}_\mathbf{k}\sigma_3T_\mathbf{k}^\dagger\sigma_3$. Therefore, the diagonal elements of $\tilde{v}_{\mathbf{k},\nu}$ are shown to be
\begin{eqnarray}
    (\tilde{v}_{\mathbf{k},\nu})_{nn}
    =(\partial_{k_\nu}\mathcal{E}_\mathbf{k})_{nn}+(T_\mathbf{k}^\dagger\sigma_3\partial_{k_\nu} T_\mathbf{k}\sigma_3\mathcal{E}_\mathbf{k})_{nn}-(\mathcal{E}_\mathbf{k}\sigma_3T_\mathbf{k}^\dagger\sigma_3\partial_{k_\nu} T_\mathbf{k})_{nn}=(\partial_\nu\mathcal{E}_\mathbf{k})_{nn};
\end{eqnarray}
thus,
\begin{equation}\label{Currentoperator}
    (\mathcal{J}_{\mathbf{k},\nu})_{nn}=\frac{1}{2}(\sigma_3\mathcal{E}_\mathbf{k})_{nn}(\partial_{k_\nu}\mathcal{E}_\mathbf{k})_{nn}.
\end{equation}
By inserting Eq.~\eqref{Currentoperator} into Eq.~\eqref{Intraband}, we arrive at
\begin{equation}
K_{\mu\nu}^{\text{intra}}=\frac{1}{V}\sum\limits_{\mathbf{k}}\sum_{n=1}^{2N}\frac{1}{2\Gamma_n}(\mathcal{A}_{\mu,\mathbf{k}})_{nn}\partial_{k_\nu}\mathcal{E}_{\mathbf{k},nn}(\sigma_3\mathcal{E}_\mathbf{k})_{nn}\partial_\varepsilon n_\mathrm{B}[(\sigma_3\mathcal{E}_\mathbf{k})_{nn}].
\end{equation}
Given the relation $n_\mathrm{B}(x)=-1-n_\mathrm{B}(-x)$, the band index can be confined to the particle space, i.e., $1\leq n\leq N$,
\begin{equation}
    K_{\mu\nu}^{\text{intra}}=\frac{1}{V}\sum\limits_{\mathbf{k}}\sum_{n=1}^{N}\frac{1}{2\Gamma_n}[(\mathcal{A}_{\mu,\mathbf{k}})_{nn}+(\mathcal{A}_{\mu,-\mathbf{k}})_{(n+N)(n+N)}]\partial_{k_\nu}\mathcal{E}_{\mathbf{k},nn}\mathcal{E}_{\mathbf{k},nn}\partial_\varepsilon n_\mathrm{B}[\mathcal{E}_{\mathbf{k},nn}].
\end{equation}
Applying particle-hole symmetry (PHS), $(\mathcal{A}_{\mu,\mathbf{k}})_{nn}=(\mathcal{A}_{\mu,-\mathbf{k}})_{(n+N)(n+N)}$, replacing $\mathcal{A}_{\mu,\mathbf{k}}$ by $\mathcal{S}_{\mu,\mathbf{k}}$ and taking $\nabla_\nu \phi=-\nabla_\nu T/T$ into account, we can obtain the intraband response Eq.~\eqref{eq:Intraband}. \\

On the other hand, by plugging the expression of $\mathcal{J}_{\nu,\mathbf{k}}$ into Eqs.~\eqref{Interband} or \eqref{Interband1}, the interband part can be reorganized as below
\begin{eqnarray}
    K_{\mu\nu}^{\text{inter}}&&=\frac{1}{V}\sum\limits_\mathbf{k}\sum\limits_{m\neq n}\frac{i}{2}(\mathcal{A}_{\mu,\mathbf{k}})_{nm}[(\sigma_3\mathcal{E}_\mathbf{k})_{mm}(v_{\nu})_{mn}+(v_{\nu})_{mn}(\sigma_3\mathcal{E}_\mathbf{k})_{nn}]\frac{\sigma_{3,mm}\sigma_{3,nn}[n_\mathrm{B}((\sigma_3\mathcal{E}_\mathbf{k})_{nn})-n_\mathrm{B}((\sigma_3\mathcal{E}_\mathbf{k})_{mm})]}{[(\sigma_3\mathcal{E}_\mathbf{k})_{mm}-(\sigma_3\mathcal{E}_\mathbf{k})_{nn}]^2},\nonumber\\
    &&=\frac{1}{V}\sum\limits_\mathbf{k}\sum_{n=1}^{2N}-(\Omega^A_{n,\mathbf{k}})_{\mu\nu}\bar{\varepsilon}_{n,\mathbf{k}}n_\mathrm{B}(\bar{\varepsilon}_{n,\mathbf{k}})-(m^A_{n,\mathbf{k}})_{\mu\nu} n_\mathrm{B}(\bar{\varepsilon}_{n,\mathbf{k}}),
\end{eqnarray}
with
\begin{eqnarray}
    &&(\Omega^A_{n,\mathbf{k}})_{\mu\nu}=\sum_{m(\neq n)}\frac{2\text{Im}[(\sigma_3 \mathcal{A}_{\mu,\mathbf{k}})_{nm}(\sigma_3 \tilde{v}_{\nu,\mathbf{k}})_{mn}]}{(\bar{\varepsilon}_{n,\mathbf{k}}-\bar{\varepsilon}_{m,\mathbf{k}})^2},\nonumber\\
    &&(m^A_{n,\mathbf{k}})_{\mu\nu}=\sum_{m(\neq n)}\frac{-\text{Im}[(\sigma_3 \mathcal{A}_{\mu,\mathbf{k}})_{nm}(\sigma_3 \tilde{v}_{\nu,\mathbf{k}})_{mn}]}{\bar{\varepsilon}_{n,\mathbf{k}}-\bar{\varepsilon}_{m,\mathbf{k}}}.
\end{eqnarray}

\section{Details of the Models}
\label{sec:AppModels}
 
\subsection{Antiferromagnetic Spin Chain}
 
We recapitulate that the Hamiltonian for the antiferromagnetic spin chain is
\begin{eqnarray}
    H&=&\sum_{i}\sum_{\delta=\pm 1}[J(\gamma S^x_{1,i}S^x_{2,i+\nu}+S^y_{1,i}S^y_{2,i+\nu}+\lambda S^z_{1,i}S^z_{2,i+\nu})+D_{12}^\nu\mathbf{e}_z\cdot(\mathbf{S}_{1,i}\times\mathbf{S}_{2,i+\nu})],
\end{eqnarray}
with exchange and DMI parameters as stated in the main text. After performing the Holstein-Primakoff transformation, the quadratic Hamiltonian written in the basis $\Psi_k=(a_{1,k},a_{2,k},a_{1,-k}^\dagger,a_{2,-k}^\dagger)^T$ reads
\begin{equation}\label{B2}
 \mathcal{H}_k=JS\left[ {\begin{array}{cccc}
 	2\lambda & 2\Delta_- cosk & 0 & 2\Delta_+\cos k+i\varrho_k\\
 	2\Delta_- cosk & 2\lambda & 2\Delta_+\cos k+i\varrho_{-k} & 0\\
 	0 & 2\Delta_+\cos k-i\varrho_k & 2\lambda &  2\Delta_- cosk\\
 	2\Delta_+\cos k-i\varrho_{-k} & 0 &  2\Delta_- cosk & 2\lambda
 	\end{array} } \right],
\end{equation}
where $\Delta_\pm=\frac{1\pm\gamma}{2}$, $\varrho_k=\sum_\nu \delta D_\nu e^{ik\nu}/J=i2D_0\sin k+2\delta_ D\cos k$, with $D_0=\frac{D_1+D_2}{2J}$ and $\delta_D=\frac{D_1-D_2}{2J}$.

\subsection{Non-coplanar Kagome Antiferromagnet}
We consider the non-coplanar kagome antiferromagnet discribed by 
\begin{eqnarray}
H=\sum_{\langle ij\rangle}J_1\mathbf{S}_i\cdot\mathbf{S}_j+\mathbf{D}_{ij}\cdot(\mathbf{S}_i\times\mathbf{S}_j)+\sum_{\langle\langle ij\rangle\rangle}J_2 \mathbf{S}_i\cdot\mathbf{S}_j,
\end{eqnarray}
where $\mathbf{D}_{ij}=\mathbf{D}_{p,ij}+D_{z,ij}\hat{z}$.
The spins cant out of the 2-D plane with a small angle $\eta$, and the spins' projection on the the $x-y$ plane form angles $\theta_i$ $(i=1,2,3)$ with respect to $x$ axis, specifically,
$\theta_1=-\pi/6$, $\theta_2=\pi/2$ and $\theta_3=7\pi/6$. For each spin $\mathbf{S}_i$, we choose a local reference frame defined as follow
\begin{eqnarray}
\mathbf{e}_{i,x}=\{\sin\theta_i\,-\cos\theta_i,0\},\qquad \mathbf{e}_{i,y}=\{\sin\eta\cos\theta_i,\sin\eta\sin\theta_i,-\cos\eta\},\qquad \mathbf{e}_{i,z}=\{\cos\eta\cos\theta_i,\cos\eta\sin\theta_i,\sin\eta\}.
\end{eqnarray}
For a given spin $\mathbf{S}_i$, in the global frame, its components can be connected to the local frame expression $\tilde{\mathbf{S}}_i$ by 
\begin{eqnarray}
S_i^\alpha=\mathbf{e}_\alpha\cdot{(\tilde{S}_i^\beta}\mathbf{e}_{i,\beta})=R_{i,\alpha\beta}\tilde{S}_i^\beta,
\end{eqnarray}
where $R_{i,\alpha\beta}=\mathbf{e}_\alpha\cdot\mathbf{e}_{i,\beta}$, or in matrix form,
\begin{eqnarray}
R_i=\left(\begin{array}{ccc}
\sin\theta_i& \sin\eta\cos\theta_i&\cos\eta\cos\theta_i\\
-\cos\theta_i& \sin\eta\sin\theta_i&\cos\eta\sin\theta_i\\
0&-\cos\eta& \sin\eta
\end{array}\right).
\end{eqnarray}
For the general spin-spin interaction a correspondence between the two frames can be written as $S_i^\alpha\Gamma_{\alpha\beta}^{ij}S_j^\beta=\tilde{S}_i^\alpha(R_i^T\Gamma^{ij}R_j)_{\alpha\beta}\tilde{S}_j^\beta$. The interaction matrices are:  $\Gamma^{ij}_{\alpha\beta}=J\delta_{\alpha\beta}$ for exchange and $\Gamma^{ij}_{\alpha\beta}=D_{ij}^\rho\epsilon^{\rho\alpha\beta}$ for DMI. Using these relations, we express the non-interacting spin wave Hamiltonian in terms of the local reference frames as
\begin{eqnarray}
H_{J_1}&&=J_1\sum_{\langle ij\rangle}\cos\theta_{ij}\tilde{\mathbf{S}}_i\cdot\tilde{\mathbf{S}}_j+2\sin^2(\theta_{ij}/2)(\cos^2\eta\tilde{S}_i^y\tilde{S}_j^y+\sin^2\eta\tilde{S}_i^z\tilde{S}_j^z)+\sin\eta\sin\theta_{ij}\hat{z}\cdot(\tilde{\mathbf{S}}_i\times\tilde{\mathbf{S}}_j),\nonumber\\
H_{J_2}&&=J_2\sum_{\langle\langle ij\rangle\rangle}\cos\theta_{ij}\tilde{\mathbf{S}}_i\cdot\tilde{\mathbf{S}}_j+2\sin^2(\theta_{ij}/2)(\cos^2\eta\tilde{S}_i^y\tilde{S}_j^y+\sin^2\eta\tilde{S}_i^z\tilde{S}_j^z)+\sin\eta\sin\theta_{ij}\hat{z}\cdot(\tilde{\mathbf{S}}_i\times\tilde{\mathbf{S}}_j),\nonumber\\
H_{D_z}&&=\sum_{\langle ij\rangle}-s_{ij}D_z[\sin\theta_{ij}(\tilde{S}_i^x\tilde{S}_j^x+\sin^2\eta\tilde{S}_i^y\tilde{S}_j^y+\cos^2\eta \tilde{S}_i^z\tilde{S}_j^z)-\sin\eta\cos\theta_{ij}\hat{z}\cdot(\tilde{\mathbf{S}}_i\times\tilde{\mathbf{S}}_j)],\nonumber\\
H_{D_p}&&=\sum_{\langle ij\rangle}-s_{ij}D_p[\sin(2\eta)\sin(\frac{\theta_{ij}}{2})(\tilde{S}_i^z\tilde{S}_j^z-\tilde{S}_i^y\tilde{S}_j^y)+\cos\eta\cos(\frac{\theta_{ij}}{2})\hat{z}\cdot(\tilde{\mathbf{S}}_i\times\tilde{\mathbf{S}}_j)],\nonumber\\
H_{D_R}&&=\sum_{\langle ij\rangle}-s_{ij}\nu_{ij} D_R[\sin(2\eta)\sin(\frac{\theta_{ij}}{2})(\tilde{S}_i^z\tilde{S}_j^z-\tilde{S}_i^y\tilde{S}_j^y)+\cos\eta\cos(\frac{\theta_{ij}}{2})\hat{z}\cdot(\tilde{\mathbf{S}}_i\times\tilde{\mathbf{S}}_j)].\nonumber\\
\end{eqnarray}
Here we used the notation that $\theta_{ij}=\theta_i-\theta_j=-s_{ij}\frac{2\pi}{3}$, $D_{z,ij}=D_zs_{ij}$ and $\mathbf{D}_{p,ij}=-s_{ij}D_p[\cos(\frac{\theta_i+\theta_j}{2})\hat{x}+\sin(\frac{\theta_i+\theta_j}{2})\hat{y}]$, where $s_{ij}$ is used to express the sign convention: $s_{ij}=1$ as the indices $i,j$ run clockwise around the triangle loop and $s_{ij}=-1$ when they run counter-clockwise. The notation $\nu_{ij}$ takes care of the opposite convention for Rashba-DMI in upward and downward triangles with $\nu_{ij}=\pm 1$ for $(ij)\in\bigtriangleup/\bigtriangledown$. Plugging in the expression of $\theta_{ij}$ and performing the Holstein-Primakoff transformation $\tilde{S}_i^x=\sqrt{\frac{S}{2}}(b^\dagger_i+b_i)$, $\tilde{S}_i^y=i\sqrt{\frac{S}{2}}(b^\dagger_i-b_i)$, $\tilde{S}_i^z=(S-b^\dagger_ib_i)$, we can obtain nearest neighbor interaction
\begin{eqnarray}
H_\mathrm{NN}&&=\frac{1}{2}S\sum_{\langle ij\rangle}[(\Delta_1^{(0)}+\nu_{ij}\Delta_{R}^{(0)}) (b^\dagger_ib_i+b^\dagger_jb_j)+(\Delta_{1,ij}+\nu_{ij}\Delta_{R,ij})b^\dagger_ib_j+h.c.+(\Delta^\prime_{1}+\nu_{ij}\Delta^\prime_{R})b_i^\dagger b_j^\dagger +h.c.]
\end{eqnarray}
with
\begin{eqnarray}
&&\Delta_1^{(0)}=J_1(1-3\sin^2\eta)-\sqrt{3}(D_z\cos^2\eta+D_p\sin(2\eta)),\nonumber\\
&&\Delta_{1,ij}=\Delta^{re}_1+is_{ij}\Delta^{im}_1,\nonumber\\
&&\Delta^{re}_1=\frac{1}{2}[(1-3\sin^2\eta)J_1+\sqrt{3}(1+\sin^2\eta)D_z-\sqrt{3}\sin(2\eta)D_p],\nonumber\\
&&\Delta^{im}_1=\cos\eta D_p+\sin\eta(D_z+\sqrt{3}J_1),\nonumber\\
&&\Delta_1^\prime=\frac{1}{2}[\cos^2\eta(\sqrt{3}D_z-3J_1)+\sqrt{3}\sin(2\eta)D_p],
\end{eqnarray}
and
\begin{eqnarray}
&&\Delta_{R}^{(0)}=-\sqrt{3}D_R\sin{2\eta},\nonumber\\
&&\Delta_{R,ij}=-\frac{\sqrt{3}}{2}D_R\sin(2\eta)+is_{ij}D_R\cos\eta,\nonumber\\
&&\Delta_{R}^\prime=\frac{\sqrt{3}}{2}\sin(2\eta)D_R.
\end{eqnarray}
In a similar way, we get second-nearest neighbor interaction, i.e. the second-nearest exchange, as
\begin{eqnarray}
H_\mathrm{NNN}&&=\frac{1}{2}S\sum_{\langle\langle ij\rangle\rangle}[\Delta_2^{(0)} (b^\dagger_ib_i+b^\dagger_jb_j)+\Delta_{2,ij}b^\dagger_ib_j+h.c.+\Delta^\prime_{2}b_i^\dagger b_j^\dagger +h.c.]
\end{eqnarray}
with
\begin{eqnarray}
&&\Delta_2^{(0)}=J_2(1-3\sin^2\eta),\nonumber\\
&&\Delta_{2,ij}=\Delta_{2}^{re}+is_{ij}\Delta_2^{im},\nonumber\\
&&\Delta_2^{re}=\frac{1}{2}(1-3\sin^2\eta)J_2,\nonumber\\
&&\Delta_2^{im}=\sqrt{3}\sin\eta J_2,\nonumber\\
&&\Delta_2^\prime=-\frac{3}{2}\cos^2\eta J_2.
\end{eqnarray}
Let's denote $H_\mathrm{NN}$ and $H_\mathrm{NNN}$ by  $H_{1}$ and $H_{2}$, respectively. The total Hamiltonian can be written as $H=H_1+H_2+H_R$. By performing Fourier transformation, $H_{m}$ ($m=1,2$) becomes 
\begin{eqnarray}
H_{m}&&=\frac{S}{2}\sum_{\mathbf{r},\alpha\beta}\sum_{\lambda=\pm 1}\frac{1}{2}\{\Delta_m^{(0)}[b^\dagger_{\alpha}(\mathbf{r})b_\alpha(\mathbf{r})+b^\dagger_\beta(\mathbf{r}+\lambda\bm{\delta}^{(m)}_{\alpha\beta})b_\beta(\mathbf{r}+\lambda\bm{\delta}^{(m)}_{\alpha\beta})]+\Delta_{m,\alpha\beta}b_\alpha^\dagger(\mathbf{r})b_\beta(\mathbf{r}+\lambda\bm{\delta}^{(m)}_{\alpha\beta})+h.c.\nonumber\\
&&+\Delta^\prime_{m}b_\alpha^\dagger(\mathbf{r})b^\dagger_\beta(\mathbf{r}+\lambda\bm{\delta}^{(m)}_{\alpha\beta})+h.c.\}\nonumber\\
&&=\frac{S}{2}\sum_{\mathbf{k},\alpha\beta}[4\Delta_{m}^{(0)}\delta_{\alpha\beta}+2\Delta_{m,\alpha\beta}\cos(\mathbf{k}\cdot\delta^{(m)}_{\alpha\beta})]b^\dagger_{\alpha,\mathbf{k}}b_{\beta,\mathbf{k}}+\Delta^\prime_{m}\cos(\mathbf{k}\cdot\delta^{(m)}_{\alpha\beta})(b^\dagger_{\alpha,\mathbf{k}}b^\dagger_{\beta,-\mathbf{k}}+b_{\alpha,\mathbf{k}}b_{\beta,-\mathbf{k}}).
\end{eqnarray}
Here $\bm{\delta}_{12}^{(1)}=\mathbf{e}_3$, $\bm{\delta}_{23}^{(1)}=\mathbf{e}_1$, $\bm{\delta}_{31}^{(1)}=\mathbf{e}_2$ and $\bm{\delta}_{12}^{(2)}=\mathbf{e}_3^\prime$, $\bm{\delta}_{23}^{(2)}=\mathbf{e}_1^\prime$, $\bm{\delta}_{31}^{(2)}=\mathbf{e}_2^\prime$. We choose $\bm{\delta}_{\alpha\beta}^{(m)}=-\bm{\delta}_{\beta\alpha}^{(m)}$ and $\mathbf{e}_1=(-\frac{1}{2},-\frac{\sqrt{3}}{2})$, $\mathbf{e}_2=(1,0)$, $\mathbf{e}_3=(-\frac{1}{2},\frac{\sqrt{3}}{2})$, $\mathbf{e}^\prime_1=\mathbf{e}_2-\mathbf{e}_3$, $\mathbf{e}^\prime_2=\mathbf{e}_3-\mathbf{e}_1$, $\mathbf{e}^\prime_3=\mathbf{e}_1-\mathbf{e}_2$. In a similar way, we can show
\begin{eqnarray}
H_R=\frac{S}{2}\sum_{\mathbf{k},\alpha\beta}i2\Delta_{R,\alpha\beta}\sin(\mathbf{k}\cdot\delta^{(1)}_{\alpha\beta})b^\dagger_{\alpha,\mathbf{k}}b_{\beta,\mathbf{k}}+i\Delta^\prime_{m}\sin(\mathbf{k}\cdot\delta^{(1)}_{\alpha\beta})(b^\dagger_{\alpha,\mathbf{k}}b^\dagger_{\beta,-\mathbf{k}}+b_{\alpha,\mathbf{k}}b_{\beta,-\mathbf{k}}).
\end{eqnarray}
Finally, the Hamiltonian is expressed in the basis $\Psi_\mathbf{k}=(b_{1,\mathbf{k}},b_{2,\mathbf{k}},b_{3,\mathbf{k}},b^\dagger_{1,-\mathbf{k}},b^\dagger_{2,-\mathbf{k}},b^\dagger_{3,-\mathbf{k}})^T$ as $H=\frac{S}{2}\sum_\mathbf{k}\Psi_\mathbf{k}\mathcal{H}_\mathbf{k}\Psi_\mathbf{k}$ with
\begin{eqnarray}
\mathcal{H}_\mathbf{k}=\left(\begin{array}{cc}
A_0+A_\mathbf{k}&B_\mathbf{k}\\
B_\mathbf{k}&A_0+A_\mathbf{k}^\ast
\end{array}\right).
\end{eqnarray}
Here, $A_0=2(\Delta_1^{(0)}+\Delta_2^{(0)})\mathbbm{1}_{3\times3}$ and  
\begin{eqnarray}
&&A_\mathbf{k}=\left(\begin{array}{ccc}
0&\cos k_3\Delta_1&\cos k_2\Delta_1^\ast\\
\cos k_3 \Delta_1^\ast &0&\cos k_1\Delta_1\\
\cos k_2\Delta_1&\cos k_1\Delta_1^\ast&0
\end{array}\right) +\left(\begin{array}{ccc}
0&\cos p_3\Delta_2&\cos p_2\Delta_2^\ast\\
\cos p_3 \Delta_2^\ast &0&\cos p_1\Delta_2\\
\cos p_2\Delta_2&\cos p_1\Delta_2^\ast&0
\end{array}\right)
+\left(\begin{array}{ccc}
0&i\sin k_3\Delta_R&-i\sin k_2\Delta_R^\ast\\
-i\sin k_3 \Delta_R^\ast &0&i\sin k_1\Delta_R\\
i\sin k_2\Delta_R&-i\sin k_1\Delta_R^\ast&0
\end{array}\right),\nonumber\\
\nonumber\\
&&B_\mathbf{k}=\Delta_1^\prime\left(\begin{array}{ccc}
0&\cos k_3&\cos k_2\\
\cos k_3 &0&\cos k_1\\
\cos k_2&\cos k_1&0
\end{array}\right) +\Delta_2^\prime\left(\begin{array}{ccc}
0&\cos p_3&\cos p_2\\
\cos p_3&0&\cos p_1\\
\cos p_2&\cos p_1&0
\end{array}\right)
+\Delta^\prime_R\left(\begin{array}{ccc}
0&i\sin k_3&-i\sin k_2\\
-i\sin k_3 &0&i\sin k_1\\
i\sin k_2&-i\sin k_1&0
\end{array}\right).	
\end{eqnarray}
We abbreviated the notations: $k_i=\mathbf{k}\cdot\mathbf{e}_i$, $p_i=\mathbf{k}\cdot\mathbf{e}_i^\prime$, $\Delta_m=\Delta_m^{re}+i\Delta_m^{im}$ ($m=1,2$), $\Delta_{R}=-\frac{\sqrt{3}}{2}D_R\sin(2\eta)+iD_R\cos\eta$ and considered the convention that $s_{12}=s_{23}=s_{31}=1$ and $s_{ij}=-s_{ji}$.

\subsection{Breathing Pyrochlore Antiferromagnet}

We consider the model 
\begin{eqnarray}\label{eq:PyrochloreHamiltonian}
H=J\sum_{\langle i j\rangle\in u}\mathbf{S}_{\mathbf{r}_i}\cdot\mathbf{S}_{\mathbf{r}_j}+J^\prime\sum_{\langle i j\rangle\in d}\mathbf{S}_{\mathbf{r}_i}\cdot\mathbf{S}_{\mathbf{r}_j}+D\sum_i(\mathbf{S}_{\mathbf{r}_i}\cdot\hat{z}_i)^2.
\end{eqnarray}
Similar to the two-dimensional model, the magnon excitation is represented via the local Holstein-Primakoff transformation as
$\mathbf{S}_\mu=(S-a^\dagger_\mu a_\mu)\hat{z}_\mu+\sqrt{\frac{S}{2}}(a_\mu+a_\mu^\dagger)\hat{x}_\mu-i\sqrt{\frac{S}{2}}(a_\mu-a_\mu^\dagger)\hat{y}_\mu$. Therefore, the exchange interaction between two neighboring spins is expressed as 
\begin{eqnarray}\label{eq:Spininteraction}
\mathbf{S}_\mu\cdot\mathbf{S}_\nu=S_\mu^cS_\nu^d\Lambda_{\mu\nu}^{cd}=S^2\Lambda_{\mu\nu}^{zz}-S(a_\mu^\dagger a_\mu+a_\nu^\dagger a_\nu)\Lambda_{\mu\nu}^{zz}+\frac{S}{2}[a^\dagger_\mu a_\nu\Gamma_{\mu\nu}+a_\mu a_\nu\Omega_{\mu\nu}+\text{H.c.}],
\end{eqnarray} 
where $\Gamma_{\mu\nu}=\Lambda_{\mu\nu}^{xx}+\Lambda_{\mu\nu}^{yy}-i\Lambda_{\mu\nu}^{xy}+i\Lambda_{\mu\nu}^{yx}$ and $\Omega_{\mu\nu}=\Lambda_{\mu\nu}^{xx}-\Lambda_{\mu\nu}^{yy}-i\Lambda_{\mu\nu}^{xy}-i\Lambda_{\mu\nu}^{yx}$. Here $\Lambda_{\mu\nu}^{cd}=\hat{c}_\mu\cdot\hat{d}_\nu$ with $\hat{c}_\mu$, $\hat{d}_\nu$ being the $c$, $d$ axis of the local frame of  $\mu$ and $\nu$ atoms, respectively, i.e., $c,d=x,y,z$ and $\mu,\nu\in (0,1,2,3)$ with $\mu\neq\nu$. We choose local frames as shown in table~\ref{table_Frame}. It can be shown by straightforward calculation that $\Lambda_{\mu\nu}^{zz}=-\frac{1}{3}$, $\Gamma_{\mu\nu}=-\frac{2}{3}$ and $\Omega_{\mu\nu}=\frac{4}{3}e^{i\phi_{\mu\nu}}$ where $\phi_{01}=\phi_{23}=-\frac{\pi}{3}$, $\phi_{02}=\phi_{13}=\frac{\pi}{3}$, $\phi_{03}=\phi_{12}=\pi$ and other terms can be generated by $\phi_{\mu\nu}=\phi_{\nu\mu}$ ($\mu\neq\nu$). By substituting the magnon representation of spin-spin interaction Eq.~\eqref{eq:Spininteraction} into Eq.~\eqref{eq:PyrochloreHamiltonian} and performing Fourier transformation, we obtain the noninteracting magnon Hamiltonian
\begin{eqnarray}
H=\sum_{\mathbf{k},\mu\nu}S[(J+J^\prime-2D)\delta_{\mu\nu}-\frac{1}{3}(J+J^\prime e^{-i\mathbf{k}\cdot\mathbf{d}_{\mu\nu}})]a^\dagger_{\mu,\mathbf{k}}a_{\nu,\mathbf{k}}+S\frac{1}{3}(J+J^\prime e^{-i\mathbf{k}\cdot\mathbf{d}_{\mu\nu}})e^{i\phi_{\mu\nu}}a_{\mu,-\mathbf{k}}a_{\nu,\mathbf{k}}+h.c.
\end{eqnarray}
where $\mathbf{d}_{\mu\nu}=\mathbf{a}_\nu-\mathbf{a}_\mu$ with $\mathbf{a}_0=(0,0,0)$, $\mathbf{a}_1=\frac{1}{2}(0,1,1)$, $\mathbf{a}_2=\frac{1}{2}(1,0,1)$, and $\mathbf{a}_3=\frac{1}{2}(1,1,0)$.
\begin{table}[ht]
	\centering
	\begin{tabular}{c| c| c| c}
		\hline\hline 
		$\mu$ & $\hat{x}_\mu$ &  $\hat{y}_\mu$ &  $\hat{z}_\mu$ \\ [0.5ex] 
		\hline 
		$0$& $\frac{1}{\sqrt{2}}(-1,1,0)$&  $\frac{1}{\sqrt{6}}(-1,-1,2)$& $\frac{1}{\sqrt{3}}(1,1,1)$\\
		$1$& $\frac{1}{\sqrt{2}}(-1,-1,0)$&  $\frac{1}{\sqrt{6}}(-1,1,-2)$& $\frac{1}{\sqrt{3}}(1,-1,-1)$\\
		$2$& $\frac{1}{\sqrt{2}}(1,1,0)$&  $\frac{1}{\sqrt{6}}(1,-1,-2)$& $\frac{1}{\sqrt{3}}(-1,1,-1)$\\
		$3$& $\frac{1}{\sqrt{2}}(1,-1,0)$&  $\frac{1}{\sqrt{6}}(1,1,2)$& $\frac{1}{\sqrt{3}}(-1,-1,1)$
		\\[1ex]
		\hline 
	\end{tabular}
	\caption{Local coordinates of AIAO breathing pyrochlore.} \label{table_Frame}
\end{table}

\end{widetext}

\bibliographystyle{apsrev}
\bibliography{Magnonedelstein}

\begin{thebibliography}{86}
\expandafter\ifx\csname natexlab\endcsname\relax\def\natexlab#1{#1}\fi
\expandafter\ifx\csname bibnamefont\endcsname\relax
  \def\bibnamefont#1{#1}\fi
\expandafter\ifx\csname bibfnamefont\endcsname\relax
  \def\bibfnamefont#1{#1}\fi
\expandafter\ifx\csname citenamefont\endcsname\relax
  \def\citenamefont#1{#1}\fi
\expandafter\ifx\csname url\endcsname\relax
  \def\url#1{\texttt{#1}}\fi
\expandafter\ifx\csname urlprefix\endcsname\relax\def\urlprefix{URL }\fi
\providecommand{\bibinfo}[2]{#2}
\providecommand{\eprint}[2][]{\url{#2}}

\bibitem[{\citenamefont{\ifmmode \check{Z}\else
  \v{Z}\fi{}uti\ifmmode~\acute{c}\else \'{c}\fi{}
  et~al.}(2004)\citenamefont{\ifmmode \check{Z}\else
  \v{Z}\fi{}uti\ifmmode~\acute{c}\else \'{c}\fi{}, Fabian, and
  Das~Sarma}}]{RevModPhys.76.323}
\bibinfo{author}{\bibfnamefont{I.}~\bibnamefont{\ifmmode \check{Z}\else
  \v{Z}\fi{}uti\ifmmode~\acute{c}\else \'{c}\fi{}}},
  \bibinfo{author}{\bibfnamefont{J.}~\bibnamefont{Fabian}}, \bibnamefont{and}
  \bibinfo{author}{\bibfnamefont{S.}~\bibnamefont{Das~Sarma}},
  \bibinfo{journal}{Rev. Mod. Phys.} \textbf{\bibinfo{volume}{76}},
  \bibinfo{pages}{323} (\bibinfo{year}{2004}).

\bibitem[{\citenamefont{Manchon et~al.}(2019)\citenamefont{Manchon,
  \ifmmode~\check{Z}\else \v{Z}\fi{}elezn\'y, Miron, Jungwirth, Sinova,
  Thiaville, Garello, and Gambardella}}]{RevModPhys.91.035004}
\bibinfo{author}{\bibfnamefont{A.}~\bibnamefont{Manchon}},
  \bibinfo{author}{\bibfnamefont{J.}~\bibnamefont{\ifmmode~\check{Z}\else
  \v{Z}\fi{}elezn\'y}}, \bibinfo{author}{\bibfnamefont{I.~M.}
  \bibnamefont{Miron}},
  \bibinfo{author}{\bibfnamefont{T.}~\bibnamefont{Jungwirth}},
  \bibinfo{author}{\bibfnamefont{J.}~\bibnamefont{Sinova}},
  \bibinfo{author}{\bibfnamefont{A.}~\bibnamefont{Thiaville}},
  \bibinfo{author}{\bibfnamefont{K.}~\bibnamefont{Garello}}, \bibnamefont{and}
  \bibinfo{author}{\bibfnamefont{P.}~\bibnamefont{Gambardella}},
  \bibinfo{journal}{Rev. Mod. Phys.} \textbf{\bibinfo{volume}{91}},
  \bibinfo{pages}{035004} (\bibinfo{year}{2019}).

\bibitem[{\citenamefont{Aronov and Lyanda-Geller}(1989)}]{aronov1989nuclear}
\bibinfo{author}{\bibfnamefont{A.}~\bibnamefont{Aronov}} \bibnamefont{and}
  \bibinfo{author}{\bibfnamefont{Y.~B.} \bibnamefont{Lyanda-Geller}},
  \bibinfo{journal}{JETP Lett.} \textbf{\bibinfo{volume}{50}},
  \bibinfo{pages}{431} (\bibinfo{year}{1989}).

\bibitem[{\citenamefont{Edelstein}(1990)}]{EDELSTEIN1990233}
\bibinfo{author}{\bibfnamefont{V.}~\bibnamefont{Edelstein}},
  \bibinfo{journal}{Solid State Commun.} \textbf{\bibinfo{volume}{73}},
  \bibinfo{pages}{233 } (\bibinfo{year}{1990}).

\bibitem[{\citenamefont{Ganichev et~al.}(2011)\citenamefont{Ganichev, Trushin,
  and Schliemann}}]{Ganichev}
\bibinfo{author}{\bibfnamefont{S.~D.} \bibnamefont{Ganichev}},
  \bibinfo{author}{\bibfnamefont{M.}~\bibnamefont{Trushin}}, \bibnamefont{and}
  \bibinfo{author}{\bibfnamefont{J.}~\bibnamefont{Schliemann}}, in
  \emph{\bibinfo{booktitle}{Handbook of spin transport and magnetism}}, edited
  by \bibinfo{editor}{\bibfnamefont{E.~Y.} \bibnamefont{Tsymbal}}
  \bibnamefont{and}
  \bibinfo{editor}{\bibfnamefont{I.}~\bibnamefont{\v{Z}uti\'c}}
  (\bibinfo{publisher}{CRC}, \bibinfo{address}{Boca Raton, FL},
  \bibinfo{year}{2011}), p. \bibinfo{pages}{487}.

\bibitem[{\citenamefont{Trushin and Schliemann}(2007)}]{PhysRevB.75.155323}
\bibinfo{author}{\bibfnamefont{M.}~\bibnamefont{Trushin}} \bibnamefont{and}
  \bibinfo{author}{\bibfnamefont{J.}~\bibnamefont{Schliemann}},
  \bibinfo{journal}{Phys. Rev. B} \textbf{\bibinfo{volume}{75}},
  \bibinfo{pages}{155323} (\bibinfo{year}{2007}).

\bibitem[{\citenamefont{Silov et~al.}(2004)\citenamefont{Silov, Blajnov,
  Wolter, Hey, Ploog, and Averkiev}}]{doi:10.1063/1.1833565}
\bibinfo{author}{\bibfnamefont{A.~Y.} \bibnamefont{Silov}},
  \bibinfo{author}{\bibfnamefont{P.~A.} \bibnamefont{Blajnov}},
  \bibinfo{author}{\bibfnamefont{J.~H.} \bibnamefont{Wolter}},
  \bibinfo{author}{\bibfnamefont{R.}~\bibnamefont{Hey}},
  \bibinfo{author}{\bibfnamefont{K.~H.} \bibnamefont{Ploog}}, \bibnamefont{and}
  \bibinfo{author}{\bibfnamefont{N.~S.} \bibnamefont{Averkiev}},
  \bibinfo{journal}{Appl. Phys. Lett.} \textbf{\bibinfo{volume}{85}},
  \bibinfo{pages}{5929} (\bibinfo{year}{2004}).

\bibitem[{\citenamefont{Yang et~al.}(2006)\citenamefont{Yang, He, Ding, Cui,
  Zeng, Wang, and Ge}}]{PhysRevLett.96.186605}
\bibinfo{author}{\bibfnamefont{C.~L.} \bibnamefont{Yang}},
  \bibinfo{author}{\bibfnamefont{H.~T.} \bibnamefont{He}},
  \bibinfo{author}{\bibfnamefont{L.}~\bibnamefont{Ding}},
  \bibinfo{author}{\bibfnamefont{L.~J.} \bibnamefont{Cui}},
  \bibinfo{author}{\bibfnamefont{Y.~P.} \bibnamefont{Zeng}},
  \bibinfo{author}{\bibfnamefont{J.~N.} \bibnamefont{Wang}}, \bibnamefont{and}
  \bibinfo{author}{\bibfnamefont{W.~K.} \bibnamefont{Ge}},
  \bibinfo{journal}{Phys. Rev. Lett.} \textbf{\bibinfo{volume}{96}},
  \bibinfo{pages}{186605} (\bibinfo{year}{2006}).

\bibitem[{\citenamefont{Stern et~al.}(2006)\citenamefont{Stern, Ghosh, Xiang,
  Zhu, Samarth, and Awschalom}}]{PhysRevLett.97.126603}
\bibinfo{author}{\bibfnamefont{N.~P.} \bibnamefont{Stern}},
  \bibinfo{author}{\bibfnamefont{S.}~\bibnamefont{Ghosh}},
  \bibinfo{author}{\bibfnamefont{G.}~\bibnamefont{Xiang}},
  \bibinfo{author}{\bibfnamefont{M.}~\bibnamefont{Zhu}},
  \bibinfo{author}{\bibfnamefont{N.}~\bibnamefont{Samarth}}, \bibnamefont{and}
  \bibinfo{author}{\bibfnamefont{D.~D.} \bibnamefont{Awschalom}},
  \bibinfo{journal}{Phys. Rev. Lett.} \textbf{\bibinfo{volume}{97}},
  \bibinfo{pages}{126603} (\bibinfo{year}{2006}).

\bibitem[{\citenamefont{Gambardella and Miron}(2011)}]{Gambardella3175}
\bibinfo{author}{\bibfnamefont{P.}~\bibnamefont{Gambardella}} \bibnamefont{and}
  \bibinfo{author}{\bibfnamefont{I.~M.} \bibnamefont{Miron}},
  \bibinfo{journal}{Philos. Trans. R. Soc. London A}
  \textbf{\bibinfo{volume}{369}}, \bibinfo{pages}{3175} (\bibinfo{year}{2011}).

\bibitem[{\citenamefont{Inoue et~al.}(2003)\citenamefont{Inoue, Bauer, and
  Molenkamp}}]{PhysRevB.67.033104}
\bibinfo{author}{\bibfnamefont{J.-i.} \bibnamefont{Inoue}},
  \bibinfo{author}{\bibfnamefont{G.~E.~W.} \bibnamefont{Bauer}},
  \bibnamefont{and} \bibinfo{author}{\bibfnamefont{L.~W.}
  \bibnamefont{Molenkamp}}, \bibinfo{journal}{Phys. Rev. B}
  \textbf{\bibinfo{volume}{67}}, \bibinfo{pages}{033104}
  (\bibinfo{year}{2003}).

\bibitem[{\citenamefont{Shen et~al.}(2014)\citenamefont{Shen, Vignale, and
  Raimondi}}]{PhysRevLett.112.096601}
\bibinfo{author}{\bibfnamefont{K.}~\bibnamefont{Shen}},
  \bibinfo{author}{\bibfnamefont{G.}~\bibnamefont{Vignale}}, \bibnamefont{and}
  \bibinfo{author}{\bibfnamefont{R.}~\bibnamefont{Raimondi}},
  \bibinfo{journal}{Phys. Rev. Lett.} \textbf{\bibinfo{volume}{112}},
  \bibinfo{pages}{096601} (\bibinfo{year}{2014}).

\bibitem[{\citenamefont{Borge et~al.}(2014)\citenamefont{Borge, Gorini,
  Vignale, and Raimondi}}]{PhysRevB.89.245443}
\bibinfo{author}{\bibfnamefont{J.}~\bibnamefont{Borge}},
  \bibinfo{author}{\bibfnamefont{C.}~\bibnamefont{Gorini}},
  \bibinfo{author}{\bibfnamefont{G.}~\bibnamefont{Vignale}}, \bibnamefont{and}
  \bibinfo{author}{\bibfnamefont{R.}~\bibnamefont{Raimondi}},
  \bibinfo{journal}{Phys. Rev. B} \textbf{\bibinfo{volume}{89}},
  \bibinfo{pages}{245443} (\bibinfo{year}{2014}).

\bibitem[{\citenamefont{Johansson et~al.}(2016)\citenamefont{Johansson, Henk,
  and Mertig}}]{PhysRevB.93.195440}
\bibinfo{author}{\bibfnamefont{A.}~\bibnamefont{Johansson}},
  \bibinfo{author}{\bibfnamefont{J.}~\bibnamefont{Henk}}, \bibnamefont{and}
  \bibinfo{author}{\bibfnamefont{I.}~\bibnamefont{Mertig}},
  \bibinfo{journal}{Phys. Rev. B} \textbf{\bibinfo{volume}{93}},
  \bibinfo{pages}{195440} (\bibinfo{year}{2016}).

\bibitem[{\citenamefont{Gorini et~al.}(2017)\citenamefont{Gorini,
  Maleki~Sheikhabadi, Shen, Tokatly, Vignale, and
  Raimondi}}]{PhysRevB.95.205424}
\bibinfo{author}{\bibfnamefont{C.}~\bibnamefont{Gorini}},
  \bibinfo{author}{\bibfnamefont{A.}~\bibnamefont{Maleki~Sheikhabadi}},
  \bibinfo{author}{\bibfnamefont{K.}~\bibnamefont{Shen}},
  \bibinfo{author}{\bibfnamefont{I.~V.} \bibnamefont{Tokatly}},
  \bibinfo{author}{\bibfnamefont{G.}~\bibnamefont{Vignale}}, \bibnamefont{and}
  \bibinfo{author}{\bibfnamefont{R.}~\bibnamefont{Raimondi}},
  \bibinfo{journal}{Phys. Rev. B} \textbf{\bibinfo{volume}{95}},
  \bibinfo{pages}{205424} (\bibinfo{year}{2017}).

\bibitem[{\citenamefont{Ganichev et~al.}(2002)\citenamefont{Ganichev, Ivchenko,
  Bel'kov, Tarasenko, Sollinger, Weiss, Wegscheider, and
  Prettl}}]{Ganichev2002}
\bibinfo{author}{\bibfnamefont{S.~D.} \bibnamefont{Ganichev}},
  \bibinfo{author}{\bibfnamefont{E.~L.} \bibnamefont{Ivchenko}},
  \bibinfo{author}{\bibfnamefont{V.~V.} \bibnamefont{Bel'kov}},
  \bibinfo{author}{\bibfnamefont{S.~A.} \bibnamefont{Tarasenko}},
  \bibinfo{author}{\bibfnamefont{M.}~\bibnamefont{Sollinger}},
  \bibinfo{author}{\bibfnamefont{D.}~\bibnamefont{Weiss}},
  \bibinfo{author}{\bibfnamefont{W.}~\bibnamefont{Wegscheider}},
  \bibnamefont{and} \bibinfo{author}{\bibfnamefont{W.}~\bibnamefont{Prettl}},
  \bibinfo{journal}{Nature} \textbf{\bibinfo{volume}{417}},
  \bibinfo{pages}{153} (\bibinfo{year}{2002}).

\bibitem[{\citenamefont{Garate and Franz}(2010)}]{PhysRevLett.104.146802}
\bibinfo{author}{\bibfnamefont{I.}~\bibnamefont{Garate}} \bibnamefont{and}
  \bibinfo{author}{\bibfnamefont{M.}~\bibnamefont{Franz}},
  \bibinfo{journal}{Phys. Rev. Lett.} \textbf{\bibinfo{volume}{104}},
  \bibinfo{pages}{146802} (\bibinfo{year}{2010}).

\bibitem[{\citenamefont{Chernyshov et~al.}(2009)\citenamefont{Chernyshov,
  Overby, Liu, Furdyna, Lyanda-Geller, and
  Rokhinson}}]{Chernyshov.OverbyNP2009}
\bibinfo{author}{\bibfnamefont{A.}~\bibnamefont{Chernyshov}},
  \bibinfo{author}{\bibfnamefont{M.}~\bibnamefont{Overby}},
  \bibinfo{author}{\bibfnamefont{X.}~\bibnamefont{Liu}},
  \bibinfo{author}{\bibfnamefont{J.~K.} \bibnamefont{Furdyna}},
  \bibinfo{author}{\bibfnamefont{Y.}~\bibnamefont{Lyanda-Geller}},
  \bibnamefont{and} \bibinfo{author}{\bibfnamefont{L.~P.}
  \bibnamefont{Rokhinson}}, \bibinfo{journal}{Nat. Phys.}
  \textbf{\bibinfo{volume}{5}}, \bibinfo{pages}{656} (\bibinfo{year}{2009}).

\bibitem[{\citenamefont{Miron et~al.}(2010)\citenamefont{Miron, Gaudin,
  Auffret, Rodmacq, Schuhl, Pizzini, Vogel, and
  Gambardella}}]{MihaiMiron.GaudinNM2010}
\bibinfo{author}{\bibfnamefont{I.~M.} \bibnamefont{Miron}},
  \bibinfo{author}{\bibfnamefont{G.}~\bibnamefont{Gaudin}},
  \bibinfo{author}{\bibfnamefont{S.}~\bibnamefont{Auffret}},
  \bibinfo{author}{\bibfnamefont{B.}~\bibnamefont{Rodmacq}},
  \bibinfo{author}{\bibfnamefont{A.}~\bibnamefont{Schuhl}},
  \bibinfo{author}{\bibfnamefont{S.}~\bibnamefont{Pizzini}},
  \bibinfo{author}{\bibfnamefont{J.}~\bibnamefont{Vogel}}, \bibnamefont{and}
  \bibinfo{author}{\bibfnamefont{P.}~\bibnamefont{Gambardella}},
  \bibinfo{journal}{Nat. Mater.} \textbf{\bibinfo{volume}{9}},
  \bibinfo{pages}{230} (\bibinfo{year}{2010}).

\bibitem[{\citenamefont{Miron et~al.}(2011)\citenamefont{Miron, Garello,
  Gaudin, Zermatten, Costache, Auffret, Bandiera, Rodmacq, Schuhl, and
  Gambardella}}]{Miron.Garello2011}
\bibinfo{author}{\bibfnamefont{I.~M.} \bibnamefont{Miron}},
  \bibinfo{author}{\bibfnamefont{K.}~\bibnamefont{Garello}},
  \bibinfo{author}{\bibfnamefont{G.}~\bibnamefont{Gaudin}},
  \bibinfo{author}{\bibfnamefont{P.-J.} \bibnamefont{Zermatten}},
  \bibinfo{author}{\bibfnamefont{M.~V.} \bibnamefont{Costache}},
  \bibinfo{author}{\bibfnamefont{S.}~\bibnamefont{Auffret}},
  \bibinfo{author}{\bibfnamefont{S.}~\bibnamefont{Bandiera}},
  \bibinfo{author}{\bibfnamefont{B.}~\bibnamefont{Rodmacq}},
  \bibinfo{author}{\bibfnamefont{A.}~\bibnamefont{Schuhl}}, \bibnamefont{and}
  \bibinfo{author}{\bibfnamefont{P.}~\bibnamefont{Gambardella}},
  \bibinfo{journal}{Nature} \textbf{\bibinfo{volume}{476}},
  \bibinfo{pages}{189} (\bibinfo{year}{2011}).

\bibitem[{\citenamefont{Pesin and MacDonald}(2012)}]{Pesin2012}
\bibinfo{author}{\bibfnamefont{D.~A.} \bibnamefont{Pesin}} \bibnamefont{and}
  \bibinfo{author}{\bibfnamefont{A.~H.} \bibnamefont{MacDonald}},
  \bibinfo{journal}{Phys. Rev. B} \textbf{\bibinfo{volume}{86}},
  \bibinfo{pages}{014416} (\bibinfo{year}{2012}).

\bibitem[{\citenamefont{Qaiumzadeh et~al.}(2015)\citenamefont{Qaiumzadeh,
  Duine, and Titov}}]{Qaiumzadeh.DuinePRB2015}
\bibinfo{author}{\bibfnamefont{A.}~\bibnamefont{Qaiumzadeh}},
  \bibinfo{author}{\bibfnamefont{R.~{\^A}.~A.} \bibnamefont{Duine}},
  \bibnamefont{and} \bibinfo{author}{\bibfnamefont{M.}~\bibnamefont{Titov}},
  \bibinfo{journal}{Phys. Rev. B} \textbf{\bibinfo{volume}{92}},
  \bibinfo{eid}{014402} (\bibinfo{year}{2015}).

\bibitem[{\citenamefont{Ado et~al.}(2017)\citenamefont{Ado, Tretiakov, and
  Titov}}]{Ado.TretiakovPRB2017}
\bibinfo{author}{\bibfnamefont{I.~A.} \bibnamefont{Ado}},
  \bibinfo{author}{\bibfnamefont{O.~A.} \bibnamefont{Tretiakov}},
  \bibnamefont{and} \bibinfo{author}{\bibfnamefont{M.}~\bibnamefont{Titov}},
  \bibinfo{journal}{Phys. Rev. B} \textbf{\bibinfo{volume}{95}},
  \bibinfo{eid}{094401} (\bibinfo{year}{2017}).

\bibitem[{\citenamefont{Belashchenko et~al.}(2019)\citenamefont{Belashchenko,
  Kovalev, and van Schilfgaarde}}]{PhysRevMaterials.3.011401}
\bibinfo{author}{\bibfnamefont{K.~D.} \bibnamefont{Belashchenko}},
  \bibinfo{author}{\bibfnamefont{A.~A.} \bibnamefont{Kovalev}},
  \bibnamefont{and} \bibinfo{author}{\bibfnamefont{M.}~\bibnamefont{van
  Schilfgaarde}}, \bibinfo{journal}{Phys. Rev. Materials}
  \textbf{\bibinfo{volume}{3}}, \bibinfo{pages}{011401} (\bibinfo{year}{2019}).

\bibitem[{\citenamefont{Dzyaloshinsky}(1958)}]{Dzyaloshinsky58}
\bibinfo{author}{\bibfnamefont{I.}~\bibnamefont{Dzyaloshinsky}},
  \bibinfo{journal}{J. Phys. Chem. Solids} \textbf{\bibinfo{volume}{4}},
  \bibinfo{pages}{241} (\bibinfo{year}{1958}).

\bibitem[{\citenamefont{Moriya}(1960)}]{Moriya60}
\bibinfo{author}{\bibfnamefont{T.}~\bibnamefont{Moriya}},
  \bibinfo{journal}{Phys. Rev.} \textbf{\bibinfo{volume}{120}},
  \bibinfo{pages}{91} (\bibinfo{year}{1960}).

\bibitem[{\citenamefont{Okuma}(2017)}]{PhysRevLett.119.107205}
\bibinfo{author}{\bibfnamefont{N.}~\bibnamefont{Okuma}},
  \bibinfo{journal}{Phys. Rev. Lett.} \textbf{\bibinfo{volume}{119}},
  \bibinfo{pages}{107205} (\bibinfo{year}{2017}).

\bibitem[{\citenamefont{Manchon et~al.}(2014)\citenamefont{Manchon, Ndiaye,
  Moon, Lee, and Lee}}]{PhysRevB.90.224403}
\bibinfo{author}{\bibfnamefont{A.}~\bibnamefont{Manchon}},
  \bibinfo{author}{\bibfnamefont{P.~B.} \bibnamefont{Ndiaye}},
  \bibinfo{author}{\bibfnamefont{J.-H.} \bibnamefont{Moon}},
  \bibinfo{author}{\bibfnamefont{H.-W.} \bibnamefont{Lee}}, \bibnamefont{and}
  \bibinfo{author}{\bibfnamefont{K.-J.} \bibnamefont{Lee}},
  \bibinfo{journal}{Phys. Rev. B} \textbf{\bibinfo{volume}{90}},
  \bibinfo{pages}{224403} (\bibinfo{year}{2014}).

\bibitem[{\citenamefont{Kovalev and Zyuzin}(2016)}]{PhysRevB.93.161106}
\bibinfo{author}{\bibfnamefont{A.~A.} \bibnamefont{Kovalev}} \bibnamefont{and}
  \bibinfo{author}{\bibfnamefont{V.}~\bibnamefont{Zyuzin}},
  \bibinfo{journal}{Phys. Rev. B} \textbf{\bibinfo{volume}{93}},
  \bibinfo{pages}{161106} (\bibinfo{year}{2016}).

\bibitem[{\citenamefont{Kovalev et~al.}(2017)\citenamefont{Kovalev, Zyuzin, and
  Li}}]{PhysRevB.95.165106}
\bibinfo{author}{\bibfnamefont{A.~A.} \bibnamefont{Kovalev}},
  \bibinfo{author}{\bibfnamefont{V.~A.} \bibnamefont{Zyuzin}},
  \bibnamefont{and} \bibinfo{author}{\bibfnamefont{B.}~\bibnamefont{Li}},
  \bibinfo{journal}{Phys. Rev. B} \textbf{\bibinfo{volume}{95}},
  \bibinfo{pages}{165106} (\bibinfo{year}{2017}).

\bibitem[{\citenamefont{Katsura et~al.}(2010)\citenamefont{Katsura, Nagaosa,
  and Lee}}]{PhysRevLett.104.066403}
\bibinfo{author}{\bibfnamefont{H.}~\bibnamefont{Katsura}},
  \bibinfo{author}{\bibfnamefont{N.}~\bibnamefont{Nagaosa}}, \bibnamefont{and}
  \bibinfo{author}{\bibfnamefont{P.~A.} \bibnamefont{Lee}},
  \bibinfo{journal}{Phys. Rev. Lett.} \textbf{\bibinfo{volume}{104}},
  \bibinfo{pages}{066403} (\bibinfo{year}{2010}).

\bibitem[{\citenamefont{Onose et~al.}(2010)\citenamefont{Onose, Ideue, Katsura,
  Shiomi, Nagaosa, and Tokura}}]{Onose297}
\bibinfo{author}{\bibfnamefont{Y.}~\bibnamefont{Onose}},
  \bibinfo{author}{\bibfnamefont{T.}~\bibnamefont{Ideue}},
  \bibinfo{author}{\bibfnamefont{H.}~\bibnamefont{Katsura}},
  \bibinfo{author}{\bibfnamefont{Y.}~\bibnamefont{Shiomi}},
  \bibinfo{author}{\bibfnamefont{N.}~\bibnamefont{Nagaosa}}, \bibnamefont{and}
  \bibinfo{author}{\bibfnamefont{Y.}~\bibnamefont{Tokura}},
  \bibinfo{journal}{Science} \textbf{\bibinfo{volume}{329}},
  \bibinfo{pages}{297} (\bibinfo{year}{2010}).

\bibitem[{\citenamefont{Matsumoto and
  Murakami}(2011{\natexlab{a}})}]{PhysRevLett.106.197202}
\bibinfo{author}{\bibfnamefont{R.}~\bibnamefont{Matsumoto}} \bibnamefont{and}
  \bibinfo{author}{\bibfnamefont{S.}~\bibnamefont{Murakami}},
  \bibinfo{journal}{Phys. Rev. Lett.} \textbf{\bibinfo{volume}{106}},
  \bibinfo{pages}{197202} (\bibinfo{year}{2011}{\natexlab{a}}).

\bibitem[{\citenamefont{Matsumoto and
  Murakami}(2011{\natexlab{b}})}]{PhysRevB.84.184406}
\bibinfo{author}{\bibfnamefont{R.}~\bibnamefont{Matsumoto}} \bibnamefont{and}
  \bibinfo{author}{\bibfnamefont{S.}~\bibnamefont{Murakami}},
  \bibinfo{journal}{Phys. Rev. B} \textbf{\bibinfo{volume}{84}},
  \bibinfo{pages}{184406} (\bibinfo{year}{2011}{\natexlab{b}}).

\bibitem[{\citenamefont{Matsumoto et~al.}(2014)\citenamefont{Matsumoto,
  Shindou, and Murakami}}]{PhysRevB.89.054420}
\bibinfo{author}{\bibfnamefont{R.}~\bibnamefont{Matsumoto}},
  \bibinfo{author}{\bibfnamefont{R.}~\bibnamefont{Shindou}}, \bibnamefont{and}
  \bibinfo{author}{\bibfnamefont{S.}~\bibnamefont{Murakami}},
  \bibinfo{journal}{Phys. Rev. B} \textbf{\bibinfo{volume}{89}},
  \bibinfo{pages}{054420} (\bibinfo{year}{2014}).

\bibitem[{\citenamefont{Laurell and Fiete}(2018)}]{PhysRevB.98.094419}
\bibinfo{author}{\bibfnamefont{P.}~\bibnamefont{Laurell}} \bibnamefont{and}
  \bibinfo{author}{\bibfnamefont{G.~A.} \bibnamefont{Fiete}},
  \bibinfo{journal}{Phys. Rev. B} \textbf{\bibinfo{volume}{98}},
  \bibinfo{pages}{094419} (\bibinfo{year}{2018}).

\bibitem[{\citenamefont{Lu et~al.}(2019)\citenamefont{Lu, Guo, Koval, and
  Jia}}]{PhysRevB.99.054409}
\bibinfo{author}{\bibfnamefont{Y.}~\bibnamefont{Lu}},
  \bibinfo{author}{\bibfnamefont{X.}~\bibnamefont{Guo}},
  \bibinfo{author}{\bibfnamefont{V.}~\bibnamefont{Koval}}, \bibnamefont{and}
  \bibinfo{author}{\bibfnamefont{C.}~\bibnamefont{Jia}},
  \bibinfo{journal}{Phys. Rev. B} \textbf{\bibinfo{volume}{99}},
  \bibinfo{pages}{054409} (\bibinfo{year}{2019}).

\bibitem[{\citenamefont{Owerre}(2017)}]{PhysRevB.95.014422}
\bibinfo{author}{\bibfnamefont{S.~A.} \bibnamefont{Owerre}},
  \bibinfo{journal}{Phys. Rev. B} \textbf{\bibinfo{volume}{95}},
  \bibinfo{pages}{014422} (\bibinfo{year}{2017}).

\bibitem[{\citenamefont{Doki et~al.}(2018)\citenamefont{Doki, Akazawa, Lee,
  Han, Sugii, Shimozawa, Kawashima, Oda, Yoshida, and
  Yamashita}}]{PhysRevLett.121.097203}
\bibinfo{author}{\bibfnamefont{H.}~\bibnamefont{Doki}},
  \bibinfo{author}{\bibfnamefont{M.}~\bibnamefont{Akazawa}},
  \bibinfo{author}{\bibfnamefont{H.-Y.} \bibnamefont{Lee}},
  \bibinfo{author}{\bibfnamefont{J.~H.} \bibnamefont{Han}},
  \bibinfo{author}{\bibfnamefont{K.}~\bibnamefont{Sugii}},
  \bibinfo{author}{\bibfnamefont{M.}~\bibnamefont{Shimozawa}},
  \bibinfo{author}{\bibfnamefont{N.}~\bibnamefont{Kawashima}},
  \bibinfo{author}{\bibfnamefont{M.}~\bibnamefont{Oda}},
  \bibinfo{author}{\bibfnamefont{H.}~\bibnamefont{Yoshida}}, \bibnamefont{and}
  \bibinfo{author}{\bibfnamefont{M.}~\bibnamefont{Yamashita}},
  \bibinfo{journal}{Phys. Rev. Lett.} \textbf{\bibinfo{volume}{121}},
  \bibinfo{pages}{097203} (\bibinfo{year}{2018}).

\bibitem[{\citenamefont{Mook et~al.}(2019{\natexlab{a}})\citenamefont{Mook,
  Henk, and Mertig}}]{PhysRevB.99.014427}
\bibinfo{author}{\bibfnamefont{A.}~\bibnamefont{Mook}},
  \bibinfo{author}{\bibfnamefont{J.}~\bibnamefont{Henk}}, \bibnamefont{and}
  \bibinfo{author}{\bibfnamefont{I.}~\bibnamefont{Mertig}},
  \bibinfo{journal}{Phys. Rev. B} \textbf{\bibinfo{volume}{99}},
  \bibinfo{pages}{014427} (\bibinfo{year}{2019}{\natexlab{a}}).

\bibitem[{\citenamefont{Mook et~al.}(2016)\citenamefont{Mook, Henk, and
  Mertig}}]{PhysRevB.94.174444}
\bibinfo{author}{\bibfnamefont{A.}~\bibnamefont{Mook}},
  \bibinfo{author}{\bibfnamefont{J.}~\bibnamefont{Henk}}, \bibnamefont{and}
  \bibinfo{author}{\bibfnamefont{I.}~\bibnamefont{Mertig}},
  \bibinfo{journal}{Phys. Rev. B} \textbf{\bibinfo{volume}{94}},
  \bibinfo{pages}{174444} (\bibinfo{year}{2016}).

\bibitem[{\citenamefont{Kim et~al.}(2019)\citenamefont{Kim, Nakata, Loss, and
  Tserkovnyak}}]{PhysRevLett.122.057204}
\bibinfo{author}{\bibfnamefont{S.~K.} \bibnamefont{Kim}},
  \bibinfo{author}{\bibfnamefont{K.}~\bibnamefont{Nakata}},
  \bibinfo{author}{\bibfnamefont{D.}~\bibnamefont{Loss}}, \bibnamefont{and}
  \bibinfo{author}{\bibfnamefont{Y.}~\bibnamefont{Tserkovnyak}},
  \bibinfo{journal}{Phys. Rev. Lett.} \textbf{\bibinfo{volume}{122}},
  \bibinfo{pages}{057204} (\bibinfo{year}{2019}).

\bibitem[{\citenamefont{Zyuzin and Kovalev}(2016)}]{PhysRevLett.117.217203}
\bibinfo{author}{\bibfnamefont{V.~A.} \bibnamefont{Zyuzin}} \bibnamefont{and}
  \bibinfo{author}{\bibfnamefont{A.~A.} \bibnamefont{Kovalev}},
  \bibinfo{journal}{Phys. Rev. Lett.} \textbf{\bibinfo{volume}{117}},
  \bibinfo{pages}{217203} (\bibinfo{year}{2016}).

\bibitem[{\citenamefont{Cheng et~al.}(2016)\citenamefont{Cheng, Okamoto, and
  Xiao}}]{PhysRevLett.117.217202}
\bibinfo{author}{\bibfnamefont{R.}~\bibnamefont{Cheng}},
  \bibinfo{author}{\bibfnamefont{S.}~\bibnamefont{Okamoto}}, \bibnamefont{and}
  \bibinfo{author}{\bibfnamefont{D.}~\bibnamefont{Xiao}},
  \bibinfo{journal}{Phys. Rev. Lett.} \textbf{\bibinfo{volume}{117}},
  \bibinfo{pages}{217202} (\bibinfo{year}{2016}).

\bibitem[{\citenamefont{Shiomi et~al.}(2017)\citenamefont{Shiomi, Takashima,
  and Saitoh}}]{PhysRevB.96.134425}
\bibinfo{author}{\bibfnamefont{Y.}~\bibnamefont{Shiomi}},
  \bibinfo{author}{\bibfnamefont{R.}~\bibnamefont{Takashima}},
  \bibnamefont{and} \bibinfo{author}{\bibfnamefont{E.}~\bibnamefont{Saitoh}},
  \bibinfo{journal}{Phys. Rev. B} \textbf{\bibinfo{volume}{96}},
  \bibinfo{pages}{134425} (\bibinfo{year}{2017}).

\bibitem[{\citenamefont{{Li} et~al.}(2019)\citenamefont{{Li}, {Sandhoefner},
  and {Kovalev}}}]{2019arXiv190710567L}
\bibinfo{author}{\bibfnamefont{B.}~\bibnamefont{{Li}}},
  \bibinfo{author}{\bibfnamefont{S.}~\bibnamefont{{Sandhoefner}}},
  \bibnamefont{and} \bibinfo{author}{\bibfnamefont{A.~A.}
  \bibnamefont{{Kovalev}}}, \bibinfo{eid}{arXiv:1907.10567}
  (\bibinfo{year}{2019}).

\bibitem[{\citenamefont{Mook et~al.}(2019{\natexlab{b}})\citenamefont{Mook,
  Neumann, Henk, and Mertig}}]{PhysRevB.100.100401}
\bibinfo{author}{\bibfnamefont{A.}~\bibnamefont{Mook}},
  \bibinfo{author}{\bibfnamefont{R.~R.} \bibnamefont{Neumann}},
  \bibinfo{author}{\bibfnamefont{J.}~\bibnamefont{Henk}}, \bibnamefont{and}
  \bibinfo{author}{\bibfnamefont{I.}~\bibnamefont{Mertig}},
  \bibinfo{journal}{Phys. Rev. B} \textbf{\bibinfo{volume}{100}},
  \bibinfo{pages}{100401} (\bibinfo{year}{2019}{\natexlab{b}}).

\bibitem[{\citenamefont{Zhang et~al.}(2018)\citenamefont{Zhang, Okamoto, and
  Xiao}}]{PhysRevB.98.035424}
\bibinfo{author}{\bibfnamefont{Y.}~\bibnamefont{Zhang}},
  \bibinfo{author}{\bibfnamefont{S.}~\bibnamefont{Okamoto}}, \bibnamefont{and}
  \bibinfo{author}{\bibfnamefont{D.}~\bibnamefont{Xiao}},
  \bibinfo{journal}{Phys. Rev. B} \textbf{\bibinfo{volume}{98}},
  \bibinfo{pages}{035424} (\bibinfo{year}{2018}).

\bibitem[{\citenamefont{Zyuzin and Kovalev}(2018)}]{PhysRevB.97.174407}
\bibinfo{author}{\bibfnamefont{V.~A.} \bibnamefont{Zyuzin}} \bibnamefont{and}
  \bibinfo{author}{\bibfnamefont{A.~A.} \bibnamefont{Kovalev}},
  \bibinfo{journal}{Phys. Rev. B} \textbf{\bibinfo{volume}{97}},
  \bibinfo{pages}{174407} (\bibinfo{year}{2018}).

\bibitem[{\citenamefont{Mook et~al.}(2018)\citenamefont{Mook, G\"obel, Henk,
  and Mertig}}]{PhysRevB.97.140401}
\bibinfo{author}{\bibfnamefont{A.}~\bibnamefont{Mook}},
  \bibinfo{author}{\bibfnamefont{B.}~\bibnamefont{G\"obel}},
  \bibinfo{author}{\bibfnamefont{J.}~\bibnamefont{Henk}}, \bibnamefont{and}
  \bibinfo{author}{\bibfnamefont{I.}~\bibnamefont{Mertig}},
  \bibinfo{journal}{Phys. Rev. B} \textbf{\bibinfo{volume}{97}},
  \bibinfo{pages}{140401} (\bibinfo{year}{2018}).

\bibitem[{\citenamefont{Nakata et~al.}(2017)\citenamefont{Nakata, Kim,
  Klinovaja, and Loss}}]{PhysRevB.96.224414}
\bibinfo{author}{\bibfnamefont{K.}~\bibnamefont{Nakata}},
  \bibinfo{author}{\bibfnamefont{S.~K.} \bibnamefont{Kim}},
  \bibinfo{author}{\bibfnamefont{J.}~\bibnamefont{Klinovaja}},
  \bibnamefont{and} \bibinfo{author}{\bibfnamefont{D.}~\bibnamefont{Loss}},
  \bibinfo{journal}{Phys. Rev. B} \textbf{\bibinfo{volume}{96}},
  \bibinfo{pages}{224414} (\bibinfo{year}{2017}).

\bibitem[{\citenamefont{Daniels et~al.}(2019)\citenamefont{Daniels, Yu, Cheng,
  Xiao, and Xiao}}]{PhysRevB.99.224433}
\bibinfo{author}{\bibfnamefont{M.~W.} \bibnamefont{Daniels}},
  \bibinfo{author}{\bibfnamefont{W.}~\bibnamefont{Yu}},
  \bibinfo{author}{\bibfnamefont{R.}~\bibnamefont{Cheng}},
  \bibinfo{author}{\bibfnamefont{J.}~\bibnamefont{Xiao}}, \bibnamefont{and}
  \bibinfo{author}{\bibfnamefont{D.}~\bibnamefont{Xiao}},
  \bibinfo{journal}{Phys. Rev. B} \textbf{\bibinfo{volume}{99}},
  \bibinfo{pages}{224433} (\bibinfo{year}{2019}).

\bibitem[{\citenamefont{Luttinger}(1964)}]{PhysRev.135.A1505}
\bibinfo{author}{\bibfnamefont{J.~M.} \bibnamefont{Luttinger}},
  \bibinfo{journal}{Phys. Rev.} \textbf{\bibinfo{volume}{135}},
  \bibinfo{pages}{A1505} (\bibinfo{year}{1964}).

\bibitem[{\citenamefont{Holstein and Primakoff}(1940)}]{Holstein1940}
\bibinfo{author}{\bibfnamefont{T.}~\bibnamefont{Holstein}} \bibnamefont{and}
  \bibinfo{author}{\bibfnamefont{H.}~\bibnamefont{Primakoff}},
  \bibinfo{journal}{Phys. Rev.} \textbf{\bibinfo{volume}{58}},
  \bibinfo{pages}{1098} (\bibinfo{year}{1940}).

\bibitem[{\citenamefont{Psaroudaki et~al.}(2017)\citenamefont{Psaroudaki,
  Hoffman, Klinovaja, and Loss}}]{PhysRevX.7.041045}
\bibinfo{author}{\bibfnamefont{C.}~\bibnamefont{Psaroudaki}},
  \bibinfo{author}{\bibfnamefont{S.}~\bibnamefont{Hoffman}},
  \bibinfo{author}{\bibfnamefont{J.}~\bibnamefont{Klinovaja}},
  \bibnamefont{and} \bibinfo{author}{\bibfnamefont{D.}~\bibnamefont{Loss}},
  \bibinfo{journal}{Phys. Rev. X} \textbf{\bibinfo{volume}{7}},
  \bibinfo{pages}{041045} (\bibinfo{year}{2017}).

\bibitem[{\citenamefont{Shitade et~al.}(2019)\citenamefont{Shitade, Daido, and
  Yanase}}]{PhysRevB.99.024404}
\bibinfo{author}{\bibfnamefont{A.}~\bibnamefont{Shitade}},
  \bibinfo{author}{\bibfnamefont{A.}~\bibnamefont{Daido}}, \bibnamefont{and}
  \bibinfo{author}{\bibfnamefont{Y.}~\bibnamefont{Yanase}},
  \bibinfo{journal}{Phys. Rev. B} \textbf{\bibinfo{volume}{99}},
  \bibinfo{pages}{024404} (\bibinfo{year}{2019}).

\bibitem[{\citenamefont{Shi et~al.}(2007)\citenamefont{Shi, Vignale, Xiao, and
  Niu}}]{PhysRevLett.99.197202}
\bibinfo{author}{\bibfnamefont{J.}~\bibnamefont{Shi}},
  \bibinfo{author}{\bibfnamefont{G.}~\bibnamefont{Vignale}},
  \bibinfo{author}{\bibfnamefont{D.}~\bibnamefont{Xiao}}, \bibnamefont{and}
  \bibinfo{author}{\bibfnamefont{Q.}~\bibnamefont{Niu}},
  \bibinfo{journal}{Phys. Rev. Lett.} \textbf{\bibinfo{volume}{99}},
  \bibinfo{pages}{197202} (\bibinfo{year}{2007}).

\bibitem[{\citenamefont{Seemann et~al.}(2015)\citenamefont{Seemann,
  K\"odderitzsch, Wimmer, and Ebert}}]{PhysRevB.92.155138}
\bibinfo{author}{\bibfnamefont{M.}~\bibnamefont{Seemann}},
  \bibinfo{author}{\bibfnamefont{D.}~\bibnamefont{K\"odderitzsch}},
  \bibinfo{author}{\bibfnamefont{S.}~\bibnamefont{Wimmer}}, \bibnamefont{and}
  \bibinfo{author}{\bibfnamefont{H.}~\bibnamefont{Ebert}},
  \bibinfo{journal}{Phys. Rev. B} \textbf{\bibinfo{volume}{92}},
  \bibinfo{pages}{155138} (\bibinfo{year}{2015}).

\bibitem[{\citenamefont{\ifmmode~\check{Z}\else \v{Z}\fi{}elezn\'y
  et~al.}(2017)\citenamefont{\ifmmode~\check{Z}\else \v{Z}\fi{}elezn\'y, Gao,
  Manchon, Freimuth, Mokrousov, Zemen, Ma\ifmmode~\check{s}\else \v{s}\fi{}ek,
  Sinova, and Jungwirth}}]{PhysRevB.95.014403}
\bibinfo{author}{\bibfnamefont{J.}~\bibnamefont{\ifmmode~\check{Z}\else
  \v{Z}\fi{}elezn\'y}}, \bibinfo{author}{\bibfnamefont{H.}~\bibnamefont{Gao}},
  \bibinfo{author}{\bibfnamefont{A.}~\bibnamefont{Manchon}},
  \bibinfo{author}{\bibfnamefont{F.}~\bibnamefont{Freimuth}},
  \bibinfo{author}{\bibfnamefont{Y.}~\bibnamefont{Mokrousov}},
  \bibinfo{author}{\bibfnamefont{J.}~\bibnamefont{Zemen}},
  \bibinfo{author}{\bibfnamefont{J.}~\bibnamefont{Ma\ifmmode~\check{s}\else
  \v{s}\fi{}ek}}, \bibinfo{author}{\bibfnamefont{J.}~\bibnamefont{Sinova}},
  \bibnamefont{and}
  \bibinfo{author}{\bibfnamefont{T.}~\bibnamefont{Jungwirth}},
  \bibinfo{journal}{Phys. Rev. B} \textbf{\bibinfo{volume}{95}},
  \bibinfo{pages}{014403} (\bibinfo{year}{2017}).

\bibitem[{\citenamefont{Kimata et~al.}(2019)\citenamefont{Kimata, Chen, Kondou,
  Sugimoto, Muduli, Ikhlas, Omori, Tomita, MacDonald, Nakatsuji
  et~al.}}]{Kimata2019}
\bibinfo{author}{\bibfnamefont{M.}~\bibnamefont{Kimata}},
  \bibinfo{author}{\bibfnamefont{H.}~\bibnamefont{Chen}},
  \bibinfo{author}{\bibfnamefont{K.}~\bibnamefont{Kondou}},
  \bibinfo{author}{\bibfnamefont{S.}~\bibnamefont{Sugimoto}},
  \bibinfo{author}{\bibfnamefont{P.~K.} \bibnamefont{Muduli}},
  \bibinfo{author}{\bibfnamefont{M.}~\bibnamefont{Ikhlas}},
  \bibinfo{author}{\bibfnamefont{Y.}~\bibnamefont{Omori}},
  \bibinfo{author}{\bibfnamefont{T.}~\bibnamefont{Tomita}},
  \bibinfo{author}{\bibfnamefont{A.~H.} \bibnamefont{MacDonald}},
  \bibinfo{author}{\bibfnamefont{S.}~\bibnamefont{Nakatsuji}},
  \bibnamefont{et~al.}, \bibinfo{journal}{Nature}
  \textbf{\bibinfo{volume}{565}}, \bibinfo{pages}{627} (\bibinfo{year}{2019}).

\bibitem[{\citenamefont{Kamra et~al.}(2017)\citenamefont{Kamra, Agrawal, and
  Belzig}}]{PhysRevB.96.020411}
\bibinfo{author}{\bibfnamefont{A.}~\bibnamefont{Kamra}},
  \bibinfo{author}{\bibfnamefont{U.}~\bibnamefont{Agrawal}}, \bibnamefont{and}
  \bibinfo{author}{\bibfnamefont{W.}~\bibnamefont{Belzig}},
  \bibinfo{journal}{Phys. Rev. B} \textbf{\bibinfo{volume}{96}},
  \bibinfo{pages}{020411} (\bibinfo{year}{2017}).

\bibitem[{\citenamefont{Ritzmann}(2015)}]{Ritzmann2015Model}
\bibinfo{author}{\bibfnamefont{U.}~\bibnamefont{Ritzmann}}, Ph.D. thesis,
  \bibinfo{school}{University of Konstanz}, \bibinfo{address}{Konstanz}
  (\bibinfo{year}{2015}).

\bibitem[{\citenamefont{Hellman et~al.}(2017)\citenamefont{Hellman, Hoffmann,
  Tserkovnyak, Beach, Fullerton, Leighton, MacDonald, Ralph, Arena, D\"urr
  et~al.}}]{RevModPhys.89.025006}
\bibinfo{author}{\bibfnamefont{F.}~\bibnamefont{Hellman}},
  \bibinfo{author}{\bibfnamefont{A.}~\bibnamefont{Hoffmann}},
  \bibinfo{author}{\bibfnamefont{Y.}~\bibnamefont{Tserkovnyak}},
  \bibinfo{author}{\bibfnamefont{G.~S.~D.} \bibnamefont{Beach}},
  \bibinfo{author}{\bibfnamefont{E.~E.} \bibnamefont{Fullerton}},
  \bibinfo{author}{\bibfnamefont{C.}~\bibnamefont{Leighton}},
  \bibinfo{author}{\bibfnamefont{A.~H.} \bibnamefont{MacDonald}},
  \bibinfo{author}{\bibfnamefont{D.~C.} \bibnamefont{Ralph}},
  \bibinfo{author}{\bibfnamefont{D.~A.} \bibnamefont{Arena}},
  \bibinfo{author}{\bibfnamefont{H.~A.} \bibnamefont{D\"urr}},
  \bibnamefont{et~al.}, \bibinfo{journal}{Rev. Mod. Phys.}
  \textbf{\bibinfo{volume}{89}}, \bibinfo{pages}{025006}
  (\bibinfo{year}{2017}).

\bibitem[{\citenamefont{Chernyshev and
  Zhitomirsky}(2014)}]{PhysRevLett.113.237202}
\bibinfo{author}{\bibfnamefont{A.~L.} \bibnamefont{Chernyshev}}
  \bibnamefont{and} \bibinfo{author}{\bibfnamefont{M.~E.}
  \bibnamefont{Zhitomirsky}}, \bibinfo{journal}{Phys. Rev. Lett.}
  \textbf{\bibinfo{volume}{113}}, \bibinfo{pages}{237202}
  (\bibinfo{year}{2014}).

\bibitem[{\citenamefont{Chernyshev and Zhitomirsky}(2015)}]{PhysRevB.92.144415}
\bibinfo{author}{\bibfnamefont{A.~L.} \bibnamefont{Chernyshev}}
  \bibnamefont{and} \bibinfo{author}{\bibfnamefont{M.~E.}
  \bibnamefont{Zhitomirsky}}, \bibinfo{journal}{Phys. Rev. B}
  \textbf{\bibinfo{volume}{92}}, \bibinfo{pages}{144415}
  (\bibinfo{year}{2015}).

\bibitem[{\citenamefont{Okuma et~al.}(2017)\citenamefont{Okuma, Yajima,
  Nishio-Hamane, Okubo, and Hiroi}}]{PhysRevB.95.094427}
\bibinfo{author}{\bibfnamefont{R.}~\bibnamefont{Okuma}},
  \bibinfo{author}{\bibfnamefont{T.}~\bibnamefont{Yajima}},
  \bibinfo{author}{\bibfnamefont{D.}~\bibnamefont{Nishio-Hamane}},
  \bibinfo{author}{\bibfnamefont{T.}~\bibnamefont{Okubo}}, \bibnamefont{and}
  \bibinfo{author}{\bibfnamefont{Z.}~\bibnamefont{Hiroi}},
  \bibinfo{journal}{Phys. Rev. B} \textbf{\bibinfo{volume}{95}},
  \bibinfo{pages}{094427} (\bibinfo{year}{2017}).

\bibitem[{\citenamefont{Toth and Lake}(2015)}]{Toth_2015}
\bibinfo{author}{\bibfnamefont{S.}~\bibnamefont{Toth}} \bibnamefont{and}
  \bibinfo{author}{\bibfnamefont{B.}~\bibnamefont{Lake}}, \bibinfo{journal}{J.
  Phys. Condens. Matter} \textbf{\bibinfo{volume}{27}}, \bibinfo{pages}{166002}
  (\bibinfo{year}{2015}).

\bibitem[{\citenamefont{Matan et~al.}(2006)\citenamefont{Matan, Grohol, Nocera,
  Yildirim, Harris, Lee, Nagler, and Lee}}]{PhysRevLett.96.247201}
\bibinfo{author}{\bibfnamefont{K.}~\bibnamefont{Matan}},
  \bibinfo{author}{\bibfnamefont{D.}~\bibnamefont{Grohol}},
  \bibinfo{author}{\bibfnamefont{D.~G.} \bibnamefont{Nocera}},
  \bibinfo{author}{\bibfnamefont{T.}~\bibnamefont{Yildirim}},
  \bibinfo{author}{\bibfnamefont{A.~B.} \bibnamefont{Harris}},
  \bibinfo{author}{\bibfnamefont{S.~H.} \bibnamefont{Lee}},
  \bibinfo{author}{\bibfnamefont{S.~E.} \bibnamefont{Nagler}},
  \bibnamefont{and} \bibinfo{author}{\bibfnamefont{Y.~S.} \bibnamefont{Lee}},
  \bibinfo{journal}{Phys. Rev. Lett.} \textbf{\bibinfo{volume}{96}},
  \bibinfo{pages}{247201} (\bibinfo{year}{2006}).

\bibitem[{\citenamefont{Kim and Tserkovnyak}(2015)}]{PhysRevB.92.020410}
\bibinfo{author}{\bibfnamefont{S.~K.} \bibnamefont{Kim}} \bibnamefont{and}
  \bibinfo{author}{\bibfnamefont{Y.}~\bibnamefont{Tserkovnyak}},
  \bibinfo{journal}{Phys. Rev. B} \textbf{\bibinfo{volume}{92}},
  \bibinfo{pages}{020410} (\bibinfo{year}{2015}).

\bibitem[{\citenamefont{Li et~al.}(2016)\citenamefont{Li, Li, Kim, Balents, Yu,
  and Chen}}]{Li2016}
\bibinfo{author}{\bibfnamefont{F.-Y.} \bibnamefont{Li}},
  \bibinfo{author}{\bibfnamefont{Y.-D.} \bibnamefont{Li}},
  \bibinfo{author}{\bibfnamefont{Y.~B.} \bibnamefont{Kim}},
  \bibinfo{author}{\bibfnamefont{L.}~\bibnamefont{Balents}},
  \bibinfo{author}{\bibfnamefont{Y.}~\bibnamefont{Yu}}, \bibnamefont{and}
  \bibinfo{author}{\bibfnamefont{G.}~\bibnamefont{Chen}},
  \bibinfo{journal}{Nat.Commun.} \textbf{\bibinfo{volume}{7}},
  \bibinfo{pages}{12691} (\bibinfo{year}{2016}).

\bibitem[{\citenamefont{Li and Chen}(2018)}]{PhysRevB.98.045109}
\bibinfo{author}{\bibfnamefont{F.-Y.} \bibnamefont{Li}} \bibnamefont{and}
  \bibinfo{author}{\bibfnamefont{G.}~\bibnamefont{Chen}},
  \bibinfo{journal}{Phys. Rev. B} \textbf{\bibinfo{volume}{98}},
  \bibinfo{pages}{045109} (\bibinfo{year}{2018}).

\bibitem[{\citenamefont{Jian and Nie}(2018)}]{PhysRevB.97.115162}
\bibinfo{author}{\bibfnamefont{S.-K.} \bibnamefont{Jian}} \bibnamefont{and}
  \bibinfo{author}{\bibfnamefont{W.}~\bibnamefont{Nie}},
  \bibinfo{journal}{Phys. Rev. B} \textbf{\bibinfo{volume}{97}},
  \bibinfo{pages}{115162} (\bibinfo{year}{2018}).

\bibitem[{\citenamefont{{Hwang} et~al.}(2017)\citenamefont{{Hwang}, {Trivedi},
  and {Randeria}}}]{2017arXiv171208170H}
\bibinfo{author}{\bibfnamefont{K.}~\bibnamefont{{Hwang}}},
  \bibinfo{author}{\bibfnamefont{N.}~\bibnamefont{{Trivedi}}},
  \bibnamefont{and}
  \bibinfo{author}{\bibfnamefont{M.}~\bibnamefont{{Randeria}}},
  \bibinfo{eid}{arXiv:1712.08170} (\bibinfo{year}{2017}).

\bibitem[{\citenamefont{Okamoto et~al.}(2013)\citenamefont{Okamoto, Nilsen,
  Attfield, and Hiroi}}]{PhysRevLett.110.097203}
\bibinfo{author}{\bibfnamefont{Y.}~\bibnamefont{Okamoto}},
  \bibinfo{author}{\bibfnamefont{G.~J.} \bibnamefont{Nilsen}},
  \bibinfo{author}{\bibfnamefont{J.~P.} \bibnamefont{Attfield}},
  \bibnamefont{and} \bibinfo{author}{\bibfnamefont{Z.}~\bibnamefont{Hiroi}},
  \bibinfo{journal}{Phys. Rev. Lett.} \textbf{\bibinfo{volume}{110}},
  \bibinfo{pages}{097203} (\bibinfo{year}{2013}).

\bibitem[{\citenamefont{Casola et~al.}(2018)\citenamefont{Casola, van~der Sar,
  and Yacoby}}]{pub.1100161418}
\bibinfo{author}{\bibfnamefont{F.}~\bibnamefont{Casola}},
  \bibinfo{author}{\bibfnamefont{T.}~\bibnamefont{van~der Sar}},
  \bibnamefont{and} \bibinfo{author}{\bibfnamefont{A.}~\bibnamefont{Yacoby}},
  \bibinfo{journal}{Nat. Rev. Mater.} \textbf{\bibinfo{volume}{3}},
  \bibinfo{pages}{17088} (\bibinfo{year}{2018}).

\bibitem[{\citenamefont{Brown}(1963)}]{Brown1963}
\bibinfo{author}{\bibfnamefont{W.~F.} \bibnamefont{Brown}},
  \bibinfo{journal}{Phys. Rev.} \textbf{\bibinfo{volume}{130}},
  \bibinfo{pages}{1677} (\bibinfo{year}{1963}).

\bibitem[{\citenamefont{Evans et~al.}(2014)\citenamefont{Evans, Fan,
  Chureemart, Ostler, Ellis, and Chantrell}}]{evans2014atomistic}
\bibinfo{author}{\bibfnamefont{R.~F.~L.} \bibnamefont{Evans}},
  \bibinfo{author}{\bibfnamefont{W.~J.} \bibnamefont{Fan}},
  \bibinfo{author}{\bibfnamefont{P.}~\bibnamefont{Chureemart}},
  \bibinfo{author}{\bibfnamefont{T.~A.} \bibnamefont{Ostler}},
  \bibinfo{author}{\bibfnamefont{M.~O.~A.} \bibnamefont{Ellis}},
  \bibnamefont{and} \bibinfo{author}{\bibfnamefont{R.~W.}
  \bibnamefont{Chantrell}}, \bibinfo{journal}{J. Phys. Condens. Matter}
  \textbf{\bibinfo{volume}{26}}, \bibinfo{pages}{103202}
  (\bibinfo{year}{2014}).

\bibitem[{\citenamefont{Barker and Bauer}(2016)}]{Barker2016}
\bibinfo{author}{\bibfnamefont{J.}~\bibnamefont{Barker}} \bibnamefont{and}
  \bibinfo{author}{\bibfnamefont{G.~E.~W.} \bibnamefont{Bauer}},
  \bibinfo{journal}{Phys. Rev. Lett.} \textbf{\bibinfo{volume}{117}},
  \bibinfo{pages}{217201} (\bibinfo{year}{2016}).

\bibitem[{\citenamefont{Inosov}(2018)}]{doi:10.1080/00018732.2018.1571986}
\bibinfo{author}{\bibfnamefont{D.}~\bibnamefont{Inosov}},
  \bibinfo{journal}{Adv. Phys.} \textbf{\bibinfo{volume}{67}},
  \bibinfo{pages}{149} (\bibinfo{year}{2018}).

\bibitem[{\citenamefont{Chernyshev and Zhitomirsky}(2009)}]{PhysRevB.79.144416}
\bibinfo{author}{\bibfnamefont{A.~L.} \bibnamefont{Chernyshev}}
  \bibnamefont{and} \bibinfo{author}{\bibfnamefont{M.~E.}
  \bibnamefont{Zhitomirsky}}, \bibinfo{journal}{Phys. Rev. B}
  \textbf{\bibinfo{volume}{79}}, \bibinfo{pages}{144416}
  (\bibinfo{year}{2009}).

\bibitem[{\citenamefont{Gvozdikova et~al.}(2011)\citenamefont{Gvozdikova,
  Melchy, and Zhitomirsky}}]{Gvozdikova_2011}
\bibinfo{author}{\bibfnamefont{M.~V.} \bibnamefont{Gvozdikova}},
  \bibinfo{author}{\bibfnamefont{P.-E.} \bibnamefont{Melchy}},
  \bibnamefont{and} \bibinfo{author}{\bibfnamefont{M.~E.}
  \bibnamefont{Zhitomirsky}}, \bibinfo{journal}{J. Phys. Condens. Matter}
  \textbf{\bibinfo{volume}{23}}, \bibinfo{pages}{164209}
  (\bibinfo{year}{2011}).

\bibitem[{\citenamefont{Svistov et~al.}(2003)\citenamefont{Svistov, Smirnov,
  Prozorova, Petrenko, Demianets, and Shapiro}}]{PhysRevB.67.094434}
\bibinfo{author}{\bibfnamefont{L.~E.} \bibnamefont{Svistov}},
  \bibinfo{author}{\bibfnamefont{A.~I.} \bibnamefont{Smirnov}},
  \bibinfo{author}{\bibfnamefont{L.~A.} \bibnamefont{Prozorova}},
  \bibinfo{author}{\bibfnamefont{O.~A.} \bibnamefont{Petrenko}},
  \bibinfo{author}{\bibfnamefont{L.~N.} \bibnamefont{Demianets}},
  \bibnamefont{and} \bibinfo{author}{\bibfnamefont{A.~Y.}
  \bibnamefont{Shapiro}}, \bibinfo{journal}{Phys. Rev. B}
  \textbf{\bibinfo{volume}{67}}, \bibinfo{pages}{094434}
  (\bibinfo{year}{2003}).

\bibitem[{\citenamefont{Hwang et~al.}(2012)\citenamefont{Hwang, Choi, Ye,
  Dela~Cruz, Xin, Zhou, and Schlottmann}}]{PhysRevLett.109.257205}
\bibinfo{author}{\bibfnamefont{J.}~\bibnamefont{Hwang}},
  \bibinfo{author}{\bibfnamefont{E.~S.} \bibnamefont{Choi}},
  \bibinfo{author}{\bibfnamefont{F.}~\bibnamefont{Ye}},
  \bibinfo{author}{\bibfnamefont{C.~R.} \bibnamefont{Dela~Cruz}},
  \bibinfo{author}{\bibfnamefont{Y.}~\bibnamefont{Xin}},
  \bibinfo{author}{\bibfnamefont{H.~D.} \bibnamefont{Zhou}}, \bibnamefont{and}
  \bibinfo{author}{\bibfnamefont{P.}~\bibnamefont{Schlottmann}},
  \bibinfo{journal}{Phys. Rev. Lett.} \textbf{\bibinfo{volume}{109}},
  \bibinfo{pages}{257205} (\bibinfo{year}{2012}).

\bibitem[{\citenamefont{Rau et~al.}(2016)\citenamefont{Rau, Wu, May, Poudel,
  Winn, Garlea, Huq, Whitfield, Taylor, Lumsden
  et~al.}}]{PhysRevLett.116.257204}
\bibinfo{author}{\bibfnamefont{J.~G.} \bibnamefont{Rau}},
  \bibinfo{author}{\bibfnamefont{L.~S.} \bibnamefont{Wu}},
  \bibinfo{author}{\bibfnamefont{A.~F.} \bibnamefont{May}},
  \bibinfo{author}{\bibfnamefont{L.}~\bibnamefont{Poudel}},
  \bibinfo{author}{\bibfnamefont{B.}~\bibnamefont{Winn}},
  \bibinfo{author}{\bibfnamefont{V.~O.} \bibnamefont{Garlea}},
  \bibinfo{author}{\bibfnamefont{A.}~\bibnamefont{Huq}},
  \bibinfo{author}{\bibfnamefont{P.}~\bibnamefont{Whitfield}},
  \bibinfo{author}{\bibfnamefont{A.~E.} \bibnamefont{Taylor}},
  \bibinfo{author}{\bibfnamefont{M.~D.} \bibnamefont{Lumsden}},
  \bibnamefont{et~al.}, \bibinfo{journal}{Phys. Rev. Lett.}
  \textbf{\bibinfo{volume}{116}}, \bibinfo{pages}{257204}
  (\bibinfo{year}{2016}).

\bibitem[{\citenamefont{Haku et~al.}(2016)\citenamefont{Haku, Kimura,
  Matsumoto, Soda, Sera, Yu, Mole, Takeuchi, Nakatsuji, Kono
  et~al.}}]{PhysRevB.93.220407}
\bibinfo{author}{\bibfnamefont{T.}~\bibnamefont{Haku}},
  \bibinfo{author}{\bibfnamefont{K.}~\bibnamefont{Kimura}},
  \bibinfo{author}{\bibfnamefont{Y.}~\bibnamefont{Matsumoto}},
  \bibinfo{author}{\bibfnamefont{M.}~\bibnamefont{Soda}},
  \bibinfo{author}{\bibfnamefont{M.}~\bibnamefont{Sera}},
  \bibinfo{author}{\bibfnamefont{D.}~\bibnamefont{Yu}},
  \bibinfo{author}{\bibfnamefont{R.~A.} \bibnamefont{Mole}},
  \bibinfo{author}{\bibfnamefont{T.}~\bibnamefont{Takeuchi}},
  \bibinfo{author}{\bibfnamefont{S.}~\bibnamefont{Nakatsuji}},
  \bibinfo{author}{\bibfnamefont{Y.}~\bibnamefont{Kono}}, \bibnamefont{et~al.},
  \bibinfo{journal}{Phys. Rev. B} \textbf{\bibinfo{volume}{93}},
  \bibinfo{pages}{220407} (\bibinfo{year}{2016}).

\bibitem[{\citenamefont{Tanaka et~al.}(2014)\citenamefont{Tanaka, Yoshida,
  Takigawa, Okamoto, and Hiroi}}]{PhysRevLett.113.227204}
\bibinfo{author}{\bibfnamefont{Y.}~\bibnamefont{Tanaka}},
  \bibinfo{author}{\bibfnamefont{M.}~\bibnamefont{Yoshida}},
  \bibinfo{author}{\bibfnamefont{M.}~\bibnamefont{Takigawa}},
  \bibinfo{author}{\bibfnamefont{Y.}~\bibnamefont{Okamoto}}, \bibnamefont{and}
  \bibinfo{author}{\bibfnamefont{Z.}~\bibnamefont{Hiroi}},
  \bibinfo{journal}{Phys. Rev. Lett.} \textbf{\bibinfo{volume}{113}},
  \bibinfo{pages}{227204} (\bibinfo{year}{2014}).

\end{thebibliography}

\end{document}